\documentclass[a4paper,fleqn]{cas-dc}

\usepackage[numbers, sort&compress]{natbib}
\usepackage[english]{babel}
\usepackage{amsmath}
\usepackage{amssymb}
\usepackage{graphicx}
\usepackage[nolist,nohyperlinks]{acronym}
\usepackage{booktabs}
\usepackage{arydshln}
\usepackage{hyperref}
\usepackage{xcolor}
\usepackage{supertabular, booktabs}

\def\tsc#1{\csdef{#1}{\textsc{\lowercase{#1}}\xspace}}
\tsc{WGM}
\tsc{QE}

\begin{document}

\let\WriteBookmarks\relax
\def\floatpagepagefraction{1}
\def\textpagefraction{.001}

\shorttitle{NS Measurements in the Multi-messenger Era}

\shortauthors{Ascenzi, Graber \& Rea}

\title [mode = title]{Neutron-star Measurements in the Multi-messenger Era}

\author[inst1, inst2, inst3]{Stefano Ascenzi}[orcid=0000-0001-5116-6789]
\cormark[1]  
\ead{stefano.ascenzi@gssi.it}

\author[inst1, inst2]{Vanessa Graber}[orcid=0000-0002-6558-1681]
\cormark[1]  
\ead{graber@ice.csic.es}

\author[inst1, inst2]{Nanda Rea}[orcid=0000-0003-2177-6388]
\ead{rea@ice.csic.es}

\affiliation[inst1]{organization={Institute of Space Sciences (ICE, CSIC)},
            addressline={Campus UAB, Carrer de Can Magrans s/n},
            city={Barcelona},
            postcode={08193},
            country={Spain}}

\affiliation[inst2]{organization={Institut d’Estudis Espacials de Catalunya (IEEC)},
            addressline={Carrer Gran Capita 2-4},
            city={Barcelona},
            postcode={08034},
            country={Spain}}

\affiliation[inst3]{organization={Gran Sasso Science Institute (GSSI)},
            addressline={Viale F. Crispi 7},
            city={L'Aquila},
            postcode={67100},
            country={Italy}}


\begin{abstract}
Neutron stars are compact and dense celestial objects that offer the unique opportunity to explore matter and its interactions under conditions that cannot be reproduced elsewhere in the Universe. Their extreme gravitational, rotational and magnetic energy reservoirs fuel the large variety of their emission, which encompasses all available multi-messenger tracers: electromagnetic and gravitational waves, neutrinos, and cosmic rays. However, accurately measuring global neutron-star properties such as mass, radius, and moment of inertia poses significant challenges. Probing internal characteristics such as the crustal composition or superfluid physics is even more complex. This article provides a comprehensive review of the different methods employed to measure neutron-star characteristics and the level of reliance on theoretical models. Understanding these measurement techniques is crucial for advancing our knowledge of neutron-star physics. We also highlight the importance of employing independent methods and adopting a multi-messenger approach to gather complementary data from various observable phenomena as exemplified by the recent breakthroughs in gravitational-wave astronomy and the landmark detection of a binary neutron-star merger. Consolidating the current state of knowledge on neutron-star measurements will enable an accurate interpretation of the current data and errors, and better planning for future observations and experiments.
\end{abstract}


\begin{keywords}
dense matter \sep equation of state \sep gravitational waves \sep multi-messenger astronomy \sep neutron stars \sep pulsars 
\end{keywords}


\maketitle


\begin{acronym}
\acrodef{NS}[NS]{neutron star}
\acrodef{BNS}[BNS]{binary neutron star}
\acrodef{GW}[GW]{gravitational wave}
\acrodef{EM}[EM]{electromagnetic}
\acrodef{WD}[WD]{white dwarf}
\acrodef{GR}[GR]{general relativity}
\acrodef{LOS}[LOS]{line of sight}
\acrodef{EOS}[EOS]{equation of state}
\acrodef{CBM}[CBM]{compact binary merger}
\acrodef{SNR}[SNR]{signal-to-noise ratio}
\acrodef{LMXB}[LMXB]{low-mass X-ray binary}
\acrodefplural{LMXB}[LMXBs]{low-mass X-ray binaries}
\acrodef{HMXB}[HMXB]{high-mass X-ray binary}
\acrodefplural{HMXB}[HMXBs]{high-mass X-ray binaries}
\acrodef{CCO}[CCO]{Central Compact Object}
\acrodef{BH}[BH]{black hole}
\acrodef{ISCO}[ISCO]{innermost stable circular orbit}
\acrodef{GRB}[GRB]{gamma-ray burst}
\acrodef{PRE}[PRE]{photospheric radius expansion}
\acrodef{MSP}[MSP]{millisecond pulsar}
\acrodef{NICER}[NICER]{Neutron Star Interior Composition Explorer}
\acrodef{KN}[KN]{kilonova}
\acrodefplural{KN}[KNe]{kilonovae}
\acrodef{RRAT}[RRAT]{Rotating Radio Transient}
\acrodef{XDINS}[XDINS]{X-ray Dim Isolated Neutron Star}
\acrodef{SuNR}[SuNR]{supernova remnant}
\acrodef{QPO}[QPO]{quasi-periodic oscillation}
\acrodef{CRSF}[CRSF]{cyclotron resonance scattering feature}
\acrodef{CFS}[CFS]{Chandrasekhar-Friedman-Schutz}
\end{acronym}


\tableofcontents


\section{Introduction}

\Acp{NS} are remarkable astrophysical objects for a number of reasons. First, they are incredibly compact, containing a mass of around $1-2\,M_\odot$ within a radius of approximately $10-12\,$km. As a result, spacetime within and around these objects is highly curved, often necessitating the use of a strong-gravity framework to model associated phenomena. Additionally, matter inside \acp{NS} exists under extreme densities, surpassing the nuclear saturation density, $\rho_0 \approx 2.8 \times 10^{14} \, {\rm g} \, {\rm cm}^{-3}$ (the density of atomic nuclei) in their cores. This unique feature makes \acp{NS} natural laboratories for investigating matter and strong interactions under conditions that cannot be replicated in controlled terrestrial environments. Furthermore, \acp{NS} possess the strongest magnetic fields in the Universe with surface field strengths ranging between $10^8\,$ to $10^{15}\,$G. This allows us to also study matter in the ultrahigh magnetic-field regime. Finally, \acp{NS} are fast and incredibly stable rotators with rotation periods ranging approximately between $1\,$ms to $10\,$s and some objects rivaling the stability of atomic clocks on Earth.

To employ \acp{NS} as cosmic laboratories, it is necessary to measure their macroscopic properties (such as the mass, radius and moment of inertia) and internal features (e.g., the composition of \ac{NS} crusts and superfluidity). These quantities provide us with a wealth of information about the phenomena occurring in the stars' interior or in their proximity, allowing us to constrain physics under extreme conditions. However, measuring these properties is inherently challenging, because they cannot be observed directly. Observable quantities often depend on several of these parameters, leading to significant uncertainties and degeneracies in their inferred values. Additionally, the methods used to measure these parameters typically rely on a range of assumptions, introducing varying degrees of model dependence. Therefore, it is of fundamental importance to develop independent methods that target given observables but are based on different assumptions. This will ultimately allow us to improve our knowledge of \acp{NS} by comparing and cross-correlating distinct independent measures.

To achieve this objective, a multi-wavelength approach is crucial. By conducting observations across the electromagnetic spectrum, we can explore the diverse population of \acp{NS} and gather complementary information about different types of sources. Furthermore, the detection of the first \ac{BNS} merger by the LIGO and Virgo advanced interferometers \citep{GW170817} marked the emergence of \acp{GW} as a new messenger for astrophysical studies. The accompanying observation of a short \ac{GRB} and a \ac{KN} \citep{GW170817_multimessenger_2017} reigned in a new era of multi-messenger astronomy, which holds great promise for revolutionizing the study of \acp{NS}.

The aim of this review is to provide an overview of the different methods employed to measure \ac{NS} properties. As fully covering the extensive literature on this topic is beyond the scope of this review, we instead highlight those approaches that we consider most relevant and promising in constraining uncertain \ac{NS} physics. In particular, we distinguish the messenger and wavelength regime of a certain measurement and assess underlying model dependencies and systematics. This way, we hope to provide the reader with a schematic overview of the different techniques used to study these extreme objects and how reliable these approaches are. We also point out that we purposefully do not aim to summarize current constraints on \ac{NS} observables. We only mention corresponding estimates sporadically, where we see appropriate, and provide ample references for the interested reader to dive deeper, instead putting our primary focus on the methodology itself.

The review is organized as follows: In Sec.~\ref{sec:state-of-the-art}, we summarize the state of the art of our \ac{NS} knowledge, focusing in particular on the description of the different subpopulations of \acp{NS} (Sec.~\ref{sec:NS-zoo}) and our current understanding of their internal structure (Sec.~\ref{sec:ns-structure}). In Sec.~\ref{sec:ns-measurements}, we discuss the different methods used to measure \ac{NS} parameters. We focus specifically on the mass (Sec.~\ref{sec:mass}), radius (Sec.~\ref{sec:radius}), moment of inertia (Sec.~\ref{sec:MoI}), tidal deformability (Sec.~\ref{sec:tidal_def}), compactness (Sec.~\ref{sec:compactness}), magnetic field (Sec.~\ref{sec:Bfields}), crustal physics (Sec.~\ref{sec:crust_prop}), superfluidity (Sec.~\ref{sec:sf}) and then touch on a few additional measurements (Sec.~\ref{sec:other_measurements}). Finally, in Sec.~\ref{sec:conclusion}, we provide a summary and discuss conclusions.


\section{State of the art}
\label{sec:state-of-the-art}

In this section, we summarize our current knowledge of \acp{NS}. We start by illustrating the different populations of \acp{NS}, describing their most important characteristics. We then review our current understanding of the internal \ac{NS} structure. Our aim here is not to provide a comprehensive and detailed description of the topic but rather introduce the reader to the general context required to better understand the following sections. For those interested in delving deeper, we refer to dedicated reviews that provide more comprehensive information on the subject matter.


\subsection{Neutron-star diversity}
\label{sec:NS-zoo}

The idea of a \ac{NS} was first proposed just a few years after the neutron’s discovery in 1932.\footnote{It is worth mentioning that there is evidence that Lev D. Landau wrote his work on dense stars \citep{Landau1932} about one year before the discovery of the neutron and delayed submission of the article for unknown reasons \citep{Yakovlev2013}.} Several physicists suggested that the supernova explosion of a massive star could signal a transition from a normal star to a compact, highly magnetic, rapidly rotating star composed mainly (but not solely) of neutrons \citep{Landau1932, BaadeZwicky1934}. It took some 30 years to confirm this notion, when in 1967 Jocelyn Bell Burnell and Antony Hewish discovered the first \ac{NS} by spotting its beamed radio emission as a repeating signal, the first \textit{pulsar}, in their radio telescope \citep{Hewish-etal1968}. Since then, many more \acp{NS} have been discovered and to date more than three thousand are known in different systems and environments: isolated or in binary systems with a large variety of companions, in double \ac{NS} systems, in the Galactic disk, in Globular clusters, and in nearby Galaxies \citep{KaspiKramer2016}.

\begin{figure}
    \centering
    \includegraphics[width = 0.48\textwidth]{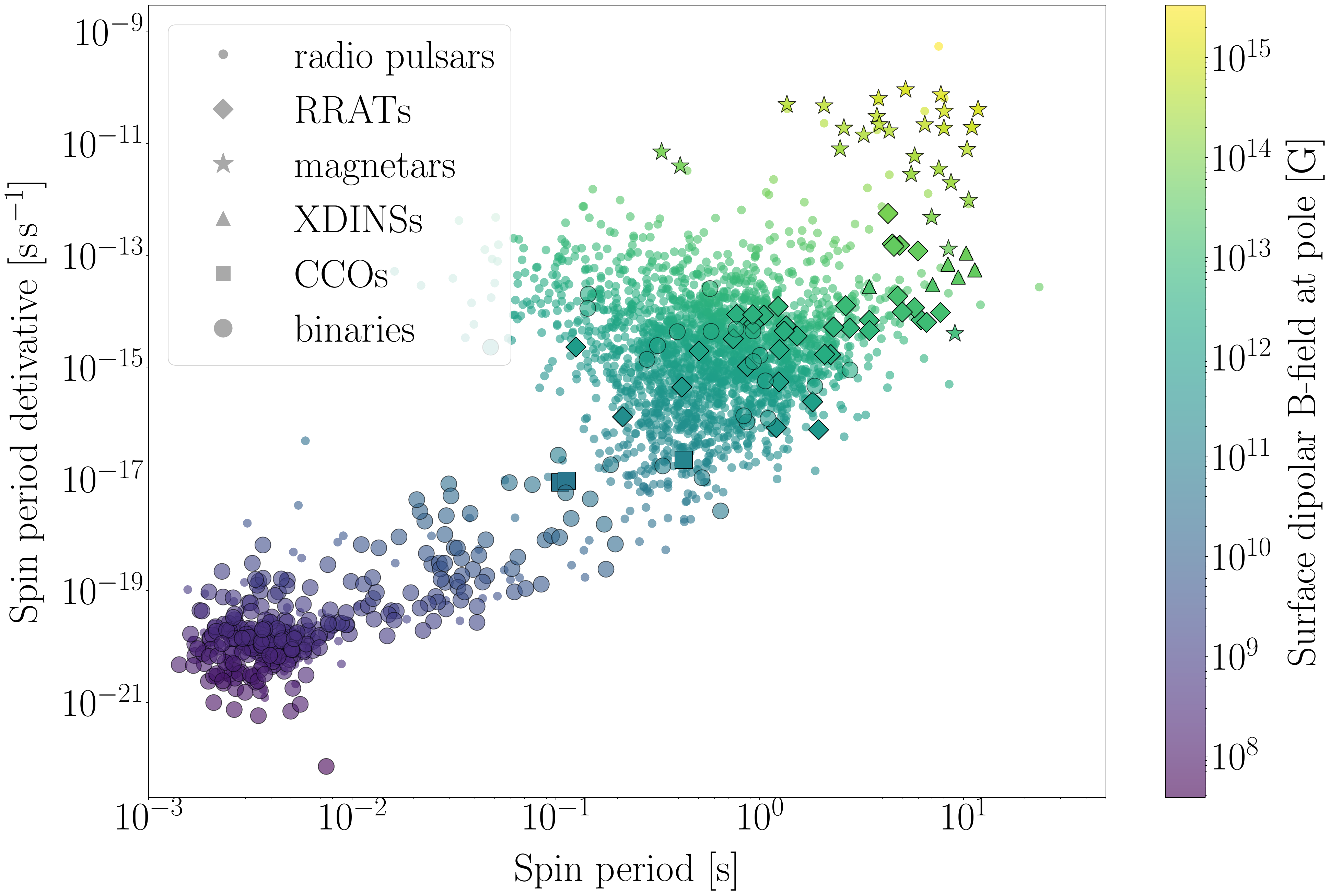}
    \caption{Period, $P$, vs period derivative, $\dot{P}$, plane for known pulsars. The bulk of the \ac{NS} population is composed of radio pulsars given as dots. We also show Rotating Radio Transients (\acp{RRAT}) as diamonds, highly magnetized magnetars as stars, X-ray Dim Isolated Neutron Stars (\acp{XDINS}) as triangles, Central Compact Objects (\acp{CCO}) as squares and spin-down driven \acp{NS} in binaries as circles. Note, however, that this figure does not include accreting X-ray binaries. Colors as indicated by the color bar on the right reflect the magnetic-field strength at the stellar pole inferred from $P$ and $\dot{P}$ measurements under the assumption of a dipolar configuration (see Sec.~\ref{sec:MD_braking} for details). Data are taken from the ATNF catalogue \citep{Manchester-etal2005, ATNF-catalogue}, the RRAT catalogue \citep{RRAT-catalogue}, the McGill Online Magnetar Catalogue \citep{OlausenKaspi2014, McGill-catalogue} and the Magnetar Outburst Online Catalog \citep{CotiZelati-etal2018, mag_outburst-catalogue}.}    \label{fig:p_pdot}
\end{figure}

In Fig.~\ref{fig:p_pdot}, we present the $P-\dot{P}$ diagram of the known pulsar population, where $P$ denotes the spin period of a \ac{NS} and $\dot{P}$ its time derivative obtained through pulsar timing. Knowledge of $P$ and $\dot{P}$, in turn, allows an estimate of the \ac{NS}'s so-called \textit{characteristic age}, $\tau_{\rm c} = P/2\dot{P}$, if the star spins down due to magnetic dipole radiation, and the surface dipolar magnetic-field strength, $B_{\rm dip}$, (see Sec.~\ref{sec:MD_braking} for details). Although the observed \ac{NS} population is dominated by radio pulsars, several extreme and puzzling sub-classes of \acp{NS} have been discovered in the last decades. These include \acp{RRAT}, magnetars, \acp{XDINS}, \acp{CCO}, and the more common X-ray \acp{NS} in binary systems accreting material from a low-mass or high-mass companion star. Despite being presumably governed by a single \ac{EOS}, \acp{NS} manifest a surprising observational diversity, which we are only just starting to understand. Moreover, the Galactic core-collapse supernova rate is unable to accommodate the formation of these different \ac{NS} classes individually \citep{KeaneKramer2008}, suggesting evolutionary links between them \citep{Kaspi2010, Vigano-etal2013, Popov2023}. We briefly summarize these different classes here.

{\bf Rotation-powered pulsars}: With thousands of objects (see dots in Fig.~\ref{fig:p_pdot}), the largest \ac{NS} class is powered by their rotational energy reservoirs, $E_{\rm rot}$, with braking due to their dipolar magnetic fields (Sec.~\ref{sec:MD_braking}) supplying luminosities of the order of $L_{\rm rot} \sim |\dot{E}_{\rm rot}| = 4\pi^2 I_{\rm NS} \dot{P} / P^{3} \approx 3.95 \times 10^{31} \, {\rm erg} \, {\rm s}^{-1}$ for a fiducial \ac{NS} moment of inertia, $I_{\rm NS}$, of $10^{45} \, {\rm g} \, {\rm cm}^2$, $P = 1 \, {\rm s}$ and $\dot{P} = 10^{-15} \, {\rm s} \, {\rm s}^{-1}$. Although emitting pulsed emission across the electromagnetic spectrum, these sources are typically observed in the radio band (see \citep{LorimerKramer2012} for a comprehensive overview). The key ingredient to activate this emission is the acceleration of charged particles, which are extracted from the star’s surface by an electrical voltage gap \citep{GoldreichJulian1969, RudermanSutherland1975}. Moreover, all isolated pulsar periods increase with time, implying a decay in their spin frequencies. Specifically, pulsars are born with fast rotation and high magnetic fields (top left in the $P - \dot{P}$ diagram~\ref{fig:p_pdot}) and evolve towards slower rotation and lower magnetic fields (bottom right). The exact trajectory and timescale differs from pulsar to pulsar and depends on their birth properties and details of the magnetic-field evolution \citep{Graber-etal2023}.

{\bf Rotating Radio Transients}: \acp{RRAT} were discovered as bright ($0.1 - 3\,$Jy), short ($2 - 30\,$ms) radio bursts that recurred randomly about every $4\,$min$\,-\,3\,$hr \citep{McLaughlin-etal2006}. A study of the arrival times of these bursts led to the discovery of common periodicities, which are interpreted as the rotational periods of underlying pulsars. Instead of constituting their own class, \acp{RRAT} are now considered an extreme form of rotation powered-pulsars that exhibit extended periods of so-called \textit{nulling}, i.e., long phases where the regular emission is switched off are interspersed with irregular, sporadic emission of radio pulses (see \citep{Keane-etal2011} for a review). Although it has been estimated that \acp{RRAT} are as numerous as the radio pulsar population \citep{McLaughlin-etal2006, KeaneKramer2008}, these sources are much harder to detect and classify because of their irregularity. \citep{RRAT-catalogue}, e.g., present a catalog of 115 \acp{RRAT} of which two thirds have $P$ and one third $\dot{P}$ measurements (see diamonds in Fig.~\ref{fig:p_pdot}).

{\bf Magnetars}: These few dozen, young objects (see stars in Fig.~\ref{fig:p_pdot}) with spin periods between $0.3-12\,$s and inferred dipole magnetic field strengths of the order of $10^{14} - 10^{15}\,$G are the strongest magnetized \acp{NS} (see \citep{Turolla-etal2015, KaspiBeloborodov2017} for two recent reviews). Magnetars (comprising the old Anomalous X-ray Pulsars and Soft Gamma Repeaters classes) were originally discovered in the 1970s (and initially misidentified as \acp{GRB}) via their powerful flares and outburst which release large amounts of energy, i.e., $10^{40} - 10^{47}$erg, over a wide range of timescales from fraction of seconds to years (see \citep{ReaEsposito2011} for a review). However, magnetars are now also known to be powerful steady high-energy emitters with luminosities of the order of $10^{34} - 10^{36}\, {\rm erg} \, {\rm s}^{-1}$. Because this phenomenology cannot be powered by the stars' rotational energy alone and no companion stars have been found, magnetar emission is generally associated with the decay and instabilities of their strong magnetic fields \citep{DuncanThompson1992, ThompsonDuncan1993, Thompson-etal2002}.

{\bf X-ray Dim Isolated Neutron Stars}: The \acp{XDINS} are a small group of seven nearby (hundreds of parsecs), thermally emitting, isolated \acp{NS}. Six of these have measured $P$ and $\dot{P}$ values (see triangles in Fig.~\ref{fig:p_pdot} and \citep{Haberl2007, Turolla2009} for reviews). They are radio quiet while relatively bright in the X-ray band with luminosities in the range $10^{31} - 10^{32}\, {\rm erg} \, {\rm s}^{-1}$ (suggesting ages of $\sim 0.5\,$Myr), and have spin periods similar to magnetars albeit with slightly lower magnetic fields on the order of $10^{13}\,$G (which are however systematically larger than those of rotation-powered pulsars). Different to most other X-ray emitting pulsars, the spectra of \acp{XDINS} are well approximated as black bodies with $k_{\rm B}T_{\rm BB} \sim 40 - 100\,$eV ($k_{\rm B}$ is the Boltzmann constant and $T_{\rm BB}$ the black-body temperature), which are superimposed with broad absorption features (see Sec.~\ref{sec:linespec-compactness}).

{\bf Central Compact Objects}: \acp{CCO} consist of radio quiet, thermally emitting X-ray sources with $k_{\rm B} T_{\rm BB} \sim 0.2-0.5\,$keV, which were discovered due to their locations at the centers of luminous shell-like \acp{SuNR} (see \citep{Pavlov-etal2004, deLuca2008} for reviews). This association implies ages of at most a few tens of kyrs. Three \acp{CCO} show X-ray pulsations with periods in the $0.1-0.4\,$s range (see squares in Fig.~\ref{fig:p_pdot}), but have very low spin-down rates despite their young ages. The corresponding rotational energy reservoirs are insufficient to produce the observed X-ray emission. Their low inferred dipolar magnetic fields between $\sim10^{10}-10^{11}\,$G, and the presence of thermally emitting hotspots are difficult to explain at the same time. This could be reconciled by strong magnetic fields buried in the interiors of CCOs following a phase of strong fallback accretion after the supernova \citep{ViganoPons2012}. This is supported by the fact that the characteristic ages, $\tau_{\rm c}$, of the three pulsed CCOs are orders of magnitude larger than the ages obtained from the \ac{SuNR} association. However, the exact nature of these sources remains uncertain.

{\bf Accreting X-ray binaries}: \acp{NS} in binary systems undergoing matter accretion from the companion star's wind or via an accretion disk have been well known since the 1970s \citep[e.g.,][]{Giacconi1974}. These are mostly observed as bright X-ray sources with luminosities typically proportional to the mass accretion rate. For a general review see \citep{Done-etal2007}. In particular, we know of accreting \acp{NS} with high-mass companions (aka \acp{HMXB}; around 300 objects known to date \citep{Fornasini-etal2023}) and those with low-mass companions (aka \acp{LMXB}; around 350 known to date \citep{Avakyan-etal2023}). Specifically, \acp{HMXB} are young systems ($\sim0.1-10\,$Myr) characterized by \ac{NS} magnetic fields of $\sim10^{12}-10^{13}$\,G, mostly accreting via wind and in wide, eccentric orbits. Their spin periods range between a few to thousands of seconds. On the other hand, \acp{LMXB} are older systems ($\sim0.1-1\,$Gyr) with a fast spinning \ac{NS} (about $1-100\,$ms), low magnetic fields of $\sim10^{8}-10^{10}\,$G and tight orbits that typically accrete via Roche-lobe overflow and form large accretion disks. After this accretion phase which leads to a significant \ac{NS} spin-up, binary pulsars often restart their rotation-powered emission with a short (recycled) spin period (see bottom-left objects depicted as circles in Fig.~\ref{fig:p_pdot}).


\begin{figure*}
    \centering
    \includegraphics[width = 0.85\textwidth]{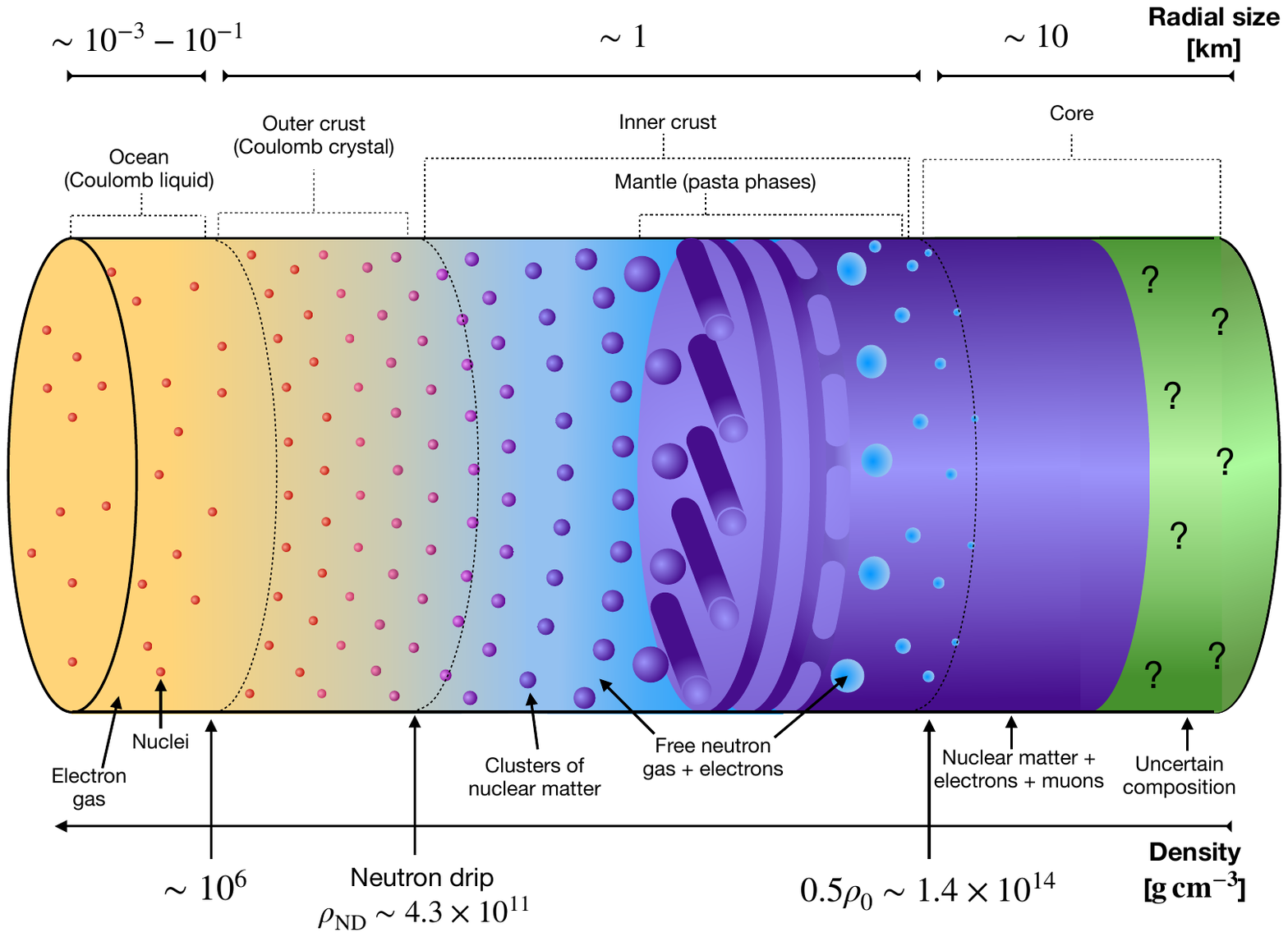}
    \caption{Artistic representation of the internal structure of a \ac{NS}. We highlight the different \ac{NS} layers as a function of the radial size (\textit{top axis}) and density (\textit{bottom axis}). The variation of color from red to purple in the crustal nuclei/nuclear-matter clusters represents an increase of neutron richness.}
    \label{fig:ns_structure}
\end{figure*}

\subsection{Neutron-star structure}
\label{sec:ns-structure}

In this section, we provide a brief overview of the current understanding of the internal \ac{NS} structure and composition, which are summarized in the sketch in Fig.~\ref{fig:ns_structure}.

Starting from the outside, the outermost \ac{NS} layer is the gaseous \emph{atmosphere} (not represented in Fig.~\ref{fig:ns_structure}) with a thickness of $\mathcal{O}(10\, \textrm{cm})$, which encloses the \emph{ocean} or \emph{envelope}. Depending on the temperature, this ocean has a characteristic depth of $\mathcal{O}(1-100 \, \textrm{m})$ \citep{Potekhin-etal2015} and is composed either of heavy elements (e.g., iron) or light elements (provided by accretion) in a Coulomb liquid state. If the magnetic field is strong and the temperature sufficiently low, a solid skin can form on top of the liquid phase (see, e.g., Fig.~10 of \citep{Potekhin-etal2015}).

Below the ocean lies a solid \emph{crust} with a total thickness of $\mathcal{O}(1 \, \textrm{km})$. The crust is divided into an \emph{outer crust} and an \emph{inner crust}. In the former, the atoms are fully ionized due to the intense pressure. The electrons form a degenerate relativistic gas, while the nuclei are organized in a Coulomb lattice. With increasing density, the nuclei become more and more neutron-rich due to electron captures. At the \emph{neutron-drip density}, $\rho_{\rm ND} \approx 4.3 \times 10^{11} \, \textrm{g} \, {\rm cm}^{-3}$, neutrons start dripping out from the nuclei, marking the transition to the \emph{inner crust}. Here, the nuclei are so neutron-rich that they are more appropriately referred to as `clusters' of nuclear matter. These spherical clusters form a lattice immersed in the gas of dripped neutrons. If the temperature is low enough, this neutron gas can experience a phase transition to a superfluid state (see Sec.~\ref{sec:sf}). Moving deeper into the star, the density increases and the clusters grow and move closer to each other. Eventually, at the bottom of the inner crust, clusters cannot retain their spherical shape and merge into cylindrical structures of nuclear matter known as \emph{spaghetti}. At even higher densities, the spaghetti merge into slab-like structures called \emph{lasagne}. Moving deeper, lasagne merges, forming a matrix of nuclear matter confining the neutron gas, first in hollow cylinders and then into bubble-like structures. Although the presence of this so-called \textit{nuclear pasta} has a negligible influence on the structure of the star, its anisotropic matter organization has a big impact on the \ac{NS} transport properties (see Secs.~\ref{sec:PPdot_cutoff} and \ref{sec:X-rayburst_cooling_crustprop}). The region at the base of the inner crust characterized by the presence of nuclear pasta is sometimes also referred to as the \emph{mantle}.

At densities of around $0.5 \rho_0$, we transition from the inner crust to a state of pure nuclear matter, which marks the beginning of the \ac{NS} \emph{core}. The core has a radial size of around $10\,$km and contains the majority of the \ac{NS} mass. This strongly degenerate liquid phase is composed primarily of neutrons with a small fraction of protons and electrons (and at higher densities also muons) to ensure charge neutrality. For \acp{NS} that are sufficiently cold, these core neutrons and protons may also undergo a superfluid and a superconducting transition, respectively (see Sec.~\ref{sec:sf}). While densities at the center of \acp{NS} could reach up to $10^{15}\, {\rm g} \, {\rm cm}^{-3}$, our understanding of many-body physics becomes more uncertain at densities above $\rho_0$ and the \ac{NS} composition, thus, unclear. However, at such extreme densities, the \textit{inner core} of \acp{NS} could contain new particle species such as hyperons or even transition to more exotic states of matter like a Bose-Einstein condensate of pions and/or kaons or a quark-gluon plasma.

Assuming spherical symmetry, the mechanical structure of a \ac{NS} is described by the Tolman-Oppenheimer-Volkoff equation \citep{OppenheimerVolkoff1939, Tolman1939}, which derives from the Einstein field equations of \ac{GR} and the continuity of the stress-energy tensor. To solve this equation, we require knowledge of the functional relation between the pressure and the density, i.e. $P(\rho)$, a relation known as the \ac{EOS}. An \ac{EOS} generally relates three thermodynamic quantities, such as pressure, density, and temperature. In most applications concerning \acp{NS}, the temperature dependence can be neglected because it is much lower than its degeneracy limit.\footnote{Exceptions are astrophysical situations in which the temperature is very high such as during \ac{NS} formation in supernovae or \ac{NS} mergers.} Such an \ac{EOS} is known as \emph{barotropic}. Roughly speaking, the \ac{EOS} describes how matter reacts under compression: \emph{Stiff} \acp{EOS} reflect those cases where an increase in density results in a significant increase in pressure, while \emph{soft} \acp{EOS} refer to the opposite case. The \ac{NS} \ac{EOS} depends both on the nuclear interactions at extremely high densities, but also on the matter composition. At present the \ac{EOS} of the outer crust is well established, while that of the inner crust and, in particular, the core \ac{EOS} are much more uncertain, as we cannot test matter under such extreme conditions on Earth.

\begin{figure}
    \centering
    \includegraphics[width = 0.48\textwidth]{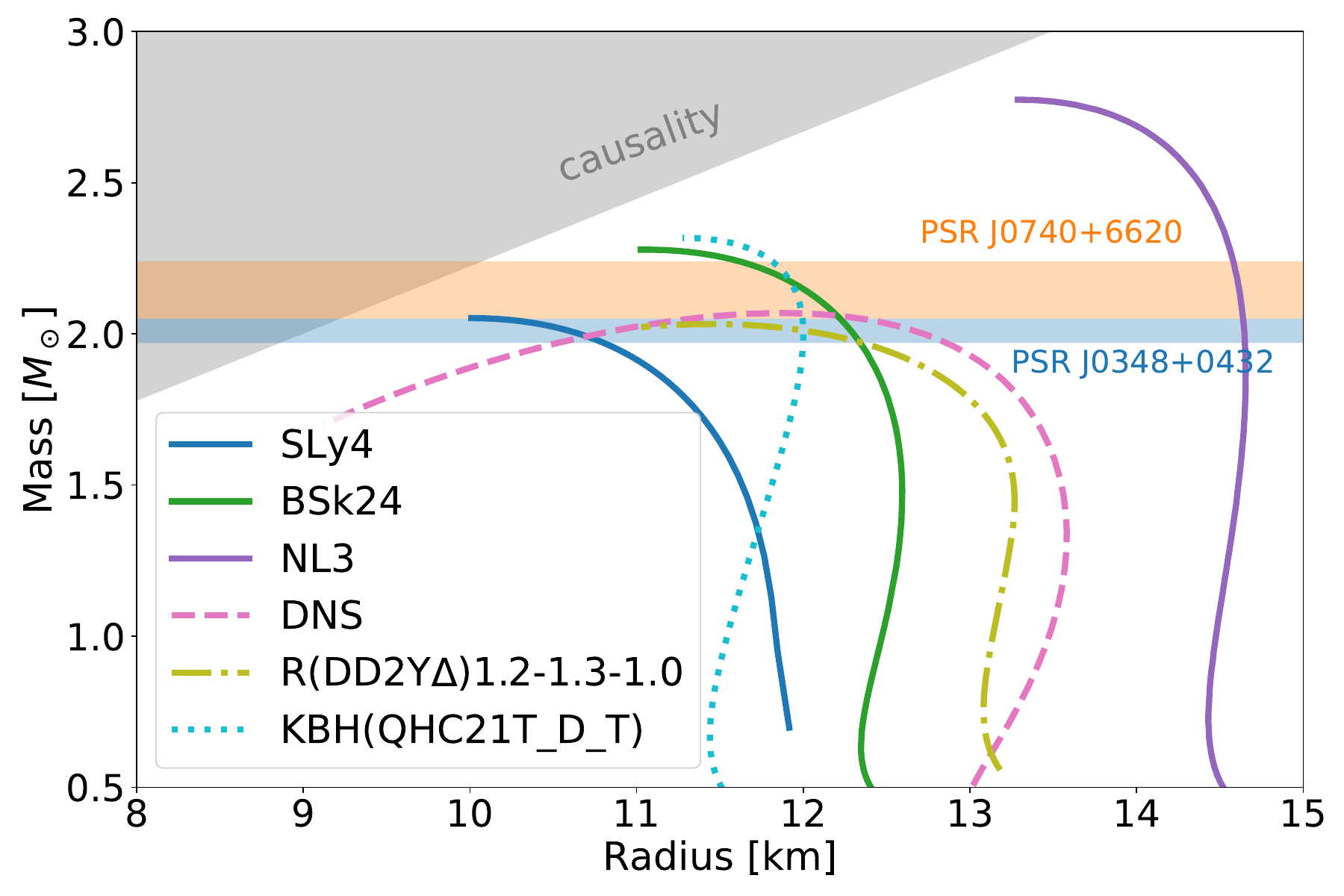}
    \caption{Mass-radius curves for a broad range of dense-matter \acp{EOS}. The solid lines show three nuclear \acp{EOS}: SLy4 \citep{Chabanat1998, DanielewiczLee2009, GulminelliRaduta2015} (blue), BSk24 \citep{Goriely2013, Pearson2018, Pearson2019, Perot2019, EOS-Potekhin} (green), and NL3 \citep{Lalazissis1997, Xia2022a, Xia2022b} (purple). The dashed line represents DNS \citep{DexheimerSchramm2008, Dexheimer2015, Dexheimer2017} (pink), containing hyperons. The dashed-dotted line shows the \ac{EOS} R(DD2Y$\Delta$) \citep{Typel2010, Raduta2020, Vinas2021} (light green), which includes hyperons and $\Delta$ resonances, while the dotted line represents KBH(QHC21T$\_$D$\_$T) \citep{Togashi2017, Kojo2022} (light blue), characterizing \acp{NS} with quark matter cores. Measurements for two heavy radio pulsars, PSR J0740+662 \citep{Fonseca2021} (orange) and PSR J0348+042 \citep{Antoniadis2013} (blue), are highlighted as horizontal bands. We also show the lower limit causality places on the \ac{NS} radius as a gray shaded region in the top-left corner. \ac{EOS} data are taken from the CompOSE database \citep{EOS-Compose}.}
    \label{fig:MR_plane}
\end{figure}

However, the dense-matter \ac{EOS}, and in particular the \ac{EOS} in the core, influences macroscopic properties of \acp{NS}, such as the star's mass, radius, moment of inertia and tidal deformability. By measuring these quantities, it is, hence, possible to constrain the \ac{EOS}. For example, the \ac{EOS} determines the maximum mass a \ac{NS} can attain, with stiff \acp{EOS} allowing for higher maximum masses. This way, measuring large \ac{NS} masses allows us to rule out the softest \acp{EOS}. Similarly, each \ac{EOS} leads to a unique relation between the mass and the stellar radius \citep{Lindblom1992}, or its moment of inertia \citep{Lattimer2005}. As an example, mass-radius relations for a range of potential nuclear \acp{EOS} are illustrated in Fig.~\ref{fig:MR_plane}. Thus, measuring two of these quantities for one or more objects allows additional \ac{EOS} constraints. Moreover, additional \ac{NS} observables are sensitive to specific internal physics, which can help us further probe the properties of dense nuclear matter, such as superfluid physics. This way, \acp{NS} can serve as cosmic laboratories to test interaction and characteristics of matter at conditions that are not reproducible on the Earth. In the following, we discuss how such measurements are achieved.

We refer those readers interested in learning more about the structure and composition of \acp{NS} to \citep{Chamel2008} for a detailed review on the composition of the crust, to \citep{Potekhin-etal2015} for a review of \ac{NS} transport properties, and \citep{LattimerPrakash2001, LattimerPrakash2016} for reviews on the \ac{NS} structure and the dense-matter \ac{EOS}.

\section{Neutron-star measurements}
\label{sec:ns-measurements}

We now turn to discussing the various methods employed to extract different \ac{NS} parameters. Each subsection is dedicated to one stellar property and details the different techniques to measure it. We summarize these techniques in a table at the beginning of each subsection and use three colors to represent the method's degree of model dependency. \emph{Green} identifies those approaches that are (relatively) model-independent, giving the most robust measurement of \ac{NS} parameters. We use \emph{yellow} to indicate methods that are somewhat model dependent and/or where the measurement of a quantity is correlated with other parameters, introducing significant uncertainty in the measured value. Finally, we use \emph{orange} to classify those methods that are strongly model dependent. These methods typically rely on a range of assumptions that are either qualitative or not always fully justifiable, but nonetheless useful in modeling \ac{NS} behavior. However, corresponding results and conclusions must be taken with caution.

We highlight that our color classification of the various methods is not fully rigorous but retains an inevitable degree of subjectiveness. However, we do provide justifications for our choice in all cases to help the reader understand our assessments.


\subsection{Mass}
\label{sec:mass}

\begin{table*}[]
    \centering
    \begin{tabular}{|c|c|c|c|c|c|c|}
        \hline
        \multicolumn{2}{|c|}{\textbf{Method}} & \textbf{GWs} & \textbf{Radio} & \textbf{Optical}  & \textbf{X-ray} & $\mathbf{\gamma}$\textbf{-ray}  \\
        \hline
        \hline
        \multirow{6}{*}{\hyperref[sec:mass_function_in_binaries]{Binary mass function}} & 1st MF  &  & \cellcolor{yellow!40} & \cellcolor{yellow!40} & &     \\
        \cline{2-7}
        & + 2nd MF &  & \cellcolor{yellow!40} & \cellcolor{yellow!40} & &     \\
        \cline{2-7}
        & + 2nd MF + optical modeling &  & & \cellcolor{orange!40} &  &     \\
        \cline{2-7}
        & + 2nd MF + eclipses &  &  &  & \cellcolor{yellow!40} & \cellcolor{yellow!40}    \\
        \cline{2-7}
        & + 1 PK parameter &  & \cellcolor{yellow!40} &  &  &     \\
        \cline{2-7}
        & + 2 PK parameters &  & \cellcolor{green!40} &  &  &     \\
        \hline
        \multicolumn{2}{|c|}{\hyperref[sec:shapiro_delay]{Shapiro delay}} &  & \cellcolor{green!40} &  & &     \\
        \hline
        \multicolumn{2}{|c|}{\hyperref[sec:gw_from_cbm]{GW in compact binary mergers}} & \cellcolor{yellow!40} & &  & &    \\
        \hline
        \multicolumn{2}{|c|}{\hyperref[sec:gw_asteroseismology_mass]{GW asteroseismology}} & \cellcolor{yellow!40} & &  & &    \\
        \hline
        \multicolumn{2}{|c|}{\hyperref[sec:ns_cooling]{Cooling}} & & & & \cellcolor{orange!40}& \\
        \hline
        \multicolumn{2}{|c|}{\hyperref[sec:gw_lensing]{Lensing in eclipsing binaries}} &  &  & \cellcolor{orange!40} & &    \\
        \hline
     \end{tabular}
    \caption{Mass measurements. Individual rows represent the methods that provide \ac{NS} masses, while columns denote different messengers/ \ac{EM} wavebands. Colored cells represent the messengers/wavelengths at which the methods operate. \textit{Green} characterizes those techniques that are (almost) model independent, \textit{yellow} those that are somewhat model dependent and/or affected by degeneracy, and \textit{orange} those methods that are strongly model dependent. Note that we split the binary-mass-function approach into different subversions, where MF = mass function and PK = post-Keplerian. All techniques are discussed in detail in the text.}
    \label{tab:mass}
\end{table*}

The first parameter of interest is the \ac{NS} mass, $M_{\rm NS}$. The mass is particularly important because its maximum value is closely connected to the \ac{EOS} of dense matter, i.e., large masses exclude those \acp{EOS} that are not sufficiently stiff to account for that measurement. Fig.~\ref{fig:MR_plane} shows the mass-radius plane and mass constraints for two radio pulsars. Moreover, combining a measured mass with a radius or moment-of-inertia measurement allows us to pin down specific \acp{EOS}.

As we discuss below, \acp{NS} in binary systems constitute the best targets to provide precise and model-independent measures for this quantity, while observables from isolated compact objects generally provide more model-dependent mass estimates.


\subsubsection{Binary mass function}
\label{sec:mass_function_in_binaries}

For \acp{NS} in binary systems, we can exploit Newtonian dynamics to constrain their masses. Using Kepler's third law, we can define the \emph{binary mass function} as \citep[e.g.,][]{Miller2021}:
\begin{equation}
    f_{\rm bin} (M_{\rm c}, M_{\rm NS}) 
        \equiv \frac{M^3_{\rm NS} \sin^3 i}{(M_{\rm c} + M_{\rm NS})^2}
    	=  \frac{K^3_{\rm c} P_{\rm orb}}{2\pi G}(1-e^2)^{3/2},
    \label{eq:mass_function}
\end{equation}
where $M_{\rm c}$ and $M_{\rm NS}$ are the mass of the binary companion and the \ac{NS}, respectively, $i$ is the angle between the \ac{LOS} and the system's orbital angular momentum (i.e., the \textit{inclination angle}), $K_{\rm c}$ is the radial velocity of the companion, $P_{\rm orb}$ and $e$ are the period and the eccentricity of the orbit, respectively, and $G$ denotes the gravitational constant. It is worth noting that the right-hand side of Eq.~\eqref{eq:mass_function} contains observable quantities, which can be measured by radio timing if the companion is a radio pulsar or by optical spectroscopy if the companion is a non-compact star or a \ac{WD}. Measuring $f_{\rm bin}$, thus, allows us to constrain $M_{\rm NS}$ (specifically obtain a lower limit) but not to determine it explicitly because of the degeneracy with $M_{\rm c}$ and the binary inclination, $i$.

The degeneracy with respect to $M_{\rm c}$ can be broken if the \ac{NS}'s radial velocity, $K_{\rm NS}$, (or equivalently the projection of the semi-major axis of the \ac{NS}'s orbit onto the observer plane) is also measured. This allows for the determination of an additional mass function, $f_{\mathrm{bin}}(M_\mathrm{NS},M_\mathrm{c})$, (namely Eq.~\eqref{eq:mass_function} after exchanging the $\mathrm{NS}$ and $\mathrm{c}$ subscripts, referred to as the 2nd mass function in Tab.~\ref{tab:mass}). The ratio of these two mass functions allows us to obtain the mass ratio, $M_{\rm c}/M_{\rm NS}$, which is equal to $K_{\rm NS}/K_{\rm c}$ following Eq.~\eqref{eq:mass_function}, and represents a further constraint for our system. We classify both these mass-function approaches as yellow in Tab.~\ref{tab:mass}.

The degeneracy with respect to the inclination angle can be resolved by measuring the so-called Shapiro delay (see Sec.~\ref{sec:shapiro_delay}), modeling of the optical lightcurve in case of a non-degenerate companion \citep{Shahbaz1998, Kerkwijk2011, Linares2018, Yap2019, Kennedy2020, Kennedy2022} or by the observation of eclipses \citep{LeachRuffini1973, Chanan1976, Rawls2011, Jonker2013, Strader2016, Clark2023}. Prime targets for the latter two kinds of studies are so-called `\emph{spiders}', \acp{MSP} (i.e., those \acp{NS} with millisecond periods, which are located in the bottom left of the $P - \dot{P}$ plane in Fig.~\ref{fig:p_pdot}) in compact binaries with $P_\mathrm{orb}\lesssim 1\, \mathrm{day}$ and orbital separation $a \sim R_\odot$, whose relativistic winds are strong enough to heat and ablate matter from their companion \citep{vandenHeuvelvanParadijs1988, Fruchter1988, Archibald2009, Roberts2013, Linares2014}. Those systems with a companion mass $M_{\rm c} = \mathcal{O}(0.01\, M_\odot)$ are called \emph{black widows} \citep[e.g.,][]{Fruchter1988}, while those with a companion mass $M_{\rm c} = \mathcal{O}(0.1\, M_\odot)$ are referred to as \emph{redbacks} \citep[e.g.,][]{Archibald2009} (see also \citep{Linares2014} for a systematic study of redbacks).

Optical lightcurve modeling exploits the fact that in a sufficiently compact binary, the companion star will be tidally deformed and heated on one side by the irradiation from the \ac{NS}. Both effects impact the optical lightcurve in an inclination-dependent way. Consequently, accurate modeling of the lightcurve allows us to break the degeneracy with $\sin i$. However, the problem with this method resides in its strong model dependency, which may introduce systematic biases due to the incompleteness of heating models \citep{Romani2015, RomaniSanchez2016, Clark2023}. For this reason, we classify the determination of the inclination through optical modeling as orange in Tab.~\ref{tab:mass}.

Nonetheless, spiders are particularly interesting targets for this kind of measurement as the \ac{NS} companion is often optically bright. This brightness allows for the determination of the companion's radial velocity via optical spectroscopy, while the \ac{NS}'s radial velocity can be obtained through pulsar timing. However, care has to be taken when measuring the companion velocity, since irradiation from the \ac{NS} heats one side of the star, shifting the center of light away from the companion's center of mass towards the binary's center of mass. If not corrected for, this effect causes the measured velocity to underestimate the true value, leading to an underestimation of the pulsar mass (see Eq.~\eqref{eq:mass_function} and \citep{Clark2023}).

In contrast, the eclipse of the \ac{NS} emission by its companion or vice versa points towards a system observed with an edge-on orientation, such that $\sin i \sim 1$, providing a less model-dependent inclination constraint. The eclipse signal is typically searched for in the X-ray and gamma-ray bands, because those energies are less affected by absorption and scattering from the inter-binary diffuse material. We, however, note that X-ray eclipses in spiders can be masked by the X-ray emission from inter-binary shocks \citep{Wadiasingh2018}, which are generated by encounters of the pulsar wind and that of the companion \citep{Clark2023}. Thus, spider systems are primarily targeted in the gamma-rays. Studying 42 spiders (plus seven redback candidates), \citep{Clark2023} for example recently identified five eclipsing systems (plus two among the candidates) and excluded the presence of gamma-ray eclipse in 32 of them. These observations led to constraints on the \ac{NS} masses in eclipsing binaries and the identification of lower \ac{NS} mass limits in non-eclipsing systems, an approach which we classify as yellow in Tab.~\ref{tab:mass}.

We also note that such measurements hint at the fact that spiders are higher in mass than other \acp{NS} in binary systems \citep{Strader2019, Linares2020}, consequently allowing for interesting constraints on the \ac{EOS}-dependent \ac{NS} maximum mass. For example, the highest masses in eclipsing systems found by \citep{Clark2023} are associated with the black widow B1957+20 \citep{Fruchter1988}, with $M_{\mathrm{NS}} = 1.67-1.94\, M_\odot$, and the redback J1816+4510, with $M_{\mathrm{NS}} =1.64-2.17\, M_\odot$ \citep{Kaplan2013}. The former mass estimate is lower than what was previously inferred from optical modeling for B1957+20 (i.e., $M_\mathrm{NS}=2.4\pm 0.1\,M_\odot$ \citep{vanKerkwijk2011}), highlighting the importance of complementary mass measurements and the quantification of underlying systematics.

Other means to break the mass-function degeneracy is by measuring two or more post-Keplerian parameters (that capture deviations from Keplerian motion as expected in the theory of \ac{GR}) through radio pulsar timing. According to \ac{GR}, post-Keplerian parameters depend on the system's mass and inclination. \Acp{MSP} are particularly promising for measuring post-Keplerian effects, as their rotational stability provides highly stable pulsar timing solutions. Besides Shapiro delay, which we dedicate the next section to, other useful post-Keplerian parameters are the periastron advance, $\dot{\omega}$ (which is typically the easiest to measure \citep{LorimerKramer2012}), the rate of orbital decay due to \ac{GW} emission, $\dot{P}_{\rm orb}$, and the Einstein delay, $\gamma$, due to time dilation and the gravitational redshift. These are defined as \citep[e.g.,][]{Miller2021}:
\begin{align}
    \dot{\omega} &= 3 \left(\frac{P_{\rm orb}}{2\pi}\right)^{-5/3} \left(1-e^2 \right)^{-1} \left(T_\odot M \right)^{2/3},
    	\label{eqn:periastron_advance} \\[1.3ex]
    \dot{P}_{\rm orb} &= - \frac{192\pi}{5} \left(\frac{P_{\rm orb}}{2\pi}\right)^{-5/3} d(e) \,T^{5/3}_\odot \frac{M_{\rm c} M_{\rm NS}}{M^{1/3}}, \\
    \gamma &= \left(\frac{P_{\rm orb}}{2\pi}\right)^{1/3} e T^{2/3}_\odot \frac{M_{\rm c}(M_{\rm NS}+ 2M_{\rm c})}{M^{4/3}},
\end{align}
where $M \equiv M_{\rm c} + M_{\rm NS}$ is the total mass of the system, measured in units of $M_\odot$. Moreover, $T_\odot \equiv G M_\odot/c^3 = 4.92549 \, \mu$s (where $c$ denotes the speed of light), and $d(e) \equiv (1 + 73e^2/24 + 37e^4/96)(1-e^2)^{-7/2}$.

The above formulae depend on \ac{GR} only, a theory whose predictions have been validated to subpercent level in several double \ac{NS} systems that contain pulsars \citep[e.g.,][]{Weisberg-etal2010, Berti-etal2015, Kramer-etal2021}. Consequently, measuring the mass function plus two (or more) of these post-Keplerian parameters provides a relatively model-independent way to measure the mass of a \ac{NS} (see classification in Tab.~\ref{tab:mass}).

We conclude by stressing that the relative uncertainty of these post-Keplerian parameters decreases with increasing observing time \citep{DamourTaylor1992}. On the contrary, a larger orbital period tends to increase the fractional uncertainties of most Keplerian parameters. For the exact scalings of various fractional uncertainties with the observation time span and the orbital period, we refer the interested reader to Tab.~2 of \citep{DamourTaylor1992} or Tab.~8.2 of \citep{LorimerKramer2012}.


\subsubsection{Shapiro delay}
\label{sec:shapiro_delay}

\begin{figure*}
    \centering
    \includegraphics[width=\textwidth]{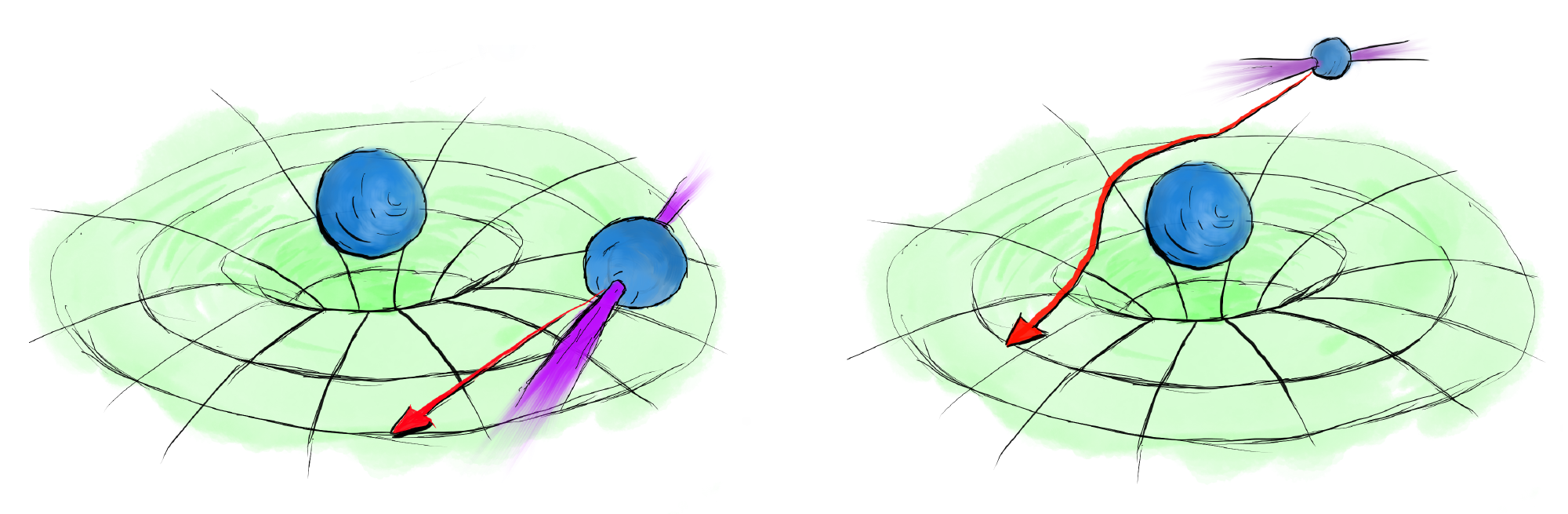}
    \caption{Artistic representation of the Shapiro delay in a binary \ac{NS} system. While the signal from a pulsar is primarily unaltered by its \ac{NS} companion in inferior conjunction (\textit{left}), the signal is affected by the companion's curvature introducing a time delay when the pulsar is in superior conjunction (\textit{right}).}
    \label{fig:shapiro_delay}
\end{figure*}

Besides the above post-Keplerian parameters, Shapiro delay provides a straightforward and precise way to measure \ac{NS} masses in a model-independent manner. Shapiro delay is a gravitational effect predicted by \ac{GR} that occurs when a signal-carrying messenger (e.g., a photon) passes near a massive object. As the messenger traverses the curved spacetime around the object, its trajectory is altered, resulting in an increased time of flight and subsequent delay in the arrival of the signal \citep{Shapiro64}. This delay is influenced by the value of $\sin i$ and the spacetime curvature, which in turn is determined primarily by the object's mass. A graphical representation of this effect is illustrated in Fig.~\ref{fig:shapiro_delay}.

Measurement of the Shapiro delay is particular effective to determine the masses of \acp{NS} in \ac{BNS} systems, where at least one of the \acp{NS} is a radio pulsar, whose radio beam passes closely by its companion during a specific orbital phase. By precisely measuring the arrival times of the radio pulses, the Shapiro delay can be accurately determined, enabling the estimation of the mass of the pulsar's companion \citep{Fonseca2016, Shamohammadi2023}. Nevertheless, the effect can also be used to determine the masses of pulsars and their companions in other kind of systems, such as those with a \ac{WD} companion \citep{RybaTaylor1991, Jacobi2003, Jacoby2005, Ferdman2010, Kaplan2014, Liu2020} (see Eqs.~\eqref{eq:Shapiro_range} and \eqref{eq:Shapiro_shape} for details).

From an observational point of view, once a model for the arrival times of the radio pulses (in absence of Shapiro delay) is defined, the Shapiro delay appears as a characteristic cusp in the residuals at superior conjunction. At this point, the pulsar is located behind its companion relative to the observer and the signal, thus, traverses the largest amount of spacetime curvature. The timing model accounting for this effect can then be modeled by two parameters, $r$ and $s$, called the Shapiro delay \emph{range} and \emph{shape}, respectively. These are functions of the companion mass, $M_{\rm c}$, and the system's inclination angle, $i$, respectively, \citep{DamourDeruelle1985, DamourDeruelle1986, LorimerKramer2012}:
\begin{align}
    r &\equiv T_\odot M_{\rm c}, \label{eq:Shapiro_range} \\[1.4ex]
    s &\equiv \sin i \nonumber \\
      & = T^{-1/3}_\odot \left(\frac{P_{\rm orb}}{2\pi}\right)^{-2/3}
      x \, \frac{ \left(M_{\rm c} + M_{\rm NS} \right)^{2/3}} {M_{\rm c}}
        \label{eq:Shapiro_shape},
\end{align}
where $M_{\rm c, NS}$ are measured in solar masses and the second equality in Eq.~\eqref{eq:Shapiro_shape} was obtained from Eq.~\eqref{eq:mass_function}\footnote{Note that we use the mass function $f_{\rm bin}(M_{\rm NS}, M_{\rm c})$ here that has been obtained from the measurement of $K_{\rm NS}$ instead of $K_{\rm c}$.} after substituting $K_{\rm NS} = a_1 \sin i (2\pi/P_{\rm orb}) (1 - e^2)^{-1/2}$. Here, $a_1$ is the orbit's semi-major axis, and we define the projected semi-major axis as $x \equiv a_1 \sin i/c$. We, however, note that the parametrization of the timing solution in terms of $r$ and $s$ is not unique \citep[e.g.,][]{FreireWex2010}.

Due to the possibility of timing radio pulsars, and the simplicity of the effect which ultimately depends only on \ac{GR}, Shapiro delay presents a precise and model-independent way to constrain the mass of a \ac{NS}. It is easiest to detect in binaries that contain heavy companions, because $M_{\rm c}$ directly affects the amplitude of the Shapiro delay, and \acp{MSP}, which can be observed with microsecond-level timing precision or better. Consequently, this approach has played a pivotal role in detecting several \acp{NS} with masses above $2 \, M_\odot$ \citep{Demorest2010, Fonseca2016, Cromartie2020}, leading to groundbreaking results in the field and the tightest constraints on the dense-matter \ac{EOS}. We stress, though, that this effect is observable primarily in systems that have a favorable orientation with respect to the observer, namely those that are close to edge-on ($\sin i \sim 1$).


\subsubsection{Gravitational waves from compact binary mergers}
\label{sec:gw_from_cbm}

Binary systems lose their orbital energy due to the emission of \acp{GW}. If a system is sufficiently massive and compact, this energy loss causes an observable shrinking of the orbit, eventually leading to the objects' merger. This can occur in binaries formed by two compact objects, such as \acp{NS}.

Such a \ac{CBM} can be formally divided into three different phases: an inspiral, a merger, and a ringdown phase. In particular, the inspiral is the phase in which the two objects are not yet in contact, and, with the exception of the final orbits, can be treated as point masses as their finite sizes play a negligible role in the dynamics. In this phase, the system's evolution is simple: the system emits \acp{GW} at a frequency, $f_{\rm GW}$, that is twice the Keplerian orbital frequency, and with an amplitude proportional to $f^{2/3}_{\rm GW}$. The orbital shrinking results in an increase of the Keplerian frequency, consequently increasing the \ac{GW} frequency and amplitude. The resulting waveform has a characteristic shape and is called \emph{chirp signal}. During this phase, the frequency evolution is described by the following equation \citep{MaggioreVol2}:
\begin{equation}
    \dot{f}_{\rm GW} = \frac{96}{5}\pi^{8/3} \left(\frac{G\mathcal{M}}{c^3}\right)^{5/3} f^{11/3}_{\rm GW},
    	\label{eqn:GW_emission}
\end{equation}
where
\begin{equation}
    \mathcal{M} \equiv \frac{(M_{\rm c} M_{\rm NS})^{3/5}}{(M_{\rm c} + M_{\rm NS})^{1/5}}
        	\label{eq:chirp}
\end{equation}
is a combination of the masses $M_{\rm NS, c}$ known as the \emph{chirp mass}.\footnote{Here, we explicitly assume that at least one of the binary components is a \ac{NS}. However, the definition~\eqref{eq:chirp} is generally valid for two inspiraling objects.} Since both $ \dot{f}_{\rm GW}$ and $f_{\rm GW}$ are measurable through the observation of \acp{GW} from a \ac{CBM}, Eq.~\eqref{eqn:GW_emission} can be used to determine the chirp mass of the system. Although measuring $\mathcal{M}$ does not allow us to deduce the mass of the \ac{NS} itself, it provides an important constraint in the $M_{\rm c}-M_{\rm NS}$ plane.

As we approach the merger, the \ac{GW} signal becomes more and more sensitive to relativistic effects, which depend on the binaries' mass ratio \citep{CutlerFlanagan1994, PoissonWill1995}. Hence, comparing a signal of sufficiently high \ac{SNR} with predicted \ac{GW} waveforms can be employed to put tighter constraints on the \ac{NS} mass \citep{GW170817}. We stress, however, that similar effects can also be mimicked by other processes related to the spins of both objects. This leads to a degeneracy between the effective spin (a weighted combination of the objects' spins and masses) and the mass ratio \citep{CutlerFlanagan1994, PoissonWill1995}. As a result, we classify this method as yellow in Tab.~\ref{tab:mass}.


\subsubsection{Gravitational-wave asteroseismology}
\label{sec:gw_asteroseismology_mass}

When perturbed from its equilibrium configuration, a \ac{NS} (approximated as a fluid) is subjected to several damped oscillatory modes, known as \emph{quasi-normal modes}. If these perturbations are non-radial and drive variations in the star's quadrupole moment, quasi-normal modes are associated with the emission of \acp{GW} \citep{LindblomDetweiler1983, MaggioreVol2}. A detection of \acp{GW} excited by such quasi-normal modes will allow us to constrain the stellar mass and radius, or any other quantity derived from both parameters, such as the compactness (see also Sec.~\ref{sec:compactness}) or the average density. This technique is known as the \textit{inverse problem} in \emph{GW asteroseismology}.

The central idea relies on the existence of several quasi-universal relations (namely, expressions that are only weakly dependent on the choice of \ac{EOS}), which link the modes' frequencies and their characteristic damping times to a given combination of the \ac{NS} mass and radius \citep{AnderssonKokkotas1998, Benhar1999, TsuiLeung2005} or other bulk quantities like the moment of inertia \citep{Lau2010} or the tidal Love number \citep{Chan2014} (for a definition of the Love number see Sec. \ref{sec:tidal_def}). These quasi-universal relations hold in particular for the so-called \textit{$f$-mode} and the \textit{$w$-modes}, which depend primarily on macroscopic \ac{NS} quantities.

The $f$-mode is the fluid's fundamental mode. It is a non-radial mode that has no nodes inside the interior of the star. Its frequency lies in the range of $1.5-3 \, \textrm{kHz}$, namely within the frequency window of terrestrial interferometers, while its damping time ranges between $0.1-0.5\, \textrm{s}$ \citep{KokkotasSchmidt1999}. The lack of nodes in the interior of the star makes the $f$-mode weakly dependent on the detailed microphysics of matter, but instead dependent on the global properties of the star. In particular, the mode frequency can be tightly connected to the average density $\bar{\rho} = M_{\rm NS}/R_{\rm NS}^3$ via:
\begin{equation}
    \omega_f \simeq a_f + b_f \left(\frac{\bar{M}_{\rm NS}}{\bar{R}_{\rm NS}^3} \right)^{1/2},
\end{equation}
where $\bar{M}_{\rm NS}$ and $\bar{R}_{\rm NS}$ are dimensionless parameters that normalize the mass and radius to $1.4\,M_\odot$ and $10\, \textrm{km}$, respectively, and $\omega_f$ is measured in $\textrm{kHz}$. \citep{AnderssonKokkotas1998} determined the values of the fitting parameters $a_f$ and $b_f$ to $0.78$ and $1.635$, respectively. We, however, highlight that the exact values of these coefficients (and those presented below) are somewhat dependent on the sample of realistic \acp{EOS} used to fit the quasi-universal relations.

We note that a similar fundamental relation also exists for the $f$-mode damping time, $\tau_f$, which can be expressed as a function of the stellar compactness (see Sec.~\ref{sec:compactness}) \citep{AnderssonKokkotas1998}:
\begin{equation}
     \frac{\bar{R}^4_{\rm NS}}{\bar{M}^3_{\rm NS}}\frac{1}{\tau_f} \simeq c_f - d_f \left(\frac{\bar{M}_{\rm NS}}{\bar{R}_{\rm NS}} \right).
        	\label{eqn-fmode_damping}
\end{equation}
Here, $c_f= 22.85$ and $d_f = 14.65$ \citep{AnderssonKokkotas1998} and the damping $\tau_f$ is measured in seconds. A combined measurement of the mode frequency and its damping time, thus allows us to determine the stellar mass and the radius.

Another set of universal relations can be found for the $w$-modes. Instead of involving fluid motions, these are oscillations of spacetime itself \citep{KokkotasSchutz1992}. As such, they are insensitive to the dense-matter \ac{EOS}. Compared to the $f$-mode, $w$-modes are characterized by higher frequencies (around $5-12\, \textrm{kHz}$ for the lowest-order modes) and shorter damping times of the order of $\mathcal{O}(0.01\, \textrm{ms})$. Their universal relations are \citep{AnderssonKokkotas1998}:
\begin{align}
   \bar{R}_{\rm NS} \omega_w &\simeq a_w - b_w \left(\frac{\bar{M}_{\rm NS}}{\bar{R}_{\rm NS}}\right),\\[1.4ex]
    \frac{\bar{M}_{\rm NS}}{\tau_w} &\simeq c_w + d_w \left(\frac{\bar{M}_{\rm NS}}{\bar{R}_{\rm NS}}\right)
        - e_w \left(\frac{\bar{M}_{\rm NS}}{\bar{R}_{\rm NS}}\right)^2,
\end{align}
where $a_w = 20.92$, $b_w = 9.14$, $c_w = 5.74$, $d_w = 103$ and $e_w = 67.45$. Here, contrary to Eq.~\eqref{eqn-fmode_damping}, the damping time $\tau_w$ is measured in milliseconds. Similar relations for the $f$-mode and the $w$-modes have been also found by several other authors \citep[e.g.,][]{Benhar1999, Benhar2004, TsuiLeung2005, Chirenti2015, PradhanChatterjee2021, Pradhan2022, Kumar2023}.

Additional modes of interest are the so-called \emph{$p$-modes} or \emph{$g$-modes}. $p$-modes, or pressure modes, are fluid modes where the gradient of the pressure acts as the restoring force. Their frequency is much more dependent on the \ac{EOS} than for the modes previously discussed, and they are not expected to lead to tight universal relations \citep{AnderssonKokkotas1998}. These modes are, however, interesting because if the mass, the radius, or another combination of these quantities are known, a measurement of the $p$-mode frequency may allow us to constrain the \ac{EOS} \citep{AnderssonKokkotas1998}. $g$-modes, or gravity modes, have buoyancy as the restoring force \citep{ReiseneggerGoldreich1992}. They are particularly sensitive to the \ac{NS} composition and the presence of discontinuities between different phases of matter \citep{Miniutti2003}. As such, $g$-mode frequency measurements may provide insights into the presence of hyperonic matter \citep[e.g.,][]{Tran2023} and the transition from hadronic to quark matter in the \ac{NS} core \citep[e.g.,][]{FloresLugones2014, Wei2020}.

The discussion so far concerns non-rotating \acp{NS}. Here, the frequency of a given mode is characterized by the spherical degree $l$ only, while there is a ($2l+1$)-degeneracy in the azimuthal order $m$, i.e., $m=-l,..., l$. Rotation has the effect of breaking this degeneracy, splitting a mode with degree $l$ into $2l+1$ modes with distinct frequencies \citep{GaertingKokkotas2008, GaertigKokkotas2011, Doneva2013}. This mode splitting is more pronounced for faster rotation, such that the mode frequency normalized to its value in the non-rotating limit can be expressed as a quadratic function of the rotational frequency \citep{GaertingKokkotas2008, GaertigKokkotas2011, Doneva2013}. Consequently, rotation introduces an additional parameter (the \ac{NS}'s angular velocity) to the inverse problem in asteroseismology.

This method is generally affected by two main issues. First, albeit the universal relations presented above holding for a large sample of hadronic and hybrid \acp{EOS}, the latter characterizing those stars where quark-matter cores are surrounded by hadronic matter, they might not be satisfied by \acp{EOS} for stars with exotic compositions, such as bare quark stars that do not contain a hadronic shell \citep[e.g.,][]{TsuiLeung2005}. For this reason, we classify this method as yellow in Tab.~\ref{tab:mass}.

The second problem concerns the detectability of \ac{NS} quasi-normal modes. Estimating the mode energy required for detection with different \ac{GW} facilities, \citep{Kokkotas2001} find that a measurement of the $f$-mode frequency with an uncertainty of $1\%$ using second-generation interferometers requires a mode energy of $\mathcal{O}(10^{-11}\,M_\odot c^2)$ for a Galactic source at a distance of $10\, \textrm{kpc}$. The required energy increases considerably to $\mathcal{O}(10^{-4}\,M_\odot c^2)$, if we also want to measure the damping time with the same precision. The situation is further complicated for $w$-modes, which are unlikely to be detected with present facilities \citep{Kokkotas2001}. Consequently, any detectable \acp{GW} associated with quasi-normal modes have to be excited by violent events, such as supernova explosions or magnetar giant flares (see Sec.~\ref{sec:QPOs-compactness}). However, the rates of these events are low if we are confined to our Galaxy \citep[e.g.,][]{Cappellaro1993, Gossan2016, Burns2021}, the most likely explanation for why we have not observed these signals yet. This situation should improve with the advanced sensitivity of next-generation facilities \citep{Punturo2010, Abbott-et-al-2017}, which will allow detection of such signals from nearby galaxies \citep{Passamonti2013}. For example, third-generation \ac{GW} detectors are expected to detect Galactic sources within $10\, \rm kpc$, if the mode energy is greater than $\mathcal{O}(10^{-12}\, M_\odot c^2)$ and within the Andromeda galaxy for mode energies above $\mathcal{O}(10^{-8}\, M_\odot c^2)$ \citep{Sathyaprakash2012}, making future detections more likely.


\subsubsection{Cooling}
\label{sec:ns_cooling}

\begin{figure}
    \centering
    \includegraphics[width = 0.48\textwidth ]{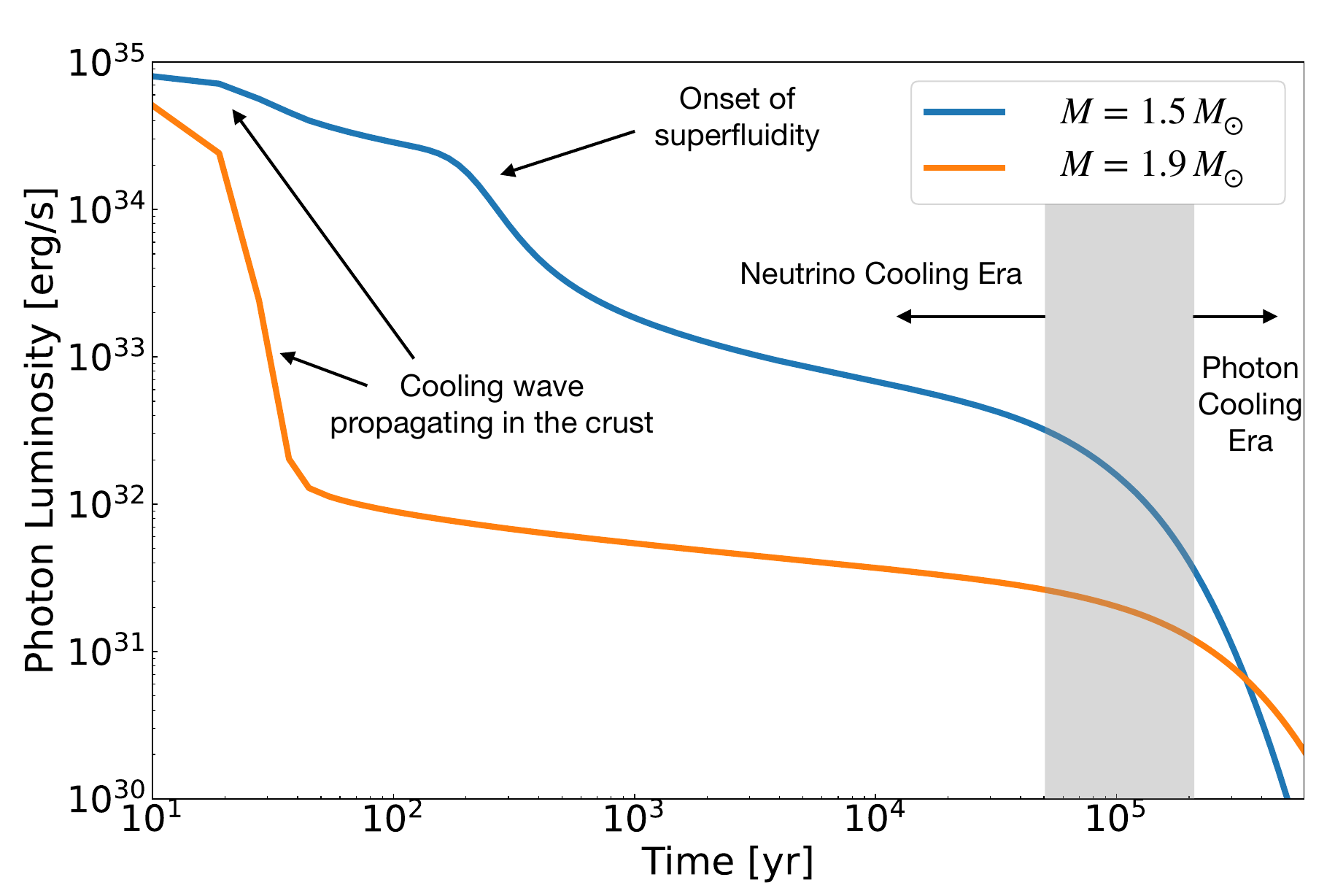}
    \caption{Comparison between two cooling curves for two \acp{NS} of different masses for the \ac{EOS} BSk24 \citep{Goriely2013, Pearson2018}. The heavier object of $1.9 \, M_{\odot}$ (orange) initially cools faster than the lighter $1.5 \, M_{\odot}$ \ac{NS} (blue) due to the fast cooling provided by direct Urca processes. We also label several additional physical processes. In particular, the gray shaded region represents the (approximate) time interval during which \ac{NS} cooling transitions from neutrino-dominated to photon-dominated.}
    \label{fig:cooling}
\end{figure}

\acp{NS} emit thermal radiation in the X-rays. By tracking their temperature evolution, we indirectly gain information on the stellar mass. Specifically, during its life, a \ac{NS} looses its thermal energy due to neutrino emission from its whole volume (which is transparent to neutrinos) and photon emission from its surface \citep{Tsuruta1998, Page2004, Page-etal2006, YakovlevPethick2004, Potekhin-etal2015} (see Fig.~\ref{fig:cooling}).

Neutrino cooling, which dominates during the first $10-100\,\mathrm{kyr}$ of a \ac{NS}'s life, is composed of different emission mechanisms, whose relevance depends on the stellar mass. In particular, one of the most efficient neutrino cooling processes is the \textit{direct Urca process} \citep{PageApplegate1992}, which consists of the following electron-capture and $\beta$-decay reactions:
\begin{align}
    &p + e^- \rightarrow n + \nu_{\rm e}, \\[1.2ex]
    &n \rightarrow p + e^{-} + \bar{\nu}_{\rm e},
    	\label{eqn:beta-decay_dURCA}
\end{align}
where the neutrino and the antineutrino escape from the star and act like an energy sink. The process strongly depends on the temperature, with a neutrino emissivity $\propto T^6$ (see, e.g., Eq.~(120) of \citep{Yakovlev-etal2001}), and may occur in the \ac{NS} core provided that the proton fraction is sufficiently high \citep{Boguta1981, Lattimer1991}. Since the proton fraction increases with increasing central pressure, and the central pressure increases with increasing mass, the presence of direct Urca ultimately depends on the stellar mass. Rapid cooling of young compact objects could, therefore, be an indication for heavy \acp{NS}.

In absence of direct Urca reactions, the main contributor to \ac{NS} cooling is the so-called \textit{modified Urca process}, which substitutes the $\beta$-decay reaction \eqref{eqn:beta-decay_dURCA} as follows:
\begin{equation}
    n + n \rightarrow n + p + e^- + \bar{\nu}_{\rm e}.
\end{equation}
This process has a steeper temperature dependence, with a neutrino emissivity scaling as $T^8$. This results in a shallower decrease of the temperature, $T$, with time, $t$, which scales as $T(t)\propto t^{-1/6}$. In contrast, direct Urca leads to $T(t)\propto t^{-1/4}$. In both cases, we are assuming an isothermal star with a heat capacity $C_v \propto T$ (see \citep{Yakovlev-etal2001} for a review).

As an example, Fig. \ref{fig:cooling} shows a comparison between the cooling curves of two \ac{NS} models characterized by the same \ac{EOS}, but two different masses. The heavier object (orange curve) is able to activate direct Urca processes causing the core to cool very fast in the first years. This results in a steep drop of almost four orders of magnitude in luminosity within the first $50\,\mathrm{yr}$. Direct Urca cooling then proceeds until at around $10^5\,\mathrm{yr}$ when the emission of photons from the surface becomes the dominant mechanism and the cooling curve steepens again. Instead, the low-mass case (blue curve) is controlled by slower modified Urca cooling and the luminosity does not suffer a sharp initial drop. Note, however, that both processes suffer strong quenching if neutrons are in a superfluid state (see Sec.~\ref{sec:cooling_sf}) \citep{Yakovlev-etal2001}.

While efficient \ac{NS} cooling due to direct Urca constrains stellar masses, we must take into account several difficulties, from the theoretical and observation sides. We first point out that the mass threshold for the activation of the direct Urca process depends on the stellar composition \citep{Miller2021}. For example, the presence of hyperons or strange-quark matter affects \ac{NS} cooling because similar, direct Urca-like processes involving these particle species could take place \citep{Iwamoto1982, Prakash1992, Haensel1994, Page-etal2006}. Moreover, other efficient neutrino-emission processes contribute to \ac{NS} cooling, such as neutrino emission due to the continuous breaking and formation of Cooper pairs (see Sec.~\ref{sec:cooling_sf}) \citep{Yakovlev-etal2001}. Finally, if \acp{NS} are strongly magnetized (as is the case for magnetars), we also have to account for additional heating due to the dissipation of electric currents. This in turn depends on the resistivity of the \ac{NS} interior and the unknown magnetic-field geometry \citep{HaenselUrkinYakovlev1990, UrpinShalybkov1995, Kaminker-etal2006, PonsVigano2019}.

From the observational side, we note that an accurate and precise determination of the \ac{NS} cooling history requires reliable age estimates. While for those sources, embedded in their \acp{SNR}, this is typically possible by tracing back the remnant's expansion history, ages are less certain for those \acp{NS} without remnants. In particular, rough estimates are obtained by identifying the \ac{NS}'s true age with the characteristic age, $\tau_{\rm c}$ (see Sec.~\ref{sec:NS-zoo}). This age measure is, however, not reliable because it does not take into account additional physics such as magnetic-field decay or fallback accretion \citep{Alpar-etal2001, Ertan-etal2009, Ronchi2022} that can impact the spin-down history of the star. Uncertain age estimates, hence, significantly hamper a reliable determination of thermal evolution. Contrasting cooling models with data is further complicated by uncertainties in distance estimates, which are reflected in luminosity uncertainties.

In conclusion, while \ac{NS} cooling is fundamental to constraining phenomena occurring in the star's interior, and can provide hints on the stellar mass, it does provide neither a precise nor an accurate, or a model-independent channel to measure the \ac{NS} mass. We hence classify this method as orange in Tab.~\ref{tab:mass}.


\subsubsection{Gravitational microlensing}
\label{sec:gw_lensing}

Compact matter distributions act as gravitational lenses and bend the light of a bright source in the background, producing distorted and magnified images of this background source \citep{Einstein1936, Liebes1964, Gould2001}. So-called \textit{gravitational microlensing} is sensitive to Galactic stellar and compact-object lenses, such as \acp{NS} \citep{Liebes1964, Paczynski1986, Horvath1996, Gould2001}. While gravitational microlensing can, in principle, be used to measure the masses of isolated \acp{NS} that lens stars in their backgrounds \citep{MaoPaczynski1996, SchwarzSeidel2002, Oslowski2008, Dai2015} such signals are weak and ridden with systematics \citep{Delplancke2001, Oslowski2008, Lem2022}, and \ac{NS} in binaries, hence, more promising targets.

Here, gravitational lensing can take place, if the system is observed edge-on (i.e., $\sin i \sim 1$ and exhibiting eclipses) and the companion emits in the optical waveband. The resulting amplification of the companion's optical lightcurve then depends on the lens mass, $M_{\rm NS}$, and the orbital separation, $a$. Specifically, the fractional amplification, defined as the ratio between the amplified flux and the companion's unlensed flux, can be written as \citep{Marsh2001}:\begin{equation}
	f_{\rm amp} = 2 \left(\frac{R_{\rm e}}{R_{\rm c}} \right)^2
    		= \frac{8GM_{\rm NS}}{c^2 a} \, \left(\frac{a}{R_{\rm c}}\right)^2,
\end{equation}
where $R_{\rm e}$ is the Einstein ring's radius and $R_{\rm c}$ the radius of the companion star. If the companion is filling its Roche lobe, the ratio $R_{\rm c}/a$ can also be expressed as a function of the system's mass ratio \citep{Marsh2001}.

Despite its apparent simplicity, the problem with this method is that in real systems the effect is expected to be small (of the order of $10^{-3}\, \mathrm{mag}$). As a result, it is typically overwhelmed by the intrinsic variability of the companion star, or by systematic uncertainties in the modeling of the optical lightcurve due to our incomplete understanding of the heating of the companion by the \ac{NS} \citep{Marsh2001, Clark2023} as already discussed earlier in Sec.~\ref{sec:mass_function_in_binaries}.

As a result, microlensing is classified as strongly model dependent in Tab.~\ref{tab:mass}. We also point out that we consider this approach the least promising of those discussed in Sec.~\ref{sec:mass} and primarily highlight it here for completeness.


\subsection{Radius}
\label{sec:radius}

\begin{table*}[]
    \centering
    \begin{tabular}{|c||c|c|c|c|c|}
        \hline
        \textbf{Method} & \textbf{GWs} & \textbf{Radio} & \textbf{Optical}  & \textbf{X-ray} & $\mathbf{\gamma}$\textbf{-ray}   \\
        \hline
        \hline
        \hyperref[sec:surface_emission_spectral_mod]{Surface emission spectral modeling} &  & &  & \cellcolor{orange!40} &    \\
         \hline
         \hyperref[sec:pre]{Photospheric radius expansion}  &  &  &  & \cellcolor{orange!40} &     \\
         \hline
         \hyperref[sec:xray_lc]{X-ray lightcurve modeling} &  &  &  & \cellcolor{yellow!40} &   \\
         \hline
         \hyperref[sec:nsbh_grb] {NS-BH merger + GRB} & \cellcolor{orange!40} &  &  & \cellcolor{orange!40} & \cellcolor{orange!40}   \\
         \hline
         \hyperref[sec:gw_asteroseismology_radius]{GW asteroseismology} & \cellcolor{yellow!40} &  &  & &    \\
         \hline
    \end{tabular}
    \caption{Radius measurements. Individual rows represent the methods that provide NS radii, while columns denote different messengers/EM wavebands. Individual colors are equivalent to those in Tab.~\ref{tab:mass}. All techniques are discussed in detail in the text.}
    \label{tab:radius}
\end{table*}

We next turn to those methods aimed at measuring the \ac{NS} radius, $R_{\rm NS}$. As discussed in Sec.~\ref{sec:ns-structure}, a radius estimate coupled with a mass measurement for the same object would allow a constraint of the dense-matter \ac{EOS} (see Fig.~\ref{fig:MR_plane}). However, despite its importance, the stellar radius is an elusive quantity and difficult to measure reliably. All techniques presented here (summarized in Tab.~\ref{tab:radius}) are characterized by a certain degree of model dependence and/or lead to results that are degenerate with other parameters. Most of the approaches involve observations in the X-ray band, which are divided primarily into spectral and energy-resolved timing methods. We will discuss why the latter generally lead to more robust $R_{\rm NS}$ estimates. Finally, we report on those methods involving \ac{GW} detections, which require detailed knowledge of the \ac{NS} structure and the processes following compact-object mergers.


\subsubsection{Spectral modeling of surface emission}
\label{sec:surface_emission_spectral_mod}

For the \acp{NS} whose X-ray fluxes are dominated by thermal black-body emission from their surfaces, we can exploit the Stefan-Boltzmann law to measure $R_{\rm NS}$ \citep{OzelFraire2016, Miller2021}. This method requires knowledge of surface temperatures, accessible through spectral analysis, along with reliable estimates of the source distances. It has been successfully applied to \acp{LMXB} in a quiescent state \citep{Brown-etal1998, Rutledge2001, Heinke2006, WebbBarret2007, Guillot2011, Guillot2013, Bogdanov2016, Marino2018, Shaw2018, dEtivaux2019, Echiburu2020}, during which we can reasonably assume that the contribution of accretion to the observed flux is absent or negligible, and \acp{CCO} \citep{Zavlin1998, Pavlov2009, Klochkov2013,  Doroshenko2022}.

The Stefan-Boltzmann law allows us to write the source's apparent radius, $R_\infty$, as
\begin{equation}
    R_\infty = \sqrt{\frac{D^2 F_\infty}{\sigma_{\rm SB} T^4_{\rm BB}}},
      \label{eq:apparent_RNS_SB}
\end{equation}
where $D$ denotes the distance, $F_\infty$ the bolometric flux, $\sigma_{\rm SB}$ the Stefan-Boltzmann constant and $T_{\rm BB}$ the black-body temperature of the star. Due to the \ac{NS}'s compactness, this apparent radius differs from the true radius as a result of light bending in the curved spacetime around the compact object:
\begin{equation}
    R_\infty = \left(1 - 2C_{\rm NS} \right)^{-1/2}R_{\rm NS},
      \label{eqn:apparent_RNS}
\end{equation}
where $C_{\rm NS}$ is the \ac{NS} \emph{compactness} (see Sec.~\ref{sec:compactness} for details) defined as follows:
\begin{equation}
    C_{\rm NS} \equiv \frac{G M_{\rm NS}}{c^2 R_{\rm NS}}.
      \label{eq:compactness}
\end{equation}
Equating relations \eqref{eq:apparent_RNS_SB} and \eqref{eqn:apparent_RNS}, thus, allows us to deduce $R_{\rm NS}$. To date, this method has been used for $\mathcal{O}(10)$ \acp{LMXB} in globular clusters \citep{Heinke2006, WebbBarret2007, Guillot2011, Bogdanov2016, Steiner2018} and a comparable number of sources in the Galactic disk \citep{Marino2018}, leading to a wide range of radius estimates between $10-14\,\rm km$ (see, e.g., Fig.~4 of \citep{OzelFraire2016}). Note, however, that the appearance of the \ac{NS} compactness in Eq.~\eqref{eqn:apparent_RNS} (see Sec.~\ref{sec:compactness} for details) introduces an additional dependence on the \ac{NS} mass \citep{Pechenick1983, Psaltis2000}. This causes a degeneracy in the radius measurement, which would be absent in a non-relativistic picture.

Although simple in concept, this technique is based on several simplifications. First, it assumes that the \ac{NS} surface has a uniform temperature, $T_{\rm BB}$. This is in general not true, especially for magnetars where the thermal conductivity is anisotropic due to the presence of strong magnetic fields (see, e.g., \citep{Potekhin-etal2015} for a review), and currents in the magnetosphere can heat the stellar surface at the footprints of the current-carrying magnetic-field bundle \citep{BeloborodovThomson2007, Beloborodov2009, Gonzalez2019}. As a result, \acp{NS} are characterized by complex surface temperature distributions. Corresponding `hot spots' are typically observed as periodic modulations in the X-ray flux at the stellar rotation frequency due to the spots' regular occultation. However, the absence of such a pulsed flux is no guarantee for a uniform temperature, because similar behavior is also possible for systems with  favorable geometry (e.g., a quasi-axisymmetric temperature map aligned with the rotation axis, or a rotation axis that is aligned with the \ac{LOS}; see, e.g., Fig.~8 of \citep{Ibrahim2023}) or highly multipolar temperature maps, where the individual contributions of hot spots are smeared out \citep{Igoshev2021}. In general, the presence of undetected hot spots leads to an underestimation of the stellar radius \citep{Elshamouty2016}.

Another assumption concerns the fact that flux contributions from the stellar magnetosphere or the accretion disk are generally negligible. Moreover, the lack of knowledge of the \ac{NS}'s atmospheric composition can be a further source of systematic error. For example, considering an atmosphere dominated by helium rather than hydrogen results in systematically larger inferred radii \citep{Servillat2012, Catuneanu2013}. Additionally, in the case of fast spinning \acp{NS}, relativistic corrections should also include the contribution from gravitomagnetic effects, which introduce dependencies on the spin, the mass and the inclination angle. For example, \citep{Baubock2015} report that for a star with $R_{\rm NS}=10 \, \rm km$ rotating at $600\,\rm Hz$, the correction is of the order of $\sim 1-4\%$ (with higher values referring to higher inclinations, i.e., an edge-on view), while it rises to $2-12\%$ for a star with $R_{\rm NS}=15 \, \rm km$. Finally, additional uncertainties may arise from difficulties in reliably measuring the distance of the source, which can be overcome by selecting sources in globular clusters, or modeling the interstellar extinction \citep{Heinke2003, WebbBarret2007}.

All these caveats discussed here lead us to classify this method as strongly model dependent (orange) in Tab.~\ref{tab:radius}.


\subsubsection{Photospheric radius expansion}
\label{sec:pre}

\begin{figure*}
    \centering
    \includegraphics[width=0.85\textwidth]{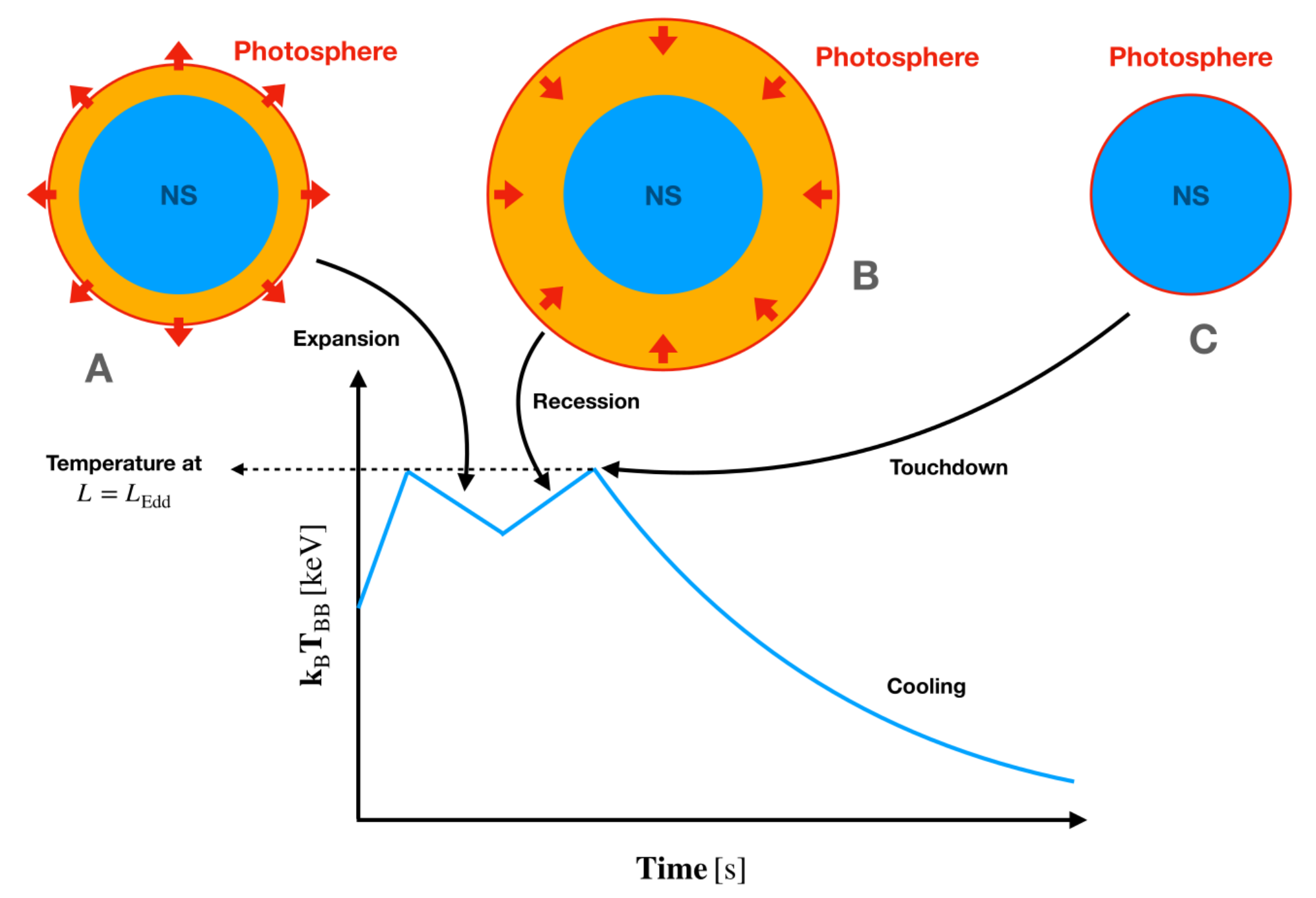}
    \caption{Illustration of the photospheric-radius-expansion scenario and the evolution of the \ac{NS}'s thermal energy with time: During a thermonuclear X-ray burst a star is heated up significantly. As a result the stellar photosphere is lifted and the temperature decreases (A). After reaching a maximum expansion, the photosphere recedes (B), causing the temperature to increase. At the final touchdown phase (C), the photosphere reaches the stellar surface (providing a measure of $R_{\rm NS}$) and the temperature is maximal. At this stage, the star has a luminosity that takes the Eddington value. Subsequently, the \ac{NS} resumes cooling without interruption. The units on the axes reflect fiducial orders of magnitude for the process, with typical temperatures of a few $\mathrm{keV}$ and durations of a few tens of seconds.}
\label{fig:pre}
\end{figure*}

Other intriguing targets for measuring \ac{NS} radii through spectral analysis are systems that undergo thermonuclear (Type-I) X-ray bursts \citep{vanParadijs1979, Lewin-etal1993, Strohmayer2006, Ozel2009, Poutanen2014, Ozel2016, Galloway-etal2020, Patruno2021}. These phenomena occur when material, accreted from a companion star, accumulates on the stellar surface until a runaway fusion reaction is ignited (see Sec.~\ref{sec:X-rayburst_cooling_crustprop} for more details).

The spectra of these X-ray bursts can be fitted with a black body, and in several cases, the temporal evolution of the temperature shows a peculiar two-peaked shape. This is characterized by an initial rise, a first decrease, a second rise, and a final decrease as illustrated in Fig.~\ref{fig:pre}. The model employed to explain this behavior is that of \ac{PRE}. In this framework, the luminosity increases above the Eddington limit during the burst, which launches an optically thick wind that lifts the photosphere. This causes the radiating area to grow and, consequently, a decrease in temperature (A in Fig.~\ref{fig:pre}). After reaching a maximum, the photosphere recedes, which drives a subsequent temperature rise (B in Fig.~\ref{fig:pre}). $T_{\rm BB}$ then reaches a maximum, i.e., the second peak, when the photosphere touches the \ac{NS} surface (C in Fig.~\ref{fig:pre}).

Touchdown is assumed to occur when the flux reaches the Eddington limit, $F_{\rm Edd}$, as, by definition, the Eddington luminosity, $L_{\rm Edd}$, represents the limit at which the outward directed radiation pressure balances the inward acting gravitational force. Thus, $F_{\rm Edd}$ can be expressed as a function of the stellar mass and radius \citep[e.g.,][]{Ozel2016})
\begin{align}
    F_{\rm Edd} &= \frac{c G M_{\rm NS}}{k D^2} \sqrt{1 - 2C_{\rm NS}}
        \left[1 + \left(\frac{k_{\rm B} T_{\rm col}}{38.8\, \mathrm{keV}}\right)^{a_{\rm g}} \right. \nonumber \\
        &\times \left. \left(1 - 2C_{\rm NS} \right)^{-a_{\rm g}/2} \right],
    \label{eq:eddflux}
\end{align}
where $k \equiv 0.2(1+X_{\rm e})\,\textrm{cm}^2 \, \textrm{g}^{-1}$ defines the electron scattering opacity with the electron fraction $X_{\rm e}$. While the square root accounts for the usual gravitational redshift (see also Eq.~\eqref{eqn:redshift}), the term in square brackets is a correction derived by \citep{Suleimanov2012} (advancing the previous result by \citep{Paczynski1983}) to take into account Compton scattering in the stellar atmosphere. Here, the parameter $a_{\rm g}$ is defined as
\begin{equation}
    a_{\rm g} \equiv 1.01 + 0.067 \log \left(\frac{g_{\rm eff}}{10^{14} \, \textrm{cm} \, {\rm s}^{-2}} \right),
\end{equation}
and is a function of the effective surface gravity
\begin{equation}
    g_{\rm eff} \equiv \frac{G M_{\rm NS}}{R_{\rm NS}^2} \left(1 - 2C_{\rm NS}\right)^{-1/2}.
\end{equation}

The degeneracy between mass and radius can be broken by measuring the apparent size, $A_\infty$, of the star during the final cooling phase, because its dependence on mass and radius differs from that in Eq.~\eqref{eq:eddflux}. In fact, we have that \citep{Miller2021}
\begin{align}
    A_\infty &= \frac{F_\infty}{\sigma_{\rm SB}T^4_{\rm col}}
        = f^{-4}_{\rm c} \left(\frac{R_\infty}{D}\right)^2 \nonumber \\
        &= f^{-4}_{\rm c} \left(\frac{R_{\rm NS}}{D}\right)^2 \left(1 - 2C_{\rm NS}\right)^{-1},
        		\label{eqn:Ainf}
\end{align}
where the quantities with the $\infty$ subscript denote those measured by an observer at infinity. We also introduced the color factor $f_{\rm c}$, which is the ratio of the color temperature $T_{\rm col}$ and the effective temperature.\footnote{We remind the reader that the effective temperature, $T_{\rm BB}$, of an object represents the temperature of a black body emitting the same power as emitted through \ac{EM} radiation by said object. The color temperature, $T_{\rm col}$, in turn denotes the temperature of a black body whose color index matches that observed for the object. In other words, $T_{\rm BB}$ is the temperature that enters the Stefan-Boltzmann law for the observed flux, while $T_{\rm col}$ is the temperature of a black body that fits the observed spectrum \citep{RybickiLightman1985}.} Equations~\eqref{eq:eddflux} and \eqref{eqn:Ainf} allow us to obtain the \ac{NS} mass and radius as a function of observable quantities.

The \ac{PRE} method involves similar assumptions to those discussed for the spectral modeling of surface emission in Sec.~\ref{sec:surface_emission_spectral_mod}. Once again, we have assumed that the emission involves the entire surface, which radiates isotropically. Next, conclusions on $R_{\rm NS}$ rely on the above picture being correct, namely that the second temperature peak indeed marks the photosphere's touchdown onto the \ac{NS} surface and that the corresponding flux takes the Eddington value. Moreover, Eq.~\eqref{eqn:Ainf} neglects corrections from gravitomagnetic effect, which are crucial if the star is fast rotating \citep{Baubock2015} (we refer the reader to the discussion at the end of Sec. \ref{sec:surface_emission_spectral_mod}). Furthermore, this approach requires good knowledge of the \ac{NS}'s atmosphere, which causes spectral distortion and influences the color factor. The atmosphere is also assumed to be unaffected by accretion during the cooling phase, which, unfortunately, is not always the case. In particular, \citep{Poutanen2014} found that atmospheric models can only be applied if Type-I X-ray bursts occur during the hard-spectral (low-accretion) state but lose their validity in the soft-spectral (high-accretion) state. Finally, we have again assumed that the entire flux originates from the stellar surface, and all other possible contributions from the system are negligible.

Due to these caveats, we classify the method as strongly model dependent (orange).


\subsubsection{Pulse profile modeling}
\label{sec:xray_lc}

Several \acp{NS} visible in the X-rays exhibit periodic flux modulations both in quiescence and during outbursts. Assuming that this periodic emission is due to one (or multiple) hot spot(s) on the stellar surface rotating rigidly with the star, we can predict the emission received by a distant observer through a technique known as \emph{ray-tracing}. This approach follows the propagation of light from the \ac{NS} surface to the observer through the curved spacetime around the star. The resulting model lightcurve is then compared to observations to constrain several parameters, including the \ac{NS} mass and radius \citep{Pechenick1983}.\footnote{Other parameters of interest are those describing the temperature map (i.e., the shape, size and temperature of the hot spots), the geometry (i.e., the orientation of the hot spots and the \ac{LOS} with respect to the rotation axis), as well as extrinsic ones such as the distance to the source and the hydrogen column density, $N_H$, that determines the interstellar absorption.} In particular, we can extract both by modeling the star's thermal emission and fitting it to the energy-resolved X-ray lightcurve \citep{Pechenick1983, Miller1998, Braje2000, Poutanen2003, PoutanenBeloborodov2006, Morsink2007, Bogdanov2019, Watts2019, Bogdanov2021}.

\begin{figure}
    \centering
    \includegraphics[width=\columnwidth]{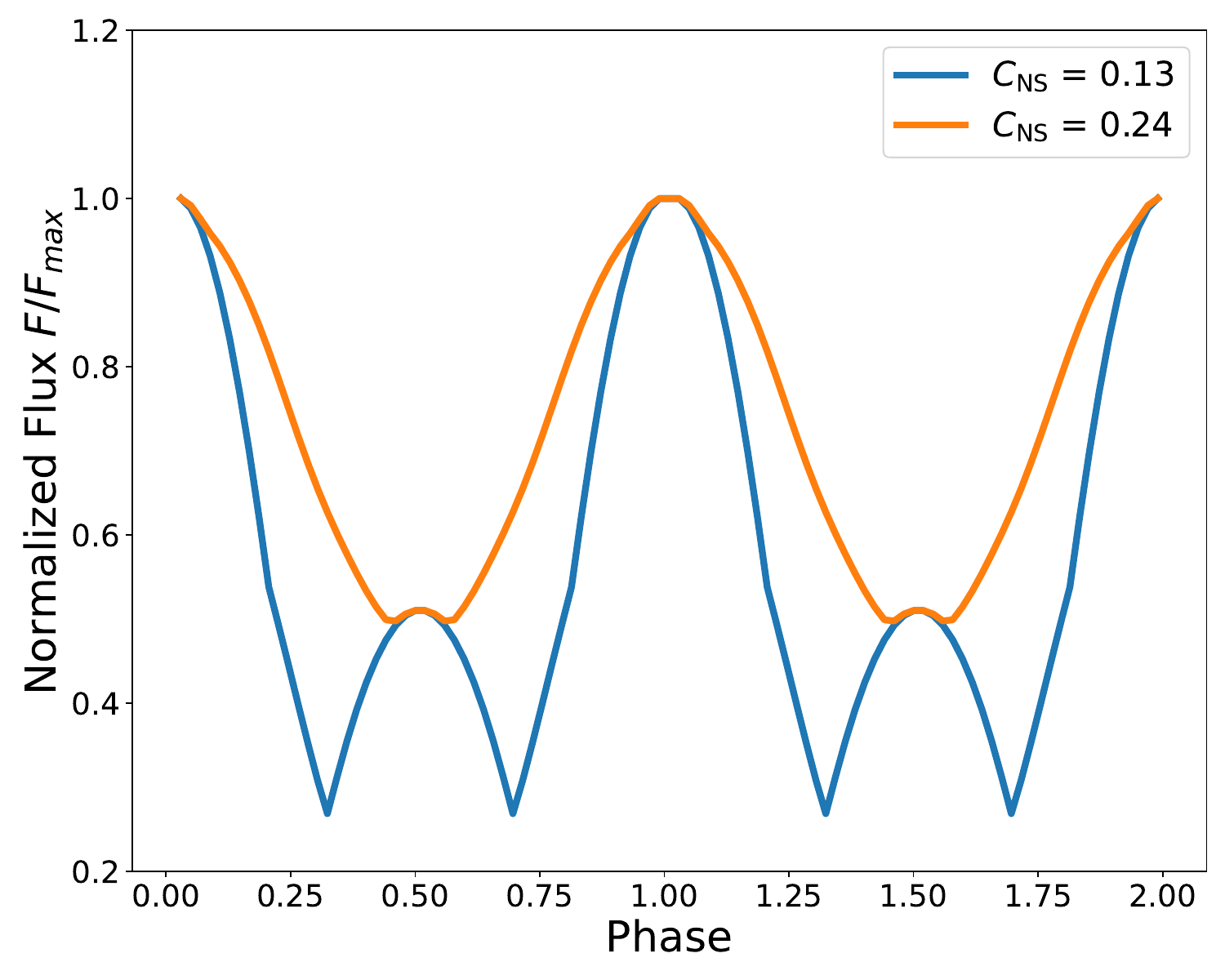}
    \caption{Comparison between the pulsed profiles of two \acp{NS} with the same radius, $R_{\rm NS}= 11\,\rm km$, but a different compactness, $C_{\rm NS}$ (see Eq.~\eqref{eq:compactness}), of $0.13$ (blue) and $0.24$ (orange), respectively. In this example, both stars are characterized by two antipodal, circular hot spots with temperatures of $1 \, \rm keV$ and sizes of $1\, \rm deg$ and $3\, \rm deg$, respectively. The larger spots generate the higher peaks in the profile, while the smaller ones result in the lower peaks. In both simulations, the \ac{LOS} and the spot centers form an angle of $90^\circ$ with respect to the rotational axis. We observe that an increase in the compactness widens the pulse profiles and decreases the amplitude of pulsations. The lightcurves were modeled with software presented in \citep{Ascenzi-in-prep}.}
    \label{fig:pulsed_profile}
\end{figure}

On one hand, the mass enters the picture through the compactness parameter (Sec.~\ref{sec:compactness}), which controls the level of light bending. Specifically, light bending allows the hot spots to be visible even if they are located on the back of the star, facing away from the observer. This effectively reduces the amplitude of the pulsed lightcurve (i.e., the difference between the peak and the minimum of the flux; see Fig. \ref{fig:pulsed_profile} for an illustration). On the other hand, the radius plays a role not only via the compactness but also through the Doppler effect, which is mass independent and causes an asymmetry in the pulse profile due to the stellar rotation. At a fixed angular velocity, a larger \ac{NS} has a higher surface velocity and, consequently, shows a stronger Doppler boost of the hot-spot emission. This influences the shape of the pulses, making them more asymmetric. Moreover, as the asymmetry becomes more pronounced for fast rotators, \acp{MSP} are the ideal candidates for constraining the radius with this kind of analysis \citep{OzelFraire2016}. We note that although this method can constrain the mass and radius at the same time, an independent measure of the mass (through any of the methods outlined in Sec.~\ref{sec:mass}) leads to tighter constraints on the radius \citep[e.g.,][]{Riley2021}.

This technique, which is based on spectro-timing observations, is more robust than the spectral approaches presented earlier, because it is less affected by systematic errors. The primary assumption made here concerns the propagation of photons in the atmosphere, which is typically assumed to be non-magnetized and composed of fully ionized hydrogen (or helium). This implies that the magnetic field has a negligible effect on the propagation of photons and that the primary sources of opacity are free-free absorption and electron scattering \citep{HoLai2001}. This is justified for \acp{MSP}, which are characterized by the weakest magnetic fields in the \ac{NS} population (see Fig.~\ref{fig:p_pdot}). Another approximation that is usually made to ray-trace the photons from the stellar surface to the observer is the so-called \emph{oblate Schwarzschild plus Doppler approximation} \citep{Morsink2007}, which improves on the Schwarzschild plus Doppler approximation \citep{Miller1998, Poutanen2003} by including the rotation-induced oblateness of the star. This is particularly important for rotational frequency above $300\, \mathrm{Hz}$, where the oblateness leaves a measurable imprint on the lightcurve \cite{Cadeau2007}. In this approximation, the rotation determines the \ac{NS} ellipticity and the Doppler boosting of light, but it has no impact on the spacetime metric, which is taken to be the Schwarzschild metric. This last assumption is justified because the \ac{NS} mass quadrupole has little effect on the metric and gravitomagnetic effects are generally negligible \citep{Braje2000, Cadeau2005, Morsink2007, Bogdanov2019}. Nevertheless, effects induced by quadrupolar deformations and rotation can also be included \citep[e.g.,][]{NattilaPihajoki2018, Riley2019}.

The main difficulty with this method lies in the degeneracy between the many parameters it employs. These include, in addition to the stellar mass and radius, the source distance, the hydrogen column density, $N_H$, as well as the parameters needed to describe the surface temperature map,\footnote{As an example, we require $\sim 4N_{\rm spot}$ parameters to describe $N_{\rm spot}$ circular hot spots of uniform temperature. Per spot, we have two parameters for the surface position, one for the spot size, and one for the temperature. More complicated shapes, such as annuli or crescents, can be obtained by superimposing circular spots (see, e.g., Figs.~4, 6, and 9 of \citep{Riley2019}).} along with the angles encoding the orientation of this map with respect to the rotational axis and the observer's \ac{LOS}. 
Low counts in the detected lightcurve or noisy data can lead to degeneracies between the model parameters or multi-modal posteriors, resulting in significant uncertainties in the mass and radius estimates \citep{Vinciguerra2023}.

These issues have significantly improved in recent years due to the unprecedented sensitivity of NASA's \ac{NICER} \citep{Gendreau2016}, installed on the \emph{International Space Station} in 2017. In particular, \ac{NICER} has allowed radius constraints for the \acp{MSP} PSR J0030+0451 \citep{Riley2019, Miller2019, Vinciguerra2023} and J0740+6620 \citep{Riley2021}. Analysis of the former led to $R_{\rm NS} = 12.71^{+1.14}_{-1.19}\, \textrm{km}$ and $M_{\rm NS} = 1.34^{+0.15}_{-0.16} \, M_{\odot}$ at $68 \%$ credibility level \citep{Riley2019} (corresponding to a relative error $<10\,\%$), which is consistent with an independent analysis by \citep{Miller2019}. Whereas PSR J0030 is an isolated \ac{NS}, PSR J0740 is in a binary system and exhibits a favorable inclination which allowed a mass measurement via Shapiro delay (see Sec.~\ref{sec:shapiro_delay}), leading to $M_{\rm NS} = 2.08^{+0.07}_{-0.07} \,M_\odot$ \citep{Fonseca2021}. As discussed above, the independent measurement of the mass allowed for an improved radius constraint for this source, resulting in $R_{\rm NS} = 12.39^{+1.30}_{-0.98}\, \textrm{km}$ \citep{Riley2021}.

Although retaining a certain level of model dependence and degeneracies between parameters (leading us to classify pulse profile modeling as yellow in Tab.~\ref{tab:radius}), we consider this technique the most promising in providing accurate and precise constraints on the \ac{NS} radius in the coming years.

We conclude by highlighting that pulse profile modeling, which also allows us to infer information about the geometry of the hot spots, encodes \ac{NS} information beyond the mass and the radius. In particular, hot-spot properties are related to the geometry of the \ac{NS}'s magnetic field \citep{PavlovZavlin1997}. Analyses of \ac{NICER} data for both \acp{MSP} discussed above disfavored a simple dipolar poloidal magnetic field configuration, which has implications for the magnetospheric structure and emission mechanisms \citep{Chen2020, SuvorovMelatos2020, Kalapotharakos2021, Carrasco2023}.


\subsubsection{Multi-messenger observations of neutron star-black hole mergers}
\label{sec:nsbh_grb}

Multi-messenger observations of \acp{CBM} provide a new channel to measure the \ac{NS} radius. For this task, \ac{NS}-\ac{BH} systems are the optimal targets due to their relative simplicity with respect to \ac{BNS} systems. Such a \ac{NS}-\ac{BH} merger has two possible outcomes. Either the \ac{NS} plunges directly into the \ac{BH}, or it is tidally disrupted outside the \ac{BH}'s \ac{ISCO} \citep{Shibata2009, Etienne2009, Chawla2010, Duez2010, Ferrari2010, PannaraleOhme2014}. In the second case, an accretion disk forms around the \ac{BH}. The resulting system may launch a relativistic jet, which is able to power a short-duration \ac{GRB}, one of the most luminous transients in the Universe \citep{Blinnikov1984, Paczynski1986b, Eichler1989} (for recent reviews about \acp{CBM} and their \ac{EM} counterparts see \citep{Nakar2020, Ascenzi2021}). \ac{GRB} emission consists of the highly variable \emph{prompt emission}, observable in gamma-rays and powered by the dissipation of kinetic energy within the jet (internal dissipation), which is followed by a multi-wavelength \emph{afterglow} due to the dissipation of jet energy in the interstellar medium (external dissipation). For an accretion-powered jet, the corresponding \ac{GRB} energy, $E_{\rm GRB}$, is proportional to the mass, $M_{\rm disk}$, of the accretion disk with a proportionality constant, $\epsilon_{\rm jet}$, that is referred to as the jet-launching efficiency. This quantity depends on the efficiency of converting the accreted mass into the jet's kinetic energy and the efficiency of converting the latter into radiation. In the case of a \ac{NS}-\ac{BH} merger, we can express $M_{\rm disk}$ as a (semi-analytical) function of relevant system parameters, specifically the masses, $M_{\rm NS}$ and $M_{\rm BH}$, of both objects, the \ac{NS} radius, $R_{\rm NS}$, and the dimensionless \ac{BH} spin parameter, $a_{\rm BH}$  \citep{Foucart2018}. As a result, we obtain
\begin{equation}
    E_{\rm GRB} = \epsilon_{\rm jet} M_{\rm disk}(M_{\rm NS}, M_{\rm BH}, a_{\rm BH}, R_{\rm NS}).
      \label{eq:nsbh_method}
\end{equation}
Whereas we can measure $E_{\rm GRB}$ from \ac{EM} observation (typically in the gamma-ray and X-ray bands for the prompt emission and afterglow, respectively), $M_{\rm NS}$, $M_{\rm BH}$ and $a_{\rm BH}$ are deduced from the parameter estimation of the \ac{GW} signal. By estimating $\epsilon_{\rm jet}$, or defining a reasonable prior distribution (typical values deduced from theoretical models \citep{McKinney2005, Hawley2007, ZalameaBeloborodov2011, TchekhovskoyGiannios2015, Christie2019, Ruiz2019, Ruiz2020} and observations of GW170817 \citep{SalafiaGiacomazzo2021} fall in the range of $10^{-4}-10^{-1}$; see Tab.~1 of \citep{SalafiaGiacomazzo2021} for a summary), we can then solve Eq.~\eqref{eq:nsbh_method} for $R_{\rm NS}$. Assuming an \ac{SNR} of $10$ for the \ac{GW} detection, \citep{Ascenzi2019} use this method to deduce that we can measure the radius with a relative uncertainty $<20\%$ at $90\%$ confidence (see \citep{Fragione2021} for a conceptually similar approach).

However, this method suffers from several assumptions. The most important one concerns our lack of detailed knowledge of the efficiency, $\epsilon_{\rm jet}$. Second, the above picture does not consider that these accretion disks can lose on the order of $10 \%$ of their mass through winds \citep[e.g.,][]{Fernandez2015, Fernandez2019}.

It is also worth adding that while \ac{NS}-\ac{BH} mergers are possible short \ac{GRB} progenitors, they are likely only responsible for a minority of the short \ac{GRB} population, whose bulk is produced by \ac{BNS} mergers \citep{Clark2015}. Moreover, \ac{NS}-\ac{BH} mergers are not particularly promising multi-messenger sources, because the tidal disruption of the \ac{NS} outside the \ac{BH} \ac{ISCO} is not particularly likely \citep{Fragione2021b}. Consequently, we consider this method of extracting $R_{\rm NS}$ as strongly model dependent, classifying it as orange in Tab.~\ref{tab:radius}. However, once the \ac{NS} radius has been tightly measured, this approach could be of great interest in constraining the uncertain jet-launching efficiency, $\epsilon_{\rm jet}$.

We finally note that the above concept can, in principle, also be used in association with \ac{BNS} mergers and exploit not only \acp{GRB} but also other associated transients such as \acp{KN} (see also Sec.~\ref{sec:kilonovae}). \citep{Dietrich2017}, for example, developed a fitting formula for the \ac{BNS}-merger dynamical ejecta (which contributes to fuel the kilonova emission) similar in concept to the formula for $M_{\rm disk}$ discussed above. However, considering \ac{BNS} systems does not reduce the level of model dependence but rather increases it due to the higher complexity of the binary.


\subsubsection{Gravitational-wave asteroseismology}
\label{sec:gw_asteroseismology_radius}

This method (classified as yellow in Tab.~\ref{tab:radius}) is the same as that presented in Sec.~\ref{sec:gw_asteroseismology_mass} to measure the mass of a \ac{NS}. We remind the reader that the main idea is to exploit quasi-universal relations linking the frequency and damping time of some particular quasi-normal mode (such as the $f$-mode or $w$-modes) with the \ac{NS} mass and radius to constrain both parameters.

In addition to the discussion in Sec.~\ref{sec:gw_asteroseismology_mass}, we present here another method to exploit asteroseismology to measure \ac{NS} radii in the context of \ac{BNS} mergers. The general idea is based on the fact that a \ac{NS} (either stable or metastable) formed as a consequence of the binary coalescence will emit \acp{GW} through the excitation of quasi-normal oscillations \citep{Stergioulas2011, BausweinJanka2012, TakamiRezzollaBaiotti2014, TakamiRezzollaBaiotti2015, RezzollaTakami2016, Kastaun2016}. Although faint, this post-merger signal should be detectable for a merger occurring at a distance of a few up to a few tens of Mpc with second-generation interferometers \citep[e.g.,][]{Clark2016, Abbott-et-al-2017b, Krolak2023, Tringali2023}, while many such detections are expected for third-generation facilities \citep[e.g.,][]{Sasli-etal2023, Walker-etal2024}. In particular, the spectrum of the post-merger signal is characterized by several peaks, whose characteristic frequencies are associated with different properties of the \ac{NS} remnant through quasi-universal relations. These relations, for example, allow us to extract the radius of a non-rotating neutron star with maximum mass \citep{BausweinJanka2012} or the stellar  compactness \citep{TakamiRezzollaBaiotti2014, TakamiRezzollaBaiotti2015, Rezzolla2018}. Therefore, detecting post-merger signals and measuring the corresponding characteristic frequencies provides a wealth of information about dense-matter physics and could, e.g., constrain possible quantum-chromodynamics phase transitions between hadronic and quark matter in hybrid stars \citep[e.g.,][]{Bauswein-etal2019, Weih-etal2020, Huang-etal2022, Prakash-etal2023}.

Moreover, \citep{Bose2018} pointed out that a measurement of the compactness via the aforementioned quasi-universal relations could be combined with a measurement of the system's total mass as derived from the inspiral phase to obtain an estimate of the \ac{NS} radius. \citep{Bose2018} underlined that, although the uncertainty on the measured frequency would be too large to obtain meaningful radius constraints for a single detection with second-generation interferometers (since sources at distances of $\sim 200\,\mathrm{Mpc}$ are characterized by \acp{SNR} $<1$), the joint uncertainty of $\sim 100$ \ac{BNS} detections could provide an average measurement on the radius with an uncertainty of $\sim 2-12 \,\%$. In this case, larger uncertainties were associated with softer \acp{EOS}.


\subsection{Moment of inertia}
\label{sec:MoI}

\begin{table*}[]
    \centering
    \begin{tabular}{|c||c|c|c|c|c|}
        \hline
        \textbf{Method} & \textbf{GWs} & \textbf{Radio} & \textbf{Optical}  & \textbf{X-ray} & $\mathbf{\gamma}$\textbf{-ray}  \\
        \hline
        \hline
         \hyperref[sec:rel_spin-orbit]{Relativistic spin-orbit coupling}  &  & \cellcolor{yellow!40} &  &  &     \\
         \hline
         \hyperref[sec:glitches-MoI]{Pulsar glitches} & & \cellcolor{orange!40} & & \cellcolor{orange!40} & \cellcolor{orange!40}   \\
         \hline
         \hyperref[sec:cgw]{Continuous \acp{GW}} & \cellcolor{yellow!40} & &  &  &    \\
         \hline
         \hyperref[sec:gw_asteroseismology_moi]{\ac{GW} asteroseismology} & \cellcolor{yellow!40} & &  &  &    \\
         \hline
    \end{tabular}
    \caption{Moment of inertia measurements. Individual rows represent the methods that provide NS moments of inertia, while columns denote different messengers/EM wavebands. Individual colors are equivalent to those in Tab.~\ref{tab:mass}.}
    \label{tab:MoI}
\end{table*}


We next turn to techniques, summarized in Tab.~\ref{tab:MoI}, that constrain the \ac{NS}'s moment of inertia, $I_{\rm NS}$. The quantity's importance lies in the fact that a measurement of $I_{\rm NS}$, coupled with a mass measurement, directly allows constraining the dense-matter \ac{EOS} and the \ac{NS} radius \citep{Lattimer2005, Raithel2016}, bypassing an independent measurement of $R_{\rm NS}$, which is generally affected by significant model dependencies and systematics (see Sec.~\ref{sec:radius}). We also point out that since $I_{\rm NS} \propto M_{\rm NS} R_{\rm NS}^2$ (i.e., the moment of inertia depends on $R_{\rm NS}$ squared), the range of moment of inertias for different \acp{EOS} is larger than the range of viable \ac{NS} radii \citep{Lattimer2005}. Consequently, $I_{\rm NS}$ measurements make constraining the \ac{EOS} easier than $R_{\rm NS}$ measurements.

We note that in what follows $I_{\rm NS}$ is associated with a fiducial moment-of-inertial value unless stated otherwise.


\subsubsection{Relativistic spin-orbit coupling}
\label{sec:rel_spin-orbit}

In \ac{GR}, the spins of two bodies in a compact binary are coupled with each other and their orbital motion \citep{Barker1975, Damour1988}. As the orbital angular momentum represents the dominant contribution to the total angular momentum, spin-spin coupling is weaker than spin-orbit coupling and can be safely neglected \citep{Lattimer2005}. However, relativistic spin-orbit coupling, also known as \textit{Lense-Thirring precession}, drives an advance of the orbit's periastron in addition to the post-Keplerian contributions introduced in Sec.~\ref{sec:mass_function_in_binaries}. Spin-orbit coupling also causes the orbital angular momentum vector to precess around the total angular momentum, which in turn drives precession of the objects' spins due to angular momentum conservation \citep{Wex1995, Lattimer2005}. As the orbital angular momentum dominates over the spins, the former contribution is generally small, while precession of the spins can be substantial. However, both effects are detectable in \ac{BNS} systems via radio timing if one or both of the stars are pulsars \citep{Kramer1998, Stairs-etal2004, Fonseca-etal2014}.

The predicted evolution of the two spins, $\mathbf{S}_i$, and orbital angular momentum, $\mathbf{L}$, due to spin-orbit coupling is \citep{Barker1975}:
\begin{align}
    \dot{\mathbf{S}}_i &= \frac{G(4M_i + 3M_{-i})}{2M_ia^3c^2(1-e^2)^{3/2}} \, \mathbf{L}\times\mathbf{S}_i,
    	\label{eqn:spin_equation} \\[1.4ex]
    \dot{\mathbf{L}} &= \sum_i \frac{G(4M_i + 3M_{-i})}{2M_ia^3c^2(1-e^2)^{3/2}} \, \left(\mathbf{S}_i -3 \frac{\mathbf{L}\cdot\mathbf{S}_i}{|\mathbf{L}|^2}\mathbf{L}\right),
    	\label{eqn:angularmom_equation}
\end{align}
where we follow \citep{Lattimer2005} and denote the two binary components with subscripts $i = {\rm A, B}$ and $-i = {\rm B, A}$, respectively. The resulting precession periods of both objects are \citep{Lattimer2005}
\begin{equation}
	P_{{\rm precession}, \, i} = \frac{2 c^2 a P_{\rm orb} M (1 - e^2)}{GM_{-i} (4 M_i + 3 M_{-i})}.
\end{equation}

Equations~\eqref{eqn:spin_equation} and \eqref{eqn:angularmom_equation} highlight that spin and orbital plane precession occur only when the spins and orbital angular momentum are misaligned. If this is the case, the change in geometry should lead to a time-varying pulse profile \citep{Hu-etal2020} (and potentially the disappearance and reappearance of the pulsar radiation \citep{Lattimer2005}) plus a variation in the orbital plane's orientation. The latter manifests itself as a change in the inclination, $i$, \cite{Damour1988} and is, thus, observable as a variation in the projected semi-major axis, provided the binary is not observed edge-on (i.e., $\sin i \nsim 1$). If, in addition, one pulsar spins much faster than the other, its spin will provide the dominant contribution to the spin-orbit coupling, and the observed effect depend directly on its moment of inertia as $|\mathbf{S}| = I_{\rm NS} \Omega$, where $\Omega = 2 \pi / P$ is the star's angular velocity.

For alignment between the spins and the orbital angular momentum, we instead obtain $\dot{\mathbf{S}}_i = 0$ and $\dot{\mathbf{L}} \propto - \sum_i \mathbf{S}_i = {\rm const}$. This implies that only the periastron motion is present, which, in turn, occurs in the opposite direction to the orbital motion. Moreover, if $|\mathbf{S}_{\rm A}| \gg |\mathbf{S}_{\rm B}|$, the resulting effect is again dependent on the moment of inertia of pulsar A and a corresponding measurement would, therefore, also constrain the \ac{NS} moment of inertia \citep[e.g.,][]{KramerWex2009, Lattimer2005, Hu-etal2020, Kramer-etal2021}.

Extracting changes in inclination angle, $i$, for known \acp{BNS} with sufficiently high precision poses significant challenges \citep[e.g.,][]{Cameron-etal2018}. As a result, $I_{\rm NS}$ constraints from the periastron advance are currently more promising. In particular, the Double Pulsar PSR J07037--3039A/B \citep{Burgay2003, Lyne2004, Kramer-etal2021}, a system whose orbit is seen almost edge-on, pulsar A rotates much faster than pulsar B and spin A is almost parallel to $\mathbf{L}$, is a promising target for this method. Accounting for future observations with MeerKat and the Square Kilometre Array, \citep{Hu-etal2020} predict a moment-of-inertia measurement for the Double Pulsar with $11\%$ accuracy by 2030.

The main limitations of this measurement are threefold. First, relativistic spin-orbit coupling depends on the individual masses of both objects. Consequently, to measure the moment of inertia through periastron advance, we require an independent measurement of the individual binary masses. Moreover, we also require knowledge of the orientation of the pulsar spin with respect to the orbital angular momentum. Third, the periastron advance has a dominant spin-independent contribution that originates from the first and second post-Newtonian correction of the orbital motion (see Sec.~\ref{sec:mass_function_in_binaries}). To extract $I_{\rm NS}$, we therefore need to be able to disentangle the spin-orbit contribution to the total periastron advance from the post-Newtonian contribution. Addressing these issues requires a precise measurement of three post-Keplerian parameters, and building up a sufficiently precise pulsar timing solution can take decades. Thus, long-term datasets such as those obtained for PSRs B1913+16 \citep{Weisberg-etal2010}, B1534+12 \citep{Fonseca-etal2014} or the Double Pulsar PSR J0737--3039A/B \citep{Kramer-etal2021} will be crucial to detect this effect (if possible at all). However, a single successful measurement would provide tight moment-of-inertia constraints.

Factoring in these caveats, we classify this method as yellow in Tab.~\ref{tab:MoI}.


\subsubsection{Pulsar glitches}
\label{sec:glitches-MoI}

Another phenomenon that indirectly probes the \ac{NS} moment of inertia are so-called \textit{pulsar glitches}, sudden spin-ups that interrupt the otherwise regular electromagnetic \ac{NS} spin-down. High-precision timing of glitching pulsars with radio telescopes allows us to quantify the size of these glitches as well as the shape of the signal on timescales of minutes to months \citep[e.g.,][]{Dodson-etal2007, Espinoza-etal2011, Yu-etal2013, Palfreyman-etal2018, Lower-etal2021, Basu-etal2022}. Glitches have also been observed in the X-rays and gamma-rays although generally with lower precision \citep[e.g.,][]{Kaspi-etal2000, DallOsso-etal2003, Gugercinoglu-etal2022}.\footnote{A notable exception are observations of glitches in the X-ray pulsar PSR J0537--6910, which plays a crucial role for glitch studies \citep{Ferdman-etal2018, Antonopoulou-etal2018, Ho-etal2020}.} As a result, we know that glitches are common in isolated pulsars and more than 650 spin-ups have now been observed in over 200 \acp{NS} \citep{Espinoza-etal2011, glitch-catalogue}, providing ample opportunity to probe related physics.

Glitch amplitudes vary between $|\Delta \Omega | / \Omega \sim 10^{-12} - 10^{-5}$, where $|\Delta \Omega|$ is the size of the glitch. Those pulsars exhibiting large glitches repeatedly (such as the Vela pulsar \citep{Dodson-etal2007, Shannon-etal2016, Palfreyman-etal2018}) are of particular interest for $I_{\rm NS}$ considerations, as sudden spin-ups in these objects are associated with the presence of internal superfluids. We refer to Sec.~\ref{sec:sf} for details on the microscopic aspects of superfluidity and focus on the macroscopic implications here.

Superfluid models explain glitches by the rapid transfer of angular momentum from an internal, differentially rotating superfluid (decoupled as a result of vortex pinning) to the rest of the star \citep[e.g.,][]{Anderson1975, Baym-etal1969b, Alpar1988, Haskell-etal2012, Graber-etal2018}. The superfluid interpretation is crucial for giant glitches, as the star-quake model (an alternative explanation based on the build up of strain in the \ac{NS} crust, analogous to Earth quakes \citep{Ruderman1969, Smoluchowski1970, Baym1971}) cannot provide sufficient stresses to explain frequent, Vela-like glitches. We will, thus, focus on the superfluid framework here.

\begin{figure}
    \centering
    \includegraphics[width = 0.43\textwidth]{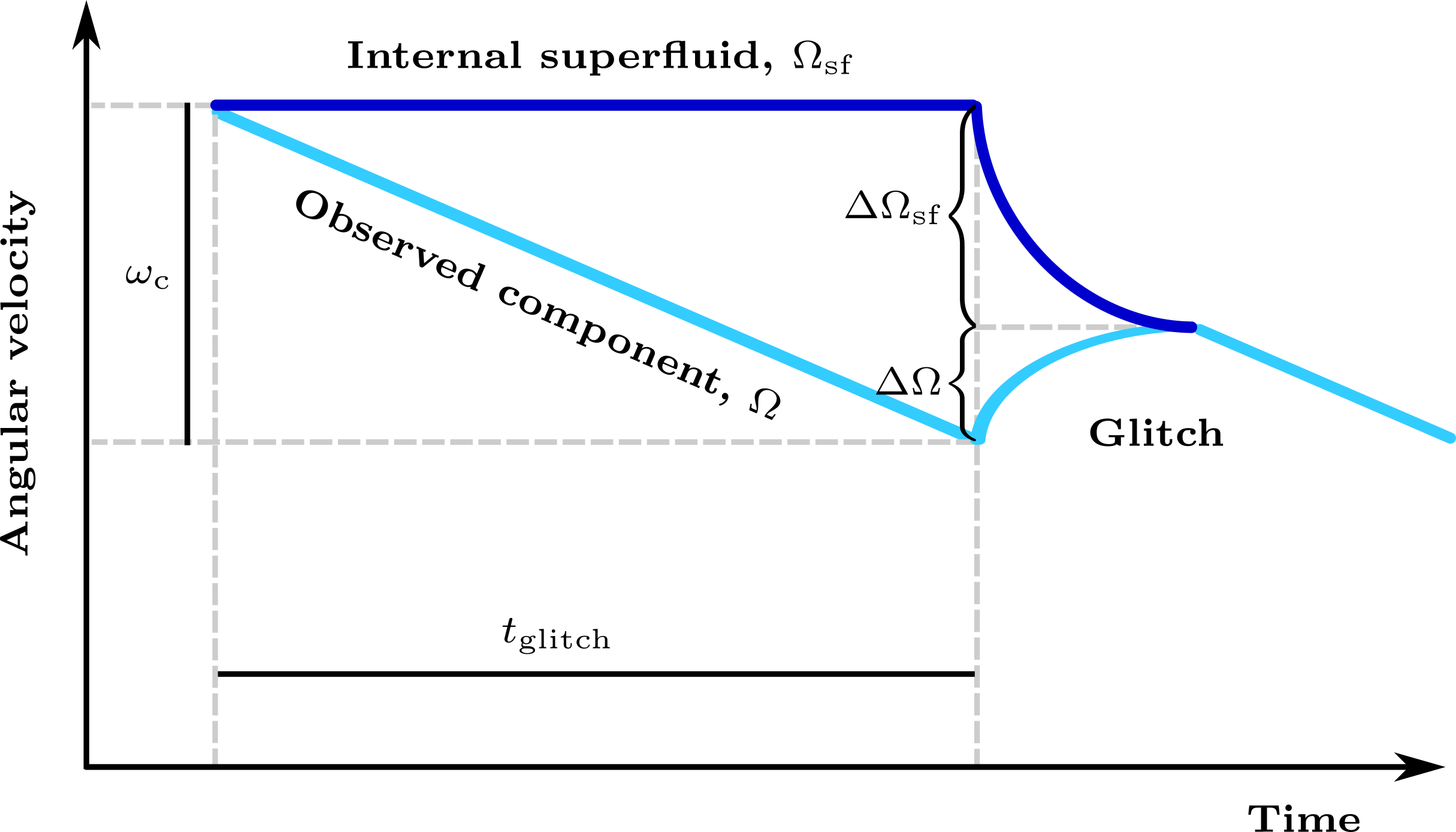}
    \hspace{0.1cm}
    \includegraphics[width = 0.47\textwidth]{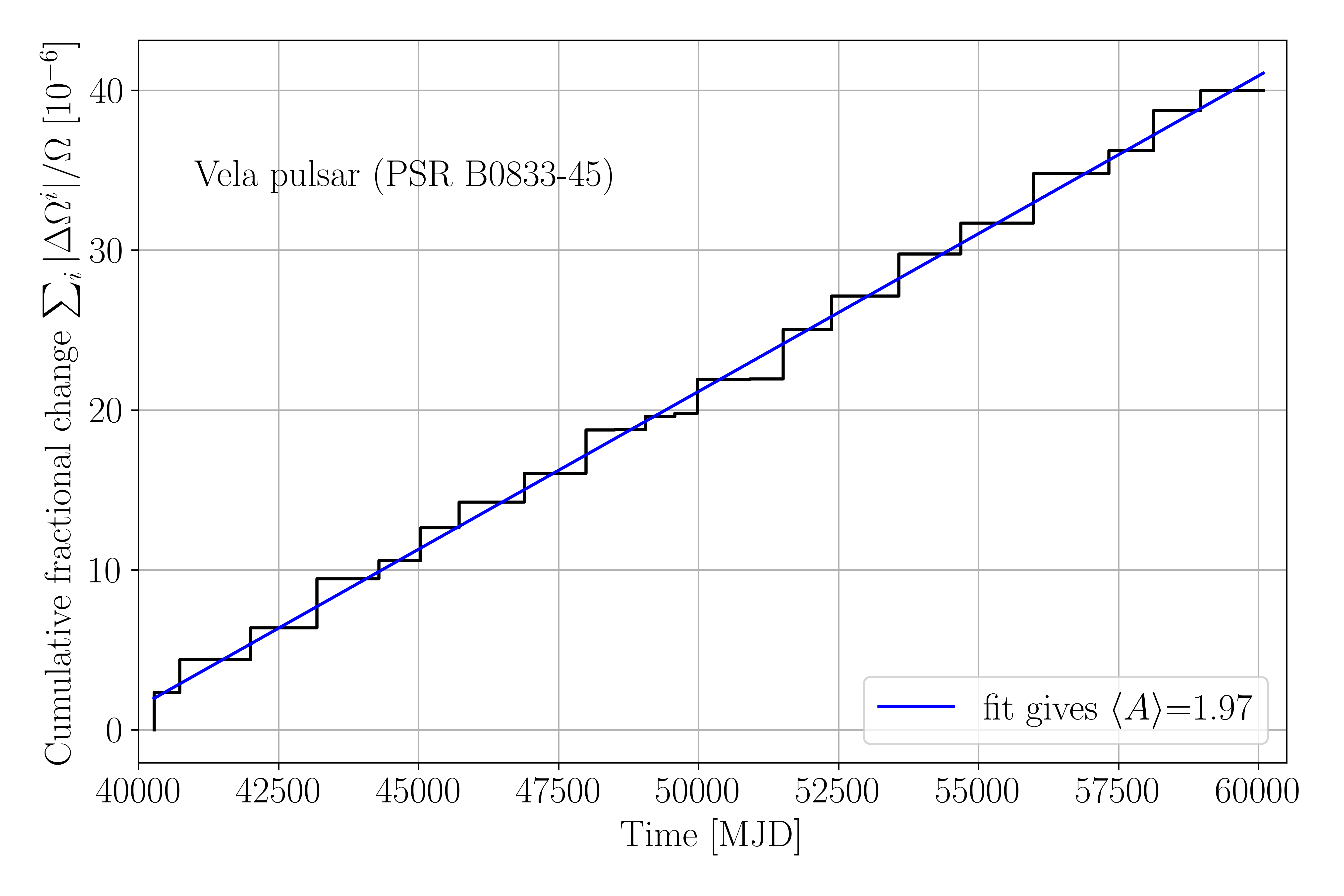}
    \caption{Illustrating the pulsar glitch phenomenon. \textit{Top panel:} Sketch of the physics underlying a glitch in the so-called \textit{two-component model} and relevant quantities used in the text. An internal superfluid is decoupled from the observed component, spinning down due to the loss of electromagnetic energy. At a critical lag, $\omega_{\rm c}$, both recouple, causing the observed component to spin up, producing the glitch. \textit{Bottom panel:} Cumulative fractional change of spin frequency caused by glitches in the Vela pulsar PSR B0833--45 over time. The straight blue line is the least-squared fit to the data, whose slope is equivalent to the average glitch activity parameter, $\langle \mathcal{A} \rangle$. The corresponding value of $\langle \mathcal{A} \rangle = 1.97$ leads to an inferred superfluid fraction of $I_{\rm sf} /I_{\rm NS} = 2 \tau_{\rm c} \langle \mathcal{A} \rangle = 1.63 \% $. Glitch data are taken from \citep{glitch-catalogue}.}
    \label{fig:glitches}
\end{figure}

Interpreting glitches in a two-component picture (shown in the top panel of Fig.~\ref{fig:glitches}) and assuming complete depletion of the superfluid reservoir, angular momentum conservation dictates
\begin{equation}
    I_{\rm sf} \Delta \Omega_{\rm sf} =  I_{\rm obs} |\Delta \Omega|.
        \label{eqn:amconservation}
\end{equation}
Here, $I_{\rm sf}$ denotes the superfluid's moment of inertia, $\Delta \Omega_{\rm sf}$ the change in its angular velocity, and $I_{\rm obs}$ the moment of inertia of the observed component undergoing the glitch, i.e., $I_{\rm sf} + I_{\rm obs} = I_{\rm NS}$. As both components recouple during the glitch and then corotate, Eq.~\eqref{eqn:amconservation} can be rewritten in terms of the velocity lag, $\omega_{\rm c} \equiv \Omega_{\rm sf} - \Omega$, at the time of the glitch, which itself depends on the interglitch time, $t_{\rm glitch}$, and characteristic age, $\tau_{\rm c}$ (see Fig.~\ref{fig:glitches} and \citep{Andersson-etal2012} for details). While this, in principle, allows us to determine $I_{\rm sf} / I_{\rm NS}$ for pulsars with at least two large glitches, interglitch times and sizes differ significantly between events for any given pulsar \citep[e.g.,][]{Espinoza-etal2011, Yu-etal2013, Lower-etal2021, Basu-etal2022}. We thus define an average glitch activity parameter, $\langle \mathcal{A} \rangle$, for pulsars exhibiting $N$ spin-ups in an observation period, $t_{\rm obs}$. $\langle \mathcal{A} \rangle$ can be directly measured from observations (see bottom panel of Fig.~\ref{fig:glitches}) and leads to
\begin{equation}
	\frac{I_{\rm sf}}{I_{\rm NS}} \approx 2 \tau_{\rm c} \, \frac{1}{t_{\rm obs}} \,
		\sum_i^N \frac{|\Delta \Omega^i|}{\Omega} \equiv 2 \tau_{\rm c} \langle \mathcal{A} \rangle.
			\label{eqn:Iratio_mod}
\end{equation}

Corresponding estimates for $I_{\rm sf}/I_{\rm NS}$ range from $0.5$ to a few percent \citep[e.g.,][]{Link-etal1999, Andersson-etal2012}, which can be compared to the moment of inertia of the neutron superfluid in the \ac{NS}'s inner crust. This region is typically considered as the origin of pulsar glitches, because vortices can directly pin to crustal lattice structures, naturally explaining the superfluid's decoupling. For standard \acp{EOS} and \ac{NS} mass ranges, the inner crust constitutes between $\approx 2 - 8 \%$ \citep[e.g.,][]{Link-etal1999, Andersson-etal2012} of the star's total moment of inertia, thus supporting the crust-origin interpretation of pulsar glitches.

Recent results \citep{Andersson-etal2012, Chamel2013}, however, suggest that Eq.~\eqref{eqn:Iratio_mod} does not account for the fact that crustal neutrons might be difficult to move due to so-called \textit{entrainment} \citep{Chamel2005, Chamel2012}, effectively reducing the available moment of inertia. While still uncertain \citep[e.g.,][]{Martin2016, Sauls-etal2020}, if entrainment is indeed present, \citep{Andersson-etal2012} estimate that the right-hand-side of Eq.~\eqref{eqn:Iratio_mod} needs to be modified by an additional factor $\sim 4.3$. For some glitching \acp{NS}, this would imply that the crustal superfluid no longer holds sufficient moment of inertia to explain large and frequent glitches, as for example observed in the Vela pulsar. This in turn can also set constraints on the underlying \acp{EOS}, \ac{NS} masses and radii \citep{Ho-etal2015}. Further assumptions on the interactions of individual vortices allow us to set additional constraints on \ac{NS} masses as, e.g., suggested by \citep{Pizzochero-etal2017}.

In conclusion, glitches provide indirect evidence on the moments of inertia of the components involved in the spin-up process and some insights into the \ac{EOS}, \ac{NS} masses, and radii. However, estimates (although readily available due to the abundance of glitch observations) are strongly model dependent as shown in Tab.~\ref{tab:MoI}. The above framework, for example, assumes that glitches are linked to the presence of superfluidity, \acp{NS} are composed of two interacting components, and the superfluid angular momentum reservoir is completely depleted during the spin-up. A better understanding of the glitch process and improved theoretical models will be needed to take further advantage of these spin-ups to provide advanced constraints on the \ac{NS} moment of inertia.


\subsubsection{Continuous gravitational waves}
\label{sec:cgw}

Rotating \acp{NS} emit \acp{GW} at the expense of their rotational energies. A time variation in the mass quadrupole moment can, for example, be produced by a tri-axial star rotating rigidly about a principle axis or an axisymmetric (bi-axial) body whose rotation axis does not coincide with the symmetry axis (see \citep{Wette2023, HaskellBejger2023} for recent reviews). Whereas in the former case, continuous \acp{GW} are emitted at a frequency $2\Omega$, the latter case also produces \ac{GW} emission at $\Omega$, which dominates over the harmonic at $2\Omega$ in the limit of small angles between the rotation and the symmetry axis \citep{Zimmermann-et-al-1979}. Emission of \acp{GW} at both $\Omega$ and $2\Omega$ can furthermore occur due to magnetic-field induced stellar distortions \citep{Bonazzola-etal1996} or the presence of superfluid components \citep{Jones2010}. Although current \ac{GW} detectors may not yet be sensitive enough to directly detect these signals (physically interesting limits on \ac{NS} properties have however been obtained with current facilities as we outline in Sec.~\ref{sec:GW-crustpropr}), third-generation interferometers have the strong potential to detect continuous \acp{GW}. We will, thus, focus on the $2 \Omega$-case, which applies to non-axial \acp{NS}, e.g., those that support a mountain (see also Sec.~\ref{sec:GW-crustpropr} for details). Such a star loses energy at the rate \citep{MaggioreVol2}
\begin{equation}
	\dot{E} \approx - \frac{32 G}{5 c^5} I_{3}^2 \varepsilon^2 \Omega^6,
		\label{eqn:cgw_Eloss}
\end{equation}
where $I_{3}$ is the moment of inertia along the rotation axis, $\varepsilon \equiv (I_2 - I_1)/I_3$ the ellipticity and $I_1, I_2$ denote the moments of inertia in the directions orthogonal to the rotational axis. Note that these moments of inertia can vary from the fiducial value, $I_{\rm NS}$, by a factor of a few. The corresponding \ac{NS} spin-down is, thus, characterized by
\begin{equation}
	\dot{\Omega} \approx - \frac{32 G}{5c^5} \Omega^5 \varepsilon^2 I_3,
		\label{eq:cgw_spindown}
\end{equation}
while the continuous \ac{GW} emission has the amplitude \citep{Jaranowski-etal1998}
\begin{equation}
	h_0 \approx \frac{4G}{c^4} \frac{1}{D} I_{3} \varepsilon \Omega^2,
		\label{eqn:GWstrain_mountain}
\end{equation}
for a source at distance, $D$.

\citep{Sieniawska2022} recently developed a method to measure the moment of inertia of a \ac{NS} from this continuous \ac{GW} emission. The method is simple in concept and exploits the quasi-standard siren nature of \ac{GW}-emitting \acp{NS}, taking advantage of the fact that Eqs.~\eqref{eq:cgw_spindown} and \eqref{eqn:GWstrain_mountain} contain the measurable quantities $h_0, \Omega$, and $\dot{\Omega}$. Although these two relations provide insufficient information to constrain the three unknowns $\varepsilon, I_3$, and $D$, we can combine both equations to eliminate the ellipticity and relate $\sqrt{I_3}/D$ to our observables:
\begin{equation}
    \frac{\sqrt{I_3}}{D} = \sqrt{\frac{2c^3}{5G}}\sqrt{\frac{\Omega}{|\dot{\Omega}|}} \, h_0.
\end{equation}

If, in addition, we have an independent way to measure $D$, we could directly constrain the moment of inertia. Unfortunately, measuring \ac{NS} distances is a difficult task \citep{OzelFraire2016}. In particular, since the \acp{NS} that we currently observe only constitute a very small fraction of the total \ac{NS} population \citep{Treves2000, Palomba2005, Riles2023}, several continuous \ac{GW} sources that will be detected by future \ac{GW} interferometers will lack an \ac{EM} counterpart. This will likely make a determination of distances even more challenging. In such cases, it might however be possible to constrain the source's distance by extracting its parallax directly from the \ac{GW} signal \citep{Sieniawska2023}. This would allow us to break the degeneracy between $I_3$ and $D$, and infer the moment of inertia using continuous \ac{GW} observations only. Our ability of measuring the \ac{GW} parallax increases with larger observation span, \ac{SNR} and \ac{NS} spin frequency \citep{Sieniawska2023}. For example, \citep{Sieniawska2023} identify a \ac{NS} rotating at $400 \, \rm Hz$ observed in the \ac{GW} band with an $S/N \sim 100$ for $1\,\rm yr$ as the best case scenario, and suggest the source's parallax can be measured if it is located within $100 \, \rm pc$ from Earth.

We can translate the concepts outlined for tri-axial stellar deformations to the continuous \acp{GW} emission by so-called \textit{$r$-modes}. $r$-modes are toroidal oscillation modes expected in rotating \acp{NS}, where the Coriolis force acts as the restoring force \citep{Andersson1998} (see \citep{Haskell2015b} for a review). They are the \ac{NS} analog of Rossby waves in the Earth's oceans \citep{Rossby1939}. One important point concerning $r$-modes is that they are expected to be subject to the so-called \ac{CFS} instability \citep{Chandrasekhar1970, FriedmanSchutz1978}. This instability occurs in rotating stars when rotation is so fast that a counter-rotating oscillation mode in the rotating frame appears to co-rotate in the laboratory frame. In this case, \ac{GW} emission subtracts angular momentum from the mode. However, as the mode's angular momentum is negative in the rotating frame, this subtraction increases the absolute value of the angular momentum of the mode. As a result, the emission of \acp{GW} does not damp the mode, but instead causes it to grow. We note, however, that while the $r$-mode satisfies the \ac{CFS} instability criterion under the assumption of a perfect fluid star, other dissipative effects such as shear or bulk viscosity can damp or even stop the instability \citep[e.g.,][]{AnderssonKokkotas2001}. As a result, complex microphysics \citep[see, e.g.,][]{IpserLindblom1991, Haskell-etal2012B} determine the highly uncertain saturation point to which the $r$-mode amplitude can grow. Nonetheless, \ac{GW} emission driven by the $r$-mode would be continuous at the expense of the star's rotational energy.

Focusing on the lowest-order Newtonian $r$-mode (characterised by the spherical harmonics $l=m=2$, where $l$ and $m$ denote the degree and order of the corresponding Legendre polynomial, $P_l^m$, respectively) for an incompressible, barotropic star, we can write the corresponding \ac{GW} strain amplitude as \citep{Owen2010}
\begin{equation}
    h_{0,r} = \sqrt{\frac{8\pi}{5}} \frac{G}{c^5}  \frac{1}{D}
        \alpha M_{\rm NS} R_{\rm NS}^3 \Tilde{J} \omega^3_r.
\end{equation}
Here, $\alpha$ is the dimensionless quantity that parameterizes the saturation amplitude \citep{Owen1998}, the dimensionless parameter $\Tilde{J}$ depends on the \ac{NS}'s mass-density profile (see, e.g., Eq.~(6) in \citep{Sieniawska2022}), and $\omega_r \simeq 4\Omega/3$ is the $r$-mode's angular frequency.

Assuming that all the energy associated with the excitation of $r$-modes is lost through \acp{GW}, we obtain the following spin-down formula:
\begin{equation}
    \dot{\Omega} = - \frac{218\pi G}{3852 c^7} \, (\alpha M_{\rm NS} R_{\rm NS}^3 \Tilde{J})^2 \, \frac{\Omega^7}{I_3}.
        \label{eq:rmode_spindown}
\end{equation}
Similar to what we discussed above, we can now combine the previous two equations to eliminate the factor $(\alpha M_{\rm NS} R_{\rm NS}^3 \Tilde{J})$ and solve for the ratio $\sqrt{I_3}/D$ to obtain:
\begin{equation}
    \frac{\sqrt{I_3}}{D} = \sqrt{\frac{8c^3}{45 G}}\sqrt{\frac{\Omega}{|\dot{\Omega}|}} \, h_{0,r}.
\end{equation}
By measuring observable quantities, we can, therefore, again constrain the stellar moment of inertia.

The underlying assumption of this technique concerns the use of Eqs.~\eqref{eq:cgw_spindown} and \eqref{eq:rmode_spindown}. Both relations assume that the observed spin-down is dominated by a specific type of continuous \ac{GW} emission. However, pulsar observations show that this is generally not satisfied, as a number of other processes can spin down a rotating \ac{NS}. These are characterized by the so-called \textit{braking index}, $n$, which encodes the exponent of $\Omega$ in relations of the form $\dot{\Omega} \propto \Omega^n$. For pulsars with precise timing solutions, it can be obtained via
\begin{equation}
	n \equiv \frac{\Omega \ddot{\Omega}}{\dot{\Omega}^2}.
		\label{eq:braking_index}
\end{equation}
See \citep{Lower-etal2021} for a recent overview of measured braking indices and corresponding systematics. From Eq.~\eqref{eq:cgw_spindown}, we, hence, deduce that $n=5$ for pulsars that slow down due to \ac{GW} emission from a mountain, and $n=7$ if \acp{GW} are emitted due to $r$-modes. In turn, $n=3$ characterizes a \ac{NS} that spins down due to magnetic dipole emission, which we discuss in more detail in Sec.~\ref{sec:MD_braking}. Consequently this method only applies to those cases, where $n=5$ or $n=7$ is observed, significantly reducing the number of potential targets for moment-of-inertia measurements, because most pulsars exhibit $n<3$. However, we note that \ac{EM} observations of the young X-ray pulsar PSR J0537--6910 hint at an $n=7$ spin-down \citep{Andersson-etal2018}. As the \ac{NS}'s distance is known, the source is a prime candidate for searches of continuous \acp{GW} from $r$-modes \citep{HaskellBejger2023}, which have started to probe relevant parameter spaces \citep{Abbott-etal2021d}.

Recently, \citep{Lu2023} extended the approach of \citep{Sieniawska2023} to also account for the dipole spin-down emission. This generalization, which introduces another parameter, i.e., the magnetic dipole moment, however requires an additional measurement to constrain the \ac{NS} moment of inertia. \citep{Lu2023} focus on the second derivative of the rotational frequency, $\ddot{\Omega}$. As a result, their method is not restricted to sources with $n=5$ only, but to those with $n\in [3,5]$. However, also in this case, knowledge of the distance is required. In particular, \citep{Lu2023} estimate that the relative uncertainty on the moment of inertia will be limited to $27\%$ with increasing observation time for a relative distance uncertainty of $20\%$.

Finally, we note that $n=5$ is not exclusive to quadrupolar \ac{GW} spin down, but also characterizes braking by magnetic quadrupole radiation. This would, however, require a magnetic-field configuration where the quadrupole component dominates over the dipolar one, for which we currently do not have observational evidence.

As a result of these restrictions, we classify this method as somewhat model dependent in Tab.~\ref{tab:MoI}.


\subsubsection{Gravitational-wave asteroseismology}
\label{sec:gw_asteroseismology_moi}

\ac{GW} asteroseismology, already discussed in Sec.~\ref{sec:gw_asteroseismology_mass} for the \ac{NS} mass, provides a valuable method to measure the moment of inertia of a \ac{NS}, which we classify as somewhat model dependent in Tab.~\ref{tab:MoI}. Besides the quasi-universal relations presented previously, \citep{Lau2010} found another universal relation for the $f$-mode that links the frequency and the damping time of the mode to the mass and moment of inertia, instead of $M_{\rm NS}$ and $R_{\rm NS}$. While the relation from \citep{Lau2010} holds for non-rotating stars, generalizations for the case of rapid rotation have also been found \citep{DonevaKokkotas2015, KrugerKokkotas2020}. We specifically note that these relations appear to be tighter for the range of possible \acp{EOS} than those relations involving mass and radius. On top of that, the relation in \citep{Lau2010} also holds for quark stars. However, as outlined in Sec.~\ref{sec:gw_asteroseismology_mass}, the main issue of this method currently concerns the detectability of \ac{GW} waves from quasi-normal modes with present facilities, and the accuracy required for measuring mode frequencies and damping times to constrain \ac{NS} parameters. Third-generation interferometers may finally allow for an exploration of this technique \citep{Maggiore2020, Andersson2021}.


\subsection{Tidal deformability}
\label{sec:tidal_def}

\begin{table*}[]
    \centering
    \begin{tabular}{|c||c|c|c|c|c|}
        \hline
        \textbf{Method} & \textbf{GWs} & \textbf{Radio} & \textbf{Optical}  & \textbf{X-ray} & $\mathbf{\gamma}$\textbf{-ray}  \\
        \hline
        \hline
            \hyperref[sec:tidal_deformability_cbm]{Tidal deformability in CBMs} & \cellcolor{yellow!40} & &  &  &   \\
            \hline
            \hyperref[sec:gw_asteroseismology_tidal]{GW asteroseismology} & \cellcolor{yellow!40} & &  &  &   \\
            \hline
    \end{tabular}
    \caption{Tidal-deformability measurements. Individual rows represent the methods that provide the tidal deformability of \acp{NS}, while columns denote different messengers/EM wavebands. Individual colors are equivalent to those in Tab.~\ref{tab:mass}.}
    \label{tab:tidal_def}
\end{table*}


Our fourth parameter of interest is the tidal deformability of \acp{NS}, a quantity which directly relates to the dense-matter \ac{EOS}. In a compact binary system composed of a \ac{NS} and a companion, which we require to be another \ac{NS} or a low-mass \ac{BH} for this parameter to be measurable (see below), the \ac{NS} is subject to the tidal field of its companion. The quadrupolar contribution to this tidal field is given by \citep{MaggioreVol2, Chatziioannou2020}:
\begin{equation}
    \mathcal{E}_{ij} = c^2 R_{0i0j},
\end{equation}
where $R_{\mu \nu \rho \sigma}$ is the Riemann tensor.\footnote{The equivalent expression in a Newtonian picture is $\mathcal{E}_{ij} = - \partial_i \partial_j U_{\rm ext}$, where $U_{\rm ext}$ is the external Newtonian gravitational potential.} This tidal field perturbs the \ac{NS} and induces a quadrupole moment, which is proportional to the field itself:
\begin{equation}
    Q_{ij} = - \lambda_{\rm NS} \mathcal{E}_{ij}.
\end{equation}
The constant of proportionality, $\lambda_{\rm NS}$, is a dimensional quantity encoding information about the \ac{NS}'s properties referred to as the \emph{tidal deformability}. It is typically expressed in terms of the \ac{NS} radius and another dimensionless quantity, the \emph{l=2 tidal Love number}, $k_2$, (where $l$ denotes the degree of the corresponding spherical harmonic) through \citep{MaggioreVol2, Chatziioannou2020}
\begin{equation}
    \lambda_{\rm NS} = \frac{2 k_2 R_{\rm NS}^5}{3G}.
\end{equation}
The tidal Love number depends on the dense-matter \ac{EOS}. For realistic choices of \ac{EOS}, it ranges between $k_2 \simeq 0.05-0.15$, with lower values corresponding to softer \acp{EOS} \citep{MaggioreVol2}. Thus, by measuring $\lambda_{\rm NS}$, it is possible to constrain the \ac{EOS}.

In the following, we will provide a brief overview of how the tidal deformability can be measured from observations of \acp{GW}. For a more detailed discussion of the topic, we refer the interested reader to \citep{DamourNagar2009, Hinderer2010, MaggioreVol2, Chatziioannou2020}.

Finally, we also mention the existence of universal relations between the moment of inertia, the tidal Love numbers, and the \ac{NS}'s quadrupole moment. These relations, known as \emph{I-Love-Q} relations, express either of these three quantities as a function of one of the other two. These relations, thus, allow us to obtain the remaining two quantities when one of them is measured \citep{YagiYunes2013a}. For example, a measurement of the moment of inertia, performed with one of the methods presented in Sec.~\ref{sec:MoI}, can be used to infer the tidal Love number through a relation between the moment of inertia and the Love number \citep{YagiYunes2013b}. Vice versa, if the tidal Love number is constrained in a compact binary merger, as explained in the next section, we can use this information to constrain the star's moment of inertia and the quadrupole moment \citep{YagiYunes2013b}.


\subsubsection{Compact-binary mergers}
\label{sec:tidal_deformability_cbm}

\ac{GW} signals from \ac{BNS} or \ac{NS}-\ac{BH} mergers are sensitive to the tidal deformability, because the quadrupole moment induced in the \ac{NS} by its companion provokes an increase in the emission of \acp{GW}. This effectively results in a faster decay of the orbit compared to a scenario that approximates the merging bodies as point masses. This effect, however, becomes relevant only when the two objects are close to merger, i.e., in the last orbits of the inspiral phase. As the orbital frequency at this stage reaches values around $200\, \textrm{Hz}$, tidal effects are generally observed only in the \ac{GW} signal above $\sim 400\, \mathrm{Hz}$, where they lead to a correction in the evolution of the \ac{GW} phase at the 5th post-Newtonian (5PN) order. However, from a numerical point of view, tidal effects in \acp{CBM} are sufficiently strong to overcome lower-order terms and are, hence, detectable \citep[e.g.,][]{Dietrich2019}.

We note, however, that the tidal deformability, $\lambda_{\rm NS}$, does not enter the 5PN-correction directly. Instead, this correction depends on the quantity
\begin{equation}
	\Tilde{\Lambda} = \frac{16}{13}\frac{(M_{\rm NS} + 12 M_{\rm c}) M^4_{\rm NS} \Lambda_{\rm NS}
		+ (M_{\rm c} + 12 M_{\rm NS}) M^4_{\rm c} \Lambda_{\rm c}} {(M_{\rm NS} + M_{\rm c})^5},
			\label{eq:lambda_tilde}
\end{equation}
where $M_{\rm c}$ is again the mass of the companion, while $\Lambda_{\rm NS, \, c}$ are dimensionless quantities related to $\lambda_{\rm NS, \, c}$ through
\begin{equation}
	\Lambda_{\rm NS, \, c} =\frac{c^{10} \lambda_{\rm NS, \, c}}{G^4 M^5_{\rm NS, \, c}}.
\end{equation}
This quantity takes values between $10-10^4$ for realistic \ac{NS} \acp{EOS} and masses, with higher values corresponding to stars of lower mass. For example, for a $1.4 \, M_\odot$ \ac{NS}, we obtain a tighter range of $10^2-10^3$ (see Fig.~1 of \citep{Chatziioannou2020}). However, if the companion is a \ac{BH}, then the tidal deformability is typically assumed to be $\lambda_{\rm c} = 0$, implying also $\Lambda_{\rm c} = 0$.\footnote{See the dedicated section in \citep{Chatziioannou2020} for a more detailed discussion about this point.} In this case, for a fixed \ac{NS} mass, $\Tilde{\Lambda}$ decreases with increasing \ac{BH} mass, suggesting that tidal effects are more pronounced, and, thus, easier to measure, when the \ac{BH} is low in mass.

Equation~\eqref{eq:lambda_tilde} shows that the measurable quantity $\Tilde{\Lambda}$ does not depend on the tidal deformability of a single object, but is instead a linear combination of $\lambda_{\rm NS}$ and $\lambda_{\rm c}$. This degeneracy can be broken by measuring higher-order (i.e., 6PN) terms in the \ac{GW} \citep{Vines2011, Damour2012}, which are however unlikely to be detectable with second-generation \ac{GW} interferometers even for high \acp{SNR} of $\sim 30$ \citep{Wade2014}.

We also note that tidal effects depend on the individual binary masses. We remind the reader that at leading order, the \ac{GW} signal provides access to a particular combination of the masses known as the chirp mass, $\mathcal{M}$ (see Sec.~\ref{sec:gw_from_cbm}). Individual masses can be constrained by measuring the next-order terms, which depend on the mass ratio that is, however, degenerate with the spin of both objects. This consequently introduces a (mild) correlation with the spins into the determination of $\Lambda_{\rm NS, \, c}$ and similarly $\lambda_{\rm NS, \, c}$. We consequently classify this method as yellow in Tab.~\ref{tab:tidal_def}.

A measurement of the tidal deformability was possible for the first time with the observation of the \ac{BNS} merger GW170817. This event allowed us to estimate the parameter $\Tilde{\Lambda}$, leading to $\Tilde{\Lambda} = 300^{+500}_{-190}$ for a low-spin prior (assuming a dimensionless spin parameter $\chi < 0.05$) or $<600$ for a high-spin prior ($\chi < 0.89$) \citep{GW170817, GW170817_improved_2019}.\footnote{These results are based on the PhenomPNRT waveform model \citep{Hannam2014}.} An alternative analysis of the same observation produced results that are consistent with these estimates \citep{De2018}. Ultimately, these insights provided constraints on the \acp{NS}' dense-matter properties, ruling out the stiffest \acp{EOS} \citep{GW170817_improved_2019}. Multiple \ac{BNS} merger detections with third-generation interferometers will improve on these \ac{NS} \ac{EOS} constraints \citep{Punturo2010, Abbott-et-al-2017}. In particular, these detections will allow measurements of $\Tilde{\Lambda}$ as well as individual masses through the \ac{GW} signal providing crucial information on the tidal deformabilities of both merging \acp{NS} \citep{Iacovelli2023}.


\subsubsection{Gravitational-wave asteroseismology}
\label{sec:gw_asteroseismology_tidal}

In Secs.~\ref{sec:gw_asteroseismology_mass}, \ref{sec:gw_asteroseismology_radius}, and \ref{sec:gw_asteroseismology_moi}, we discussed how universal relations enable us to infer the stellar mass, radius and moment of inertia, respectively, from quasi-normal mode frequencies of an oscillating, \ac{GW} emitting \ac{NS}. Additional universal relations involving oscillation frequencies and the star's tidal deformability also exist, allowing us to measure the latter once the corresponding \ac{GW} signal is detected. In particular, \citep{Chan2014} find several relations between the frequency of an $f$-mode of degree $l$ and the tidal deformability of degree $l'$ for both \acp{NS} and quark stars. Specifically, the cases $l=l'$ lead to the most universal, i.e., most \ac{EOS}-insensitive, relationships. Similar relations have also been discovered in subsequent studies \citep[e.g.,][]{SotaniKumar2021, Lioutas2021, ZhaoLattimer2022, Pradhan2023}. Their origin lies in the combination of the relation between the $f$-mode frequency and the moment of inertia from Sec.~\ref{sec:gw_asteroseismology_moi} and the above mentioned \emph{I-Love-Q} relations \citep{Chan2014}. Following the same logic as presented in the previous sections on \ac{GW} asteroseismology, we also classify this method as yellow here.


\subsection{Compactness}
\label{sec:compactness}


\begin{table*}[]
    \centering
    \begin{tabular}{|c||c|c|c|c|c|}
        \hline
        \textbf{Method} & \textbf{GWs} & \textbf{Radio} & \textbf{Optical}  & \textbf{X-ray} & $\mathbf{\gamma}$\textbf{-ray}  \\
        \hline
            \hyperref[sec:linespec-compactness]{Line spectroscopy (redshift)} & & &  & \cellcolor{green!40}  &  \\
              \hline
            \hyperref[sec:xray_ppm_compactness]{Pulse profile modeling} &  &  &  & \cellcolor{yellow!40} &  \\
            \hline
            \hyperref[sec:QPOs-compactness]{Giant-flare QPOs} & & &  & \cellcolor{orange!40}  &   \\
            \hline
    \end{tabular}
    \caption{Compactness measurements. Individual rows represent the methods that provide the \ac{NS} compactness, while columns denote different messengers/EM wavebands. Individual colors are equivalent to those in Tab.~\ref{tab:mass}.}
    \label{tab:compactness}
\end{table*}


The compactness, $C_{\rm NS}$ (defined in Eq.~\eqref{eq:compactness}), measures how densely matter is compressed within the \ac{NS} volume. Even for the most extreme \acp{EOS}, $C_{\rm NS}$ ranges between $0.1$ and $0.3$ \citep{LattimerPrakash2001}, whereas causality, i.e., the requirement that the speed of sound must be lower than the speed of light, sets a maximum limit of $C_{\rm NS} = 0.354$ \citep{Haensel1999} (see also Fig.~\ref{fig:MR_plane}).

The main observable imprints (summarized in Tab.~\ref{tab:compactness}) directly dependent on this quantity are the gravitational redshift of spectral features and its influence on the bending of photon trajectories. As a result, several methods discussed previously, in particular many of those that provide access to the stellar radius, also lead to compactness constraints. While we will not repeat a discussion of these methods in detail, we briefly mention one which is particularly useful to constrain the compactness. Additionally, we discuss another phenomenon in strongly magnetized \ac{NS} that might provide independent information on $C_{\rm NS}$.


\subsubsection{Line spectroscopy}
\label{sec:linespec-compactness}

Observed X-ray spectra of \acp{NS} can be permeated by spectral lines due to atomic transitions in regions close to the stellar surface. In particular, in isolated \acp{NS} such lines are generated due to absorption in the thin stellar atmospheres. Bright thermally emitting \acp{NS}, such as \acp{XDINS}, are the most promising targets for such observations (see, e.g., \citep{Turolla2009} and references therein). Spectral lines are also observed in accreting \acp{NS}, where (depending on the viewing geometry) emission and absorption processes in the accretion column or the inner edge of the accretion disk can explain the observed spectral features \citep{OzelPsaltis2003, Bhattacharyya2010, Baubock-etal2013}.

Due to the emission/absorption region's close proximity to the stellar surface, spectral lines can be gravitationally redshifted as a result of the \ac{NS}'s compactness. More precisely, the gravitational redshift, $z$, is directly related to the stellar  compactness by the following equation:
\begin{equation}
    1 + z \equiv \left(1 - 2\frac{R_{\rm NS}}{r_{\rm line}}C_{\rm NS}\right)^{-1/2}
      \simeq \bigl(1 - 2C_{\rm NS}\bigr)^{-1/2}.
        \label{eqn:redshift}
\end{equation}
Here $r_{\rm line}$ is the distance of the line's origin from the center of the star and the last equality holds if $r_{\rm line}\simeq R_{\rm NS}$. Extracting the redshift via line spectroscopy, thus, in principle, provides a clean and direct way to measure the \ac{NS} compactness.

However, two issues arise for accreting systems. First, since these sources are usually fast spinning, spectral lines are subject to the Doppler effect, which results in the broadening and distortion of the line profile into an asymmetric shape \citep{OzelPsaltis2003}. This introduces an important systematic in measuring the compactness, which can be as high as $10\%$ \citep{OzelPsaltis2003}. Second, to associate the redshift with the compactness, we need to be able to identify the region that produces the spectral lines. However, this is not always possible for accreting sources. For example, \citep{Cottam2002} claimed the detection of narrow absorption lines in the burst spectra of the \ac{LMXB} EXO 0748--676, which they associated with an atmospheric origin, a claim that was dismissed by subsequent analyses. Nonetheless, for accreting sources, approximate knowledge of $r_{\rm line}$ still allows us to use Eq.~\eqref{eqn:redshift} to put a lower limit on the compactness, even without knowing $R_{\rm NS}/r_{\rm line}$.

As the aforementioned systematics only concern accreting systems and not \acp{XDINS}, the latter are better suited to measuring the gravitational redshift and extracting the compactness. We, therefore, classify this method as green in Tab. \ref{tab:compactness}. We do, however, point out that a definite detection of gravitationally redshifted atomic lines from either class of \acp{NS} has not been possible to date.

We finally note that \citep{Paerels1997} suggest that gravitationally shifted spectral lines from \ac{XDINS} atmospheres could be used not only to measure the compactness but provide further information that would allow us to break the degeneracy between $M_{\rm NS}$ and $R_{\rm NS}$. The idea is based on the assumption that spectral lines can change their shapes due to so-called pressure broadening, namely due to the Stark shift induced by the transient electric fields generated by charges in the proximity of the absorbing atoms. This effect depends primarily on the density of charges in the \ac{NS} atmosphere, which in turn is controlled by the gravitational acceleration on the surface. This quantity is $\propto M_{\rm NS}/R_{\rm NS}^2$. If a measurement of the gravitational acceleration supplements a measurement of the compactness obtained from the line's redshift, we thus would have sufficient constraints to solve for the mass as well as the radius of the \ac{NS} \citep{Paerels1997}.

In addition to the lack of actual measurements of gravitationally redshifted lines, we also note that pressure broadening is not the only line-broadening phenomenon taking place around \acp{NS}. Other contributions are provided by rotational broadening, thermal Doppler broadening, and magnetic broadening \citep{Ozel2013, Paerels1997}. The former is caused by the Doppler shift resulting from fast stellar rotation and does not affect slowly rotating \acp{XDINS}, whereas the second effect is due to the thermal motion of the atoms in the \ac{NS} atmosphere. The last mechanism is caused by the Zeeman effect. While \citep{Paerels1997} find that thermal Doppler broadening is subdominant to pressure broadening, magnetic broadening can become comparable to the pressure effect for magnetic fields above $10^9 \, \textrm{G}$ ($10^{10} \, \textrm{G}$) for the oxygen Ly$\alpha$ (Ly$\beta$) spectral line. The degeneracy between different broadening effects would significantly complicate the inference of the gravitational surface acceleration and corresponding mass and radius constraints for future line measurements.


\subsubsection{Pulsed profile modeling}
\label{sec:xray_ppm_compactness}

We discussed this relativistic ray-tracing technique in the context of constraining the \ac{NS} radius in Sec.~\ref{sec:xray_lc}. We remind that the compactness controls how strongly the trajectories of photons emitted from the star's surface are bent. Larger $C_{\rm NS}$ lead to a lower amplitudes of the pulsed emission (see Fig.~\ref{fig:pulsed_profile}), because light-bending causes a higher fraction of the stellar surface to be visible, preventing the occultation of hot spots as the star rotates. However, in contrast to the radius measurement discussed previously, we do not need to supplement the compactness information by the Doppler effect to break a degeneracy. However, other systematics outlined in Sec.~\ref{sec:xray_lc} still apply. Using this technique for \ac{NICER} observations of the \ac{MSP} PSR J0030+0451, \citep{Riley2019, Miller2019}, e.g., derive a compactness of $C_{\rm NS} \simeq 0.15 - 0.17$. 

As we do not have a measurement of the Doppler effect as an additional requirement,  slowly rotating \acp{NS} would also be useful targets for the determination of the compactness. We note, however, that slow rotators are often associated with strong magnetic fields (see Fig.~\ref{fig:p_pdot} and Sec.~\ref{sec:MD_braking}), which complicates the modeling of photon trajectories, introducing additional measurement uncertainties. As a result, we also classify this method as somewhat model dependent in Tab.~\ref{tab:compactness}.


\subsubsection{Quasi-periodic oscillations in magnetar giant flares}
\label{sec:QPOs-compactness}

As outlined in Sec.~\ref{sec:NS-zoo}, magnetars are strongly magnetized, highly variable \acp{NS}. Three such Galactic sources have shown so-called \textit{giant flares}, extremely powerful events which exhibit an initial ($< 1 \, {\rm s}$) spike during which X-ray luminosities can reach up to $10^{47} \, {\rm erg} \, {\rm s}^{-1}$, followed by a decay in brightness over a few hundred seconds. In these tails, strong \acp{QPO} with frequencies ranging from tens of Hz to kHz (generally grouped into low- and high-frequency modes below and above $\sim 250 \,$Hz, respectively) have been detected across various timescales \citep{Barat-etal1983, Israel-etal2005, StrohmayerWatts2005, WattsStrohmayer2006, StrohmayerWatts2006}. In addition, two broad kHz \acp{QPO} have also been observed in the main peak of an extragalactic gamma-ray giant flare \citep{Castro-Tirado-etal2021}. Although associated with a magnetar, the event's origin is somewhat uncertain as giant flares outside the Milky Way \citep[see also][]{Trigg-etal2023, Mereghetti-etal2023} are difficult to distinguish from regular short \acp{GRB}.

The exact physical mechanisms responsible for the generation of magnetar giant flares and subsequent \acp{QPO} are not yet fully understood. However, the burst trigger is typically associated with the slow build up of magnetic stresses and their subsequent catastrophic release, either via large-scale magnetohydrodynamic instabilities or crust quakes in the \ac{NS} interior \citep{ThompsonDuncan1995} or via magnetospheric reconnections in the stellar exterior \citep[e.g.,][]{Thompson-etal2002, Beloborodov2009}. The resulting giant flare then excites oscillations in the star \citep{Duncan1998}. As the corresponding mode frequencies are sensitive to a number of \ac{NS} properties, magnetar \acp{QPO} complement the \ac{GW} asteroseismology angle discussed in Secs.~\ref{sec:gw_asteroseismology_mass}, \ref{sec:gw_asteroseismology_radius}, \ref{sec:gw_asteroseismology_moi}, and \ref{sec:gw_asteroseismology_tidal}.

In particular, magnetar \acp{QPO} were initially interpreted as (discrete) torsional shear modes of the \ac{NS} crust \citep[e.g.,][]{Duncan1998, Messios-etal2001, Piro2005b, SamuelssonAndersson2007, Sotani-etal2008, SteinerWatts2009, Sotani-etal2016} or (continuum) Alfv\'en oscillations of the strongly magnetized fluid core \citep[e.g.,][]{CerdaDuran-etal2009, Sotani-etal2008, Colaiuda-etal2009}, thus providing access to crustal physics and the stellar magnetic field (for details see Secs.~\ref{sec:giantflares_qpo_crust} and \ref{sec:QPOs-Bfield}, respectively). However, \citep{Levin2006, Levin2007, Glampedakis-etal2006} found that for magnetar field strengths, oscillations in both \ac{NS} layers are effectively coupled, leading to a revised interpretation of \acp{QPO} as global magneto-elastic torsional oscillations \citep[e.g.,][]{Gabler-etal2011, vanHovenLevin2011}. Although this picture accounts for the low-frequency modes, additional physics are needed to explain those at high frequencies. The presence of superfluidity could play a crucial role here \citep{Gabler-etal2013, PassamontiLander2014, Gabler-etal2016} (see Sec.~\ref{sec:magnetarqpos_sf}). We also note that \acp{QPO} not only probe nucleonic physics but can also provide insights into the presence of other exotic phases of matter in \ac{NS} cores \citep[e.g.,][]{WattsReddy2007, Li-etal2023}.

Moreover, realistic mode calculations also need to account for relativistic effects, which introduces additional dependencies on the stellar mass and radius, specifically the compactness, $C_{\rm NS}$ \citep{SamuelssonAndersson2007, Sotani-etal2008}. Invoking a global magneto-elastic picture, \citep{Gabler-etal2016, Gabler-etal2018} find that the fundamental mode in the low-frequency regime satisfies the following expression (see also \citep{Sotani-etal2008})
 \begin{align}
 	f \simeq 2.5 f_0 \left[1 - 4.58 \frac{M_{\rm NS}}{R_{\rm NS}}
        + 6.06 \left( \frac{M_{\rm NS}}{R_{\rm NS}}\right)^2 \right],
 \end{align}
where $f_0$ is a reference frequency that encodes all other relevant physics (see Eq.~\eqref{eqn:me_fundamental} for details).

Although relations of this form are, in principle, straightforward to apply to observed \ac{QPO} frequencies, we point out that the key issue remains the correct association of observed modes with the underlying oscillation physics. Identifying a \ac{QPO} frequency at $29 \,$Hz in the 2004 giant flare of the magnetar SGR 1806--20 \citep{StrohmayerWatts2006} with the above expression, \citep{Gabler-etal2018} use a Bayesian approach to find that their model favors a relatively small \ac{NS} compactness of $C_{\rm NS} \lesssim 0.19$ at $68$\% credibility level. Further observational data is needed to confirm this trend. However, the extreme rarity of giant flares as well as difficulties in accurately identifying \ac{QPO} frequencies and their decay timescales \citep[e.g.,][]{Pumpe-etal2018, Miller-etal2019b} significantly hamper further studies. As a result of these limitations, we classify this method as orange in Tab.~\ref{tab:compactness}.


\subsection{Magnetic field}
\label{sec:Bfields}


\begin{table*}[]
    \centering
    \begin{tabular}{|c||c|c|c|c|c|}
        \hline
        \textbf{Method} & \textbf{GWs} & \textbf{Radio} & \textbf{Optical}  & \textbf{X-ray} & $\mathbf{\gamma}$\textbf{-ray}  \\
        \hline
        \hline
            \hyperref[sec:MD_braking]{Magnetic dipole breaking} & &\cellcolor{yellow!40} & \cellcolor{yellow!40} & \cellcolor{yellow!40}   & \cellcolor{yellow!40} \\
            \hline
            \hyperref[sec:cyclotron]{Cyclotron lines} & & &  & \cellcolor{yellow!40}    & \\
            \hline
            \hyperref[sec:linespec-Bfield]{Line spectroscopy (line broadening)} & & &  & \cellcolor{orange!40}  &   \\
            \hline
            \hyperref[sec:QPOs-Bfield]{Giant-flare \acp{QPO}} & & &  & \cellcolor{orange!40}    &  \\
            \hline
    \end{tabular}
    \caption{Magnetic-field measurements. Individual rows represent the methods that probe the \ac{NS} magnetic field, while columns denote different messengers/EM wavebands. Individual colors are equivalent to those in Tab.~\ref{tab:mass}.}
    \label{tab:mf}
\end{table*}


Having provided an overview of measurements of global parameters, i.e., those that take a single value for a given \ac{NS}, we now turn to quantities that can vary across the star.

We first focus on the stellar magnetic field, which plays a significant role in shaping the dynamics and the emission properties of \acp{NS}. Below, we primarily address those methods (see Tab.~\ref{tab:mf}) that probe the strength of the magnetic induction, $B$, discuss the stellar regions that these estimates apply to and highlight difficulties in constraining our understanding of the magnetic-field geometry. In addition to focusing on the field on (or close to) the stellar surface, we also outline a potential approach to constrain the magnetic properties of the \ac{NS} interior, which are still highly uncertain.


\subsubsection{Magnetic-dipole braking}
\label{sec:MD_braking}

As highlighted in Sec.~\ref{sec:NS-zoo}, pulsar timing enables reliable determination of the stellar spin period, $P$, and its period derivative, $\dot{P}$. Knowledge of both quantities allows us to group these compact objects into distinct classes and analyze their properties. In particular, pulsar periods are observed to increase with time suggesting the loss of rotational energy, which in turn powers the star's electromagnetic radiation.

In the simplest picture, we can consider the star as a rotating magnetic dipole in vacuum, whose magnetic axis is misaligned with the rotation axis by an angle, $\psi$. Such a configuration has a time-varying dipole moment as seen from infinity and classical electrodynamics dictates the radiation of energy with frequency, $\Omega$, at the following rate \citep{ShapiroTeukolsky1983}:
\begin{equation}
	\dot{E} \approx - \frac{B_{\rm dip}^2 R_{\rm NS}^6}{6 c^3} \,  \Omega^4 \sin^2 \psi.
		\label{eq:Edot_vacuum}
\end{equation}
Here, $B_{\rm dip}$ is the magnetic-field strength at the magnetic pole on the stellar surface. This dictates $\dot{E} = 0$ for an aligned rotator with $\psi = 0$, suggesting that such a pulsar cannot emit \ac{EM} radiation. However, in reality, \acp{NS} are not surrounded by a vacuum but with plasma-filled magnetospheres \citep{GoldreichJulian1969}. Corresponding three-dimensional simulations have shown that the rotational evolution of realistic \acp{NS} is instead better approximated by the following two equations \citep{Spitkovsky2006}:
\begin{align}
	\dot{\Omega} &\approx - \frac{B_{\rm dip}^2 R_{\rm NS}^6}{4 c^3 I_{\rm NS}} \, \Omega^3
			\left( \kappa_0 + \kappa_1 \sin^2 \psi \right),
		\label{eq:Omegadot_realistic} \\[1.4ex]
	\dot{\psi} &\approx - \frac{B_{\rm dip}^2 R_{\rm NS}^6}{4 c^3 I_{\rm NS}} \, \Omega^2
			\kappa_2 \sin \psi \cos \psi,
		\label{eq:chidot_realistic}
\end{align}
where $\kappa_0 \simeq \kappa_1 \simeq \kappa_2 \simeq 1$ for typical \ac{NS} magnetospheres \citep{Philippov-etal2014}. The vacuum solution is recovered for $\kappa_0 =0$, $\kappa_1 = 2/3$ and $\kappa_2 = 1$. Invoking Eq.~\eqref{eq:braking_index} for the braking index, we thus obtain $n=3$ for a spin-down driven by magnetic dipole braking as already discussed in Sec.~\ref{sec:cgw}. In this case, the characteristic age, defined as $\tau_{\rm c} \equiv - \Omega / [(n-1)\dot{\Omega}]$ reduces to $\tau_{\rm c} = -\Omega / 2 \dot{\Omega} = P / 2 \dot{P}$ as mentioned in Sec.~\ref{sec:NS-zoo}. The second equation suggests that the misalignment angle between the rotation and the magnetic field axis decreases with time, i.e., pulsars typically evolve towards alignment.

Focusing, e.g., on $\psi = 0$ for simplicity and using $\Omega = 2 \pi / P$, we can rewrite Eq.~\eqref{eq:Omegadot_realistic} to relate the dipolar magnetic-field strength at the pole to observable quantities as follows:
\begin{align}
	B_{\rm dip} &\approx \sqrt{ \frac{c^3 I_{\rm NS}} {\pi^2 R_{\rm NS}^6} P \dot{P}} \nonumber \\[1.4ex]
		& \approx 1.7 \times 10^{12} \left( \frac{R_{\rm NS}}{10 \, {\rm km}} \right)^{-3}
			\left( \frac{I_{\rm NS}}{10^{45} \, {\rm g} \, {\rm cm}^2} \right)^{1/2}  \nonumber \\[1.4ex]
		&\times \left( \frac{\dot{P}}{10^{-15} \, {\rm s} \, {\rm s}^{-1}} \right)^{1/2}
			\left( \frac{P}{1 \, {\rm s}} \right)^{1/2} \, {\rm G}.
		\label{eq:dipol_estim}
\end{align}
We used this relation, which is in principle applicable across the \ac{EM} spectrum, to estimate the magnetic-field strengths of known pulsars as represented by the color in the $P-\dot{P}$ diagram shown in Fig.~\ref{fig:p_pdot}.

Note that the focus on the dipolar field component is certainly a strong simplification. However, it likely provides an acceptable description for radio pulsars whose radio beams are typically associated with lighthouse-like radiation, characteristic for a dipolar field arrangement \citep{LorimerKramer2012}. For magnetars, which generally are not observed as regular radio emitters, the magnetic-field geometry is likely more complex \citep{Vigano-etal2013, GourgouliatosCumming2014, WoodHollerbach2015, Vigano-etal2021, DeGrandis-etal2021, Dehman-etal2023}. Nonetheless, we point out that higher-order field components decay faster than the dipolar component as we move away from the stellar surface. As a result, a dipolar magnetic field should dominate the field geometry at large distances even for magnetars. Consequently, Eq.~\eqref{eq:dipol_estim} is useful for giving a first indication of typical magnetic fields strengths but should not be identified with the true \ac{NS} field. Moreover, additional uncertainties around the nature of pulsar magnetospheres, the presence of pulsar winds \citep[e.g.,][]{GaenslerSlane2006} and the angle between the rotation and the magnetic axis \citep{TaurisManchester1998, Rookyard-etal2015} might easily change the estimate \eqref{eq:dipol_estim} by a factor of $\sim 2-5$. Note also that consideration of the surface dipole field at the equatorial belt (and not the \ac{NS} pole) would reduce the above estimate by another factor of $2$.

As a result of these caveats and the additional dependencies on the uncertain values of $I_{\rm NS}$ and $R_{\rm NS}$, we classify this method as yellow in Tab.~\ref{tab:mf}.


\subsubsection{Cyclotron lines}
\label{sec:cyclotron}

Other means to estimate the \ac{NS} magnetic-field strength employ \textit{cyclotron lines}. These spectral features, typically observed in absorption, are also known as \acp{CRSF} and have been discovered in the X-ray emission of magnetars, \acp{XDINS} and accreting X-ray pulsars \citep{Rea-etal2008, Ferrigno-etal2011, Borghese-etal2015, Staubert-etal2019, Diez-etal2022}. By comparing observed cyclotron lines with model predictions, we can refine our understanding of the \ac{NS} magnetic field. In particular, the energies of charged particles moving in strong magnetic fields are quantized (so-called \textit{Landau levels}), so that resonant scattering of photons on these particles leads to characteristic absorption features in the X-ray spectra at the resonance energies. Corresponding models can, thus, predict expected locations and profiles of the cyclotron lines based on the field strength and other parameters (see, e.g., \citep{Schwarm-etal2017a, Schwarm-etal2017b} for recent work on modeling \acp{CRSF}). Consequently, the energies of cyclotron lines, for a given charged particle species, are directly related to the magnetic field in the vicinity of the line formation region.

More precisely, in the case of an electronic origin of the cyclotron resonance, the energy of the centroid of the line, which is typically fitted with a Gaussian profile, is related to the (local) $B$-field strength as \citep[e.g.,][]{Staubert-etal2019}:
\begin{align}
  E_{\rm cyc} &= \frac{\hbar e}{m_{\rm e} c} n_L \, \frac{1}{1+z} \, B  \nonumber \\[1.4ex]
		&\approx  11.6\,n_L \, \frac{1}{1+z} \left( \frac{B}{10^{12} \, {\rm G}} \right) {\rm keV},
        \label{eqn:E_cycl}
\end{align}
where $\hbar$ is the reduced Planck constant, $e$ the electric charge, $m_{\rm e}$ the mass of the electron, $z$ the gravitational redshift due to the \ac{NS}'s compactness defined in Eq.~\eqref{eqn:redshift} and $1/(1+z)$ a term of order 1. Moreover, $n_L$ denotes the number of the Landau level involved in the interaction, e.g., $n_L = 1$ represents the case of scattering from the ground level to the first excited Landau level. The resulting line is referred to as the \textit{fundamental line}; higher $n_L$ are referred to as \textit{harmonics}.

The same physics also applies to other charged particles. For protons, for example, we have to rescale the mass in Eq.~\eqref{eqn:E_cycl}, which reduces the separation of the Landau levels, and, hence, the line's centroid energy, $E_{\rm cyc}$, by three orders of magnitude (assuming the same $n_L$, $M_{\rm NS}$, $R_{\rm NS}$ and $B$) \citep{Zane-etal2001}. Proton lines are, therefore, likely only observable in the X-ray emission of magnetars and \acp{XDINS}, whose fields are stronger than those of canonical pulsars \citep{Staubert-etal2019}.

Applying this approach to \ac{NS} observations provides the most direct way of measuring magnetic-field strengths to date, and, for non-accreting pulsars, has led to estimates in line with those obtained through the dipolar braking method discussed in Sec.~\ref{sec:MD_braking} \citep{Borghese-etal2015, Borghese-etal2017, Staubert-etal2019}. We highlight, however, that cyclotron lines, in principle, also offer information about the local magnetic-field strength and, therefore, the field geometry which typically remains elusive \citep[see, however, e.g.,][]{Petri-etal2023}. \citep{Borghese-etal2015, Borghese-etal2017}, e.g., analyzed narrow cyclotron features in two \acp{XDINS} that showed strong dependence on the pulsars' rotational phase and were, therefore, associated with confined magnetic structures close to the stellar surface.

However, note that although this approach is relatively straightforward once cyclotron lines are detected, the main detection bias lies in identifying the resonance's origin, i.e., whether photons are scattered by electrons or protons. This is particularly important for strongly magnetized \acp{NS}, where field estimates differ by several orders of magnitude depending on the underlying charged particles \citep{Rea-etal2008}. In strongly magnetized sources, the situation is further complicated because other effects such as complex inhomogeneous temperature distribution on the stellar surface \citep{Vigano-etal2014}, or photo-ionization close to the star \citep{Hambaryan-etal2009} can also lead to absorption.

As a result, we classify this method as yellow in Tab.~\ref{tab:mf}.


\subsubsection{Line spectroscopy}
\label{sec:linespec-Bfield}

In Sec.~\ref{sec:linespec-compactness}, we mentioned how Stark effect-induced line broadening can provide access to the gravitational surface acceleration of a star. When combined with its compactness, this measurement enables us to deduce both the star's mass and radius. However, we acknowledged the existence of other spectral broadening effects, in particular the Zeeman effect. Following a similar rationale, we can also estimate the \ac{NS}'s (local) magnetic field in the emission region if we attribute line broadening to the Zeeman effect \citep{SarazinBachall1977, Paerels1997}. In this case, the line broadens according to $\Delta E_{\rm Zeeman} = \mu_B B$, with $\mu_B$ representing the Bohr magneton \citep{Paerels1997}.

However, this method faces the same constraints as outlined above. We need to establish the region around the \ac{NS} from which the line emission originates and understand the physical mechanism responsible for the broadening. Furthermore, as previously mentioned, to date, no unambiguous observation of a gravitationally redshifted atomic line originating from close to a \ac{NS} has been made. We, hence, classify this approach again as strongly model dependent in Tab.~\ref{tab:mf}.


\subsubsection{Quasi-periodic oscillations in magnetar giant flares}
\label{sec:QPOs-Bfield}

The magnetic field plays an important role in magnetar \acp{QPO} (see Sec.~\ref{sec:QPOs-compactness} for an overview of the phenomenon). Specifically, \acp{QPO} have been associated with the excitation of torsional Alfv\'en oscillations in the magnetized fluid core of \acp{NS}. These vibrations do not form a discrete set of modes as magnetic fields lines have a continuous range of lengths but \citep{ThompsonDuncan2001} roughly estimate the frequency of the fundamental mode as
\begin{align}
	f_{\rm A} \simeq \frac{v_{\rm A}}{4 R_{\rm NS}}
		&\approx 7.1 \left( \frac{\rho_{\rm c}}{10^{14} \, {\rm g} \, {\rm cm}^{-3}} \right)^{-1/2} \nonumber \\[1.4ex]
		&\times  \left( \frac{B}{10^{14} \, {\rm G}} \right)
		\left( \frac{R_{\rm NS}}{10 \, {\rm km}}\right)^{-1}  \, {\rm Hz},
			\label{eqn:Alfven-QPO}
\end{align}
in the range of observed low-frequency \acp{QPO} \citep{Barat-etal1983, Israel-etal2005, StrohmayerWatts2005, WattsStrohmayer2006, StrohmayerWatts2006}. Here, $v_{\rm A} \equiv B / \sqrt{4 \pi \rho_{\rm c}}$ is the Alfv\'en speed, which introduces a dependence on the (mean) internal magnetic field, $B$, and the mass density, $\rho_{\rm c}$, of the charged particles in the stellar interior. Note that the latter constitutes only a few percent of the total core mass. However, the presence of superfluidity (see Sec.~\ref{sec:magnetarqpos_sf}) might couple the protons and neutrons, thus increasing the mass fraction involved in the oscillations \citep{Turolla-etal2015}. This complicates $B$-field estimates from measured \ac{QPO} frequencies using Eq.~\eqref{eqn:Alfven-QPO} or similar relations.

Moreover, while Alfv\'en oscillations confined to the \ac{NS} core might lead to the formation of long-lived \acp{QPO} \citep{Levin2007, Sotani-etal2008, CerdaDuran-etal2009, Colaiuda-etal2009}, the picture is more complex because a global magnetic field tightly couples the \ac{NS} crust and core, involving the entire star in the torsional vibrations. The presence of a $B$-field has two additional effects: First, a strong magnetic field shifts the crustal oscillations (see Sec.~\ref{sec:giantflares_qpo_crust}) to higher frequencies \citep{Sotani-etal2008, Gabler-etal2018} according to
\begin{equation}
	f_{\rm crust} \simeq f_{\rm crust, 0} \left[ 1 + \zeta \left( \frac{B}{10^{14} \, {\rm G}} \right)^2 \right]^{1/2},
        \label{eqn:Bfield_mod_fcrust}
\end{equation}
where $f_{\rm crust, 0}$ is a crustal mode frequency in the absence of magnetic fields and the coefficient $\zeta$ (of order $10^{-2}$ \citep{Gabler-etal2018}) is obtained from fits to numerical simulations. This process could explain the observed high-frequency \acp{QPO} provided that several high overtones are excited. However, this does not answer why only a few of these modes are observed and not a whole spectrum. Superfluidity has been suggested to play a key role here \citep{Gabler-etal2013, PassamontiLander2014, Gabler-etal2016} (see Sec.~\ref{sec:magnetarqpos_sf}). The magnetic field further strongly affects whether global magneto-elastic oscillations reach the \ac{NS} surface. In particular, \citep{Gabler-etal2012, Gabler-etal2018} find that magneto-elastic \acp{QPO} are primarily confined to the \ac{NS} core for fields below $B \simeq 10^{15} \, {\rm G}$. Only above this threshold are \acp{QPO} with sufficient (detectable) amplitudes found on the surface. If the observed \acp{QPO} are indeed correctly described by such a model, this would provide a lower bound on the (mean) magnetic field in the magnetar interior. Detailed estimates of the \textit{breakout field} are, however, also dependent on superfluid effects and the nuclear \ac{EOS} \citep{Gabler-etal2018}.

Finally, note that mode calculations rely on the assumption of specific magnetic field geometries, and are often restricted to dipolar configurations for simplicity (see however \citep{Messios-etal2001, Sotani-etal2008b, Sotani2015, LinkvanEysden2016, deSouzaChirenti2019}). This is particularly apparent for Alfv\'en oscillations depending crucially on the length of $B$-field lines, and, hence, the field geometry. As a result, \ac{QPO} frequency relations extracted from theoretical calculations or numerical simulations are generally only valid for specific field configurations. Although promising, this currently hinders us from fully exploring magnetar \ac{QPO} asteroseismology to pin down the highly uncertain \ac{NS} field geometry.

As already highlighted in Sec.~\ref{sec:QPOs-compactness}, the main issue of this method is, thus, the identification of the underlying physics for a given mode frequency. In addition, the magnetic-field strength and its geometry are highly degenerate with other stellar properties further complicating this approach. Consequently, we classify the use of magnetar \acp{QPO} as probes for the \ac{NS} magnetic field as orange in Tab.~\ref{tab:mf}.


\subsection{Crust physics}
\label{sec:crust_prop}


\begin{table*}[]
    \centering
    \begin{tabular}{|c||c|c|c|c|c|}
        \hline
        \textbf{Method} & \textbf{GWs} & \textbf{Radio} & \textbf{Optical}  & \textbf{X-ray} & $\mathbf{\gamma}$\textbf{-ray} \\
        \hline
        \hline
             \hyperref[sec:PPdot_cutoff]{Magnetar spin-period limit} &   & \cellcolor{orange!40} &  & &  \\
            \hline
             \hyperref[sec:X-rayburst_cooling_crustprop]{Thermonuclear X-ray bursts} &   & &  & \cellcolor{orange!40}  & \\
            \hline
            \hyperref[sec:GW-crustpropr]{GW from mountains} & \cellcolor{orange!40}   & &  &  & \\
            \hline
            \hyperref[sec:resonant_shattering_flares]{Resonant shattering flares} & \cellcolor{orange!40}  & &  &   & \cellcolor{orange!40}  \\
            \hline
            \hyperref[sec:giantflares_qpo_crust]{Giant-flare QPOs} &   & &  & \cellcolor{orange!40}  &    \\
            \hline
            \hyperref[sec:glitches-crustprop]{Pulsar glitches} & & \cellcolor{orange!40} &  & \cellcolor{orange!40} & \cellcolor{orange!40} \\
             \hline
    \end{tabular}
    \caption{Crustal measurements. Individual rows represent the methods that probe the \ac{NS} crust physics, while columns denote different messengers/EM wavebands. Individual colors are equivalent to those in Tab.~\ref{tab:mass}.}
    \label{tab:crust_comp}
\end{table*}


This section is dedicated to the different techniques that constrain the composition and the properties of the outer \ac{NS} layer. For a comprehensive review of crustal physics, briefly summarized in Sec.~\ref{sec:ns-structure} and Fig.~\ref{fig:ns_structure}, see, e.g., \citep{Chamel2008}.

Observational information comes from across the \ac{EM} spectrum as well as \acp{GW}. As highlighted in Tab.~\ref{tab:crust_comp}, it is difficult to derive model-independent limits on crustal properties and corresponding constraints are primarily qualitative in nature. As a result, all methods are classified as orange in Tab.~\ref{tab:crust_comp}. We will outline underlying model assumptions and systematics individually. Furthermore, we note that some of the techniques addressed below are also sensitive to the properties of the stellar core, albeit with lower significance due to larger uncertainties around the core's composition and properties. Whenever this is the case, we, however, highlight this connection for the benefit of the reader.


\subsubsection{Spin-period limit in magnetars}
\label{sec:PPdot_cutoff}

As highlighted in the $P-\dot{P}$ diagram shown in Fig.~\ref{fig:p_pdot}, the spin-period distribution of the pulsar population appears to exhibit a cut-off at around $10\,$s. While in the radio band such a period limit can be explained by both a limitation in the radio emission mechanism \citep[e.g.,][]{ChenRuderman1993, Zhang-etal2000} and observational biases in detecting long-period radio signals \citep{Morello-etal2020, Singh-etal2022},\footnote{We note that the recent discovery of several coherent periodic radio sources with ultra-long periods \citep{Caleb-etal2022, HurleyWalker-etal2022, HurleyWalker-etal2023} has begun to open up the parameter space beyond periods of $60 \,$s. However, the nature of these sources remains uncertain and their properties difficult to reconcile with our current understanding of \ac{NS} evolution \citep[e.g.,][]{Rea-etal2023}.} these limitations do not apply to magnetars \citep{Pons-etal2013}. Instead, the lack of detections of isolated X-ray pulsars with periods above $12\,$s has been associated with internal \ac{NS} physics and interpreted as intrinsic to the magnetar population.

In particular, extensive 2D and 3D simulations of \ac{NS} magnetic-field evolution \citep[e.g.,][]{Vigano-etal2013, GourgouliatosCumming2014, WoodHollerbach2015, Vigano-etal2021, DeGrandis-etal2021, Dehman-etal2023} have shown how the crustal composition and the field geometry affect the spin-period evolution of highly magnetized \acp{NS}. In these objects, magnetic fields are thought to have evolved in such a way that after $\sim 10^5 \,$yrs, electric currents are primarily located close to the crust-core interface. The resistive properties of this region, thus, control the dissipation of magnetic energy. Specifically, a highly resistive layer, as typically expected for amorphous, impure nuclear pasta (see Sec.~\ref{sec:ns-structure}), drives fast magnetic-field decay. The interplay between $B$ and $P$ via Eq.~\eqref{eq:Omegadot_realistic}, thus, implies a natural limit for the spin period, which \citep{Pons-etal2013} found to be in the $10 - 20\,$s range, in accordance with the observations.

Although this hints at the presence of a nuclear pasta phase in the inner crust, constraints remain strongly model dependent. In addition to uncertainties around the magnetic-field geometry in the \ac{NS} interior, which strongly affects the above picture, we also point out that transport properties of nuclear pasta are still uncertain \citep[e.g.,][]{Horowitz2008, Yakovlev2015, Nandi2016}, complicating quantitative constraints on the crust's composition. As a result, we classify this method as orange in Tab.~\ref{tab:crust_comp}.


\subsubsection{Thermonuclear X-ray bursts}
\label{sec:X-rayburst_cooling_crustprop}

As mentioned in Sec.~\ref{sec:pre}, accreting pulsars exhibit thermonuclear X-ray bursts due to the unstable burning of matter in the outer \ac{NS} layers following extended periods of accretion \citep[e.g.,][]{vanParadijs1979, Lewin-etal1993, Strohmayer2006, Poutanen2014, Ozel2016, Patruno2021}. The phenomenon is common and several thousand of these Type-I X-ray bursts have now been observed in around 100 sources \citep{Galloway-etal2020}, demonstrating a large diversity. Analysing the emission and contrasting data with theoretical models, thus, provides a plethora of information on the \ac{NS} crust.

Burst oscillations with frequencies between $\sim 10 - 600 \, {\rm Hz}$ are often observed in the X-ray fluxes during the rise and decay phases of bursts \citep[e.g.,][]{Watts2012, Bilous2019}. These frequencies are typically closely associated with the \ac{NS} rotation rate (and thus also used to measure the stellar spins), likely originating from inhomogeneous structures on the stars' surface \citep{Chakrabarty-etal2003, Spitkovsky-etal2002}. Two classes of models exist for these high-frequency oscillations, both relying on the fact that small temperature anomalies on the \ac{NS} surface cause localized ignition and subsequent spreading of the flame front. However, in \textit{hot-spot models}, the resulting burning is confined to a small region of the star due to the stalling of the flame. This is potentially connected to the crustal magnetic field \citep[e.g.,][]{Brown1998, Cavecchi-etal2011} and naturally explains oscillations at the spin frequency. In contrast, in \textit{global-mode models}, the flame spreads over the entire stellar surface subsequently exciting shallow (but large-scale) waves in the \ac{NS} crust. In this picture, the association between burst-oscillation and spin frequencies sets tight constraints on the kinds of modes that can be excited \citep[e.g.,][]{Heyl2004, Piro2005}, which in turn depend strongly on the composition of crustal matter \citep{Chamel2008, Watts2012}.

Overall, ignition conditions, burst characteristics and the stars' temperature evolution are strongly dependent on the thermal properties of matter in the crust (and the core) \citep{Colpi-etal2001, Cumming-etal2006, Meisel-etal2018}. As the composition of burning ashes left behind from thermonuclear explosions is not accessible through observations due to the star's strong surface gravity and continued accretion burying any burnt material, original crust compositions cannot be directly inferred. Instead, a comparison between burst light curves and theoretical models, requiring detailed knowledge of nuclear reaction chains \citep{Schatz2006, Parikh-etal2013}, can provide insights into the make up of the outer \ac{NS} layers.

Tracking thermal emission post-outburst plays a crucial role in probing additional physics \citep[e.g.,][]{Wijnands-etal2017}. In this quiescent state, luminosities drop by several orders of magnitude compared to the burst but are much higher than in isolated, old \acp{NS}, because burning ashes move inwards and react with the existing crust \citep[e.g.,][]{Sato1979, Haensel1990}, providing an additional heating source \citep{Brown-etal1998}. We can specifically compare models of this \textit{deep crustal heating process} to the observed behavior of transient sources \citep[e.g.,][]{Campana-etal1998, Yakovlev-etal2003, Wijnands-etal2003}, where accretion episodes last a few days to weeks, and quasi-persistent sources \citep[e.g.,][]{Rutledge-etal2002, Cackett-etal2006, Shternin-etal2007}, where accretion can last for years up to decades. In the former case, \acp{NS} are generally in a steady-state, where cooling processes (neutrino emission from the entire \ac{NS} volume and photon emission from the surface; see Sec.~\ref{sec:ns_cooling}) are balanced by deep crustal heating. The heat generated by the nuclear explosions is completely radiated away \citep{Chamel2008}:
\begin{equation}
    L_{\rm neutrino} + L_{\rm photon} = L_{\rm crustal \, heating} = Q \, \frac{\langle \dot{M} \rangle}{m_{\rm u}},
\end{equation}
where $Q$ is the total heat release per accreted nucleon, $\langle \dot{M} \rangle$ the time-averaged accretion rate and $m_{\rm u}$ the atomic mass unit. Combining observations with reasonable estimates for $\langle \dot{M} \rangle$ and $Q$, thus allows a constraint on the neutrino cooling rate, or vice versa \citep[e.g.,][]{YakovlevPethick2004, Wijnands-etal2017}. This approach provided, e.g., a strong hint for the presence of direct Urca reactions in the core of the first transiently accreting \ac{MSP} \citep{Heinke-etal2009}.

For quasi-persistent sources, extended accretion periods lead to more significant heating in the stellar crust, bringing it out of equilibrium with the core \citep[e.g.,][]{Page2013}. Observing the subsequent cooling behavior in quiescence allows the detailed study of thermal crust properties. This approach has highlighted, for example, the need for a primarily pure crust with high thermal conductivity (required due to short inferred cooling timescales) \citep[e.g.,][]{Brown2009, Roggero2016}, which also suggests the presence of a purification process because burning ashes themselves are highly impure \citep{Meisel-etal2018}. Moreover, several cooling curves are better explained with a localised, high-resistivity region in the inner crust, effectively providing an insulating layer \citep{Horowitz-etal2015, Deibel-etal2017}. As discussed in Sec.~\ref{sec:PPdot_cutoff}, this is naturally explained by a nuclear pasta layer (although transport properties remain uncertain \citep[e.g.,][]{Horowitz2008, Yakovlev2015, Nandi2016}). Finally, within our current knowledge of nuclear reaction rates, observations of high temperatures at the onset of quiescence suggest the need for an additional heating source, so-called \textit{shallow heating}, at densities $\rho \lesssim 10^{11} \, {\rm g \, cm}^{-3}$ \citep[e.g.,][]{Brown2009, Turlione-etal2015, Parikh-etal2017}. Its physical origin remains uncertain but highlights the importance of quasi-persistent sources in improving our knowledge of \ac{NS} crust physics. For a separate discussion of implications for superfluidity see Sec.~\ref{sec:accretingNScooling_sf}.

In summary, Type-I X-ray bursts in accreting \acp{NS} have strong potential to probe a large number of crustal properties. However, as outlined above, theoretical models are complex and a number of open problems and systematics (like uncertain distances, unknown envelope compositions and variability in the accretion rate) make it difficult to set quantitative constraints on the \ac{NS} crust, despite plenty of burst observations. We thus classify this method as orange.


\subsubsection{Continuous gravitational-wave emission from mountains}
\label{sec:GW-crustpropr}

\begin{figure}
    \centering
    \includegraphics[width = 0.47\textwidth]{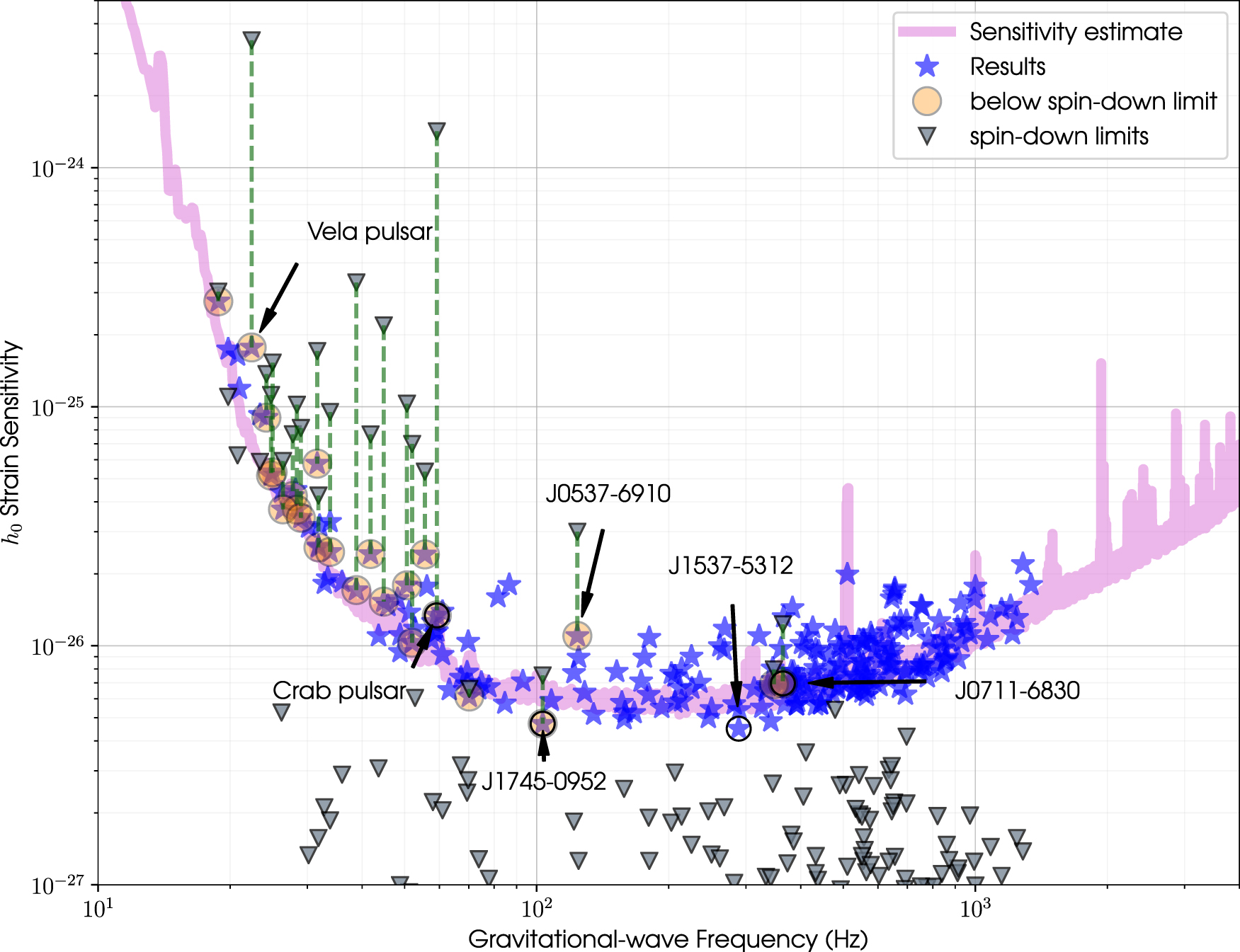}
    \caption{Upper limits on the \ac{GW} strain, $h_0$, for 237 known pulsars obtained through targeted searches in the cumulative O1-O3 data of the LIGO and Virgo \ac{GW} observatories \citep{Abbott-etal2022b}. Spin-down limits are shown as gray triangles, while blue stars highlight $95\%$ credible upper limits on $h_0$. Those pulsars where the \ac{GW} information surpasses the spin-down limit are shown with orange circles. The pink curve represents the expected sensitivity of all three \ac{GW} detectors during O3. Image reproduced with permission from \citep{Abbott-etal2022b}. Copyright by AAS.}
    \label{fig:GW_strains}
\end{figure}

As outlined in Sec.~\ref{sec:cgw}, rotating \acp{NS} with asymmetric deformations emit quasi-chromatic, continuous \acp{GW}. For small deformations and rigid rotation with angular frequency, $\Omega$, about a principle axis, a scenario typical for a \ac{NS} with a mountain, the star loses energy according to Eq.~\eqref{eqn:cgw_Eloss} and emits \acp{GW} at $2\Omega$ with an amplitude given by Eq.~\eqref{eqn:GWstrain_mountain}. Corresponding \ac{NS} observations constrain the ellipticity, $\varepsilon$.

First, because the \ac{GW} luminosity can be at most equivalent to the loss of kinetic energy caused by the pulsar's \ac{GW} spin-down, pulsar timing data provide an upper limit on the ellipticity. The most stringent constraints on the maximum ellipticity that a \ac{NS} can support come from young, energetic objects, because older \ac{NS} are expected to retain little of their initial asymmetry unless they accrete matter from a companion (see below). For the Crab pulsar, we, for example, obtain the following \textit{spin-down limit} from Eq.~\eqref{eq:cgw_spindown} assuming that the pulsar's spin down is entirely driven by \ac{GW} emission:
\begin{align}
	\varepsilon &< 7.4 \times 10^{-4} \left( \frac{I_3}{10^{45} \, {\rm g} \, {\rm cm}^2} \right)^{-1/2}
			\left( \frac{P}{33 \, {\rm ms}} \right)^{3/2} \nonumber\\[1ex]
			&\times \left( \frac{\dot{P}}{4.21 \times 10^{-13} \, {\rm s} \, {\rm s}^{-1}} \right)^{1/2} .
		\label{eqn:spindownlimit_Crab}
\end{align}
For a pulsar with a radius of $R_{\rm NS} = 10\,$km, this corresponds to a mountain height of $\sim \varepsilon R_{\rm NS} \lesssim 7.4\,$m.

Complementary constraints, beating the spin-down limits in several cases, come from the non-detection of continuous \ac{GW} signals from \acp{NS} with advanced \ac{GW} detectors (for recent reviews see \citep{Riles2023, Wette2023, HaskellBejger2023}). These searches lead to upper limits on the maximum strain, $h_0$ (see Fig.~\ref{fig:GW_strains}), which in turn sets constraints on $\varepsilon$ via Eq.~\eqref{eqn:GWstrain_mountain}. Specifically
\begin{align}
	\varepsilon &< 2.4 \times 10^{-5} \left( \frac{I_3}{10^{45} \, {\rm g} \, {\rm cm}^2} \right)^{-1}
			\nonumber\\[1ex]
			&\times \left( \frac{D}{1 \, {\rm kpc}} \right) \left( \frac{P}{10 \, {\rm ms}} \right)^{2} \left( \frac{h_0}{10^{-24}} \right).
\end{align}

For `blind' all-Sky searches of unknown sources, current ellipticity limits span $\varepsilon \simeq 10^{-4} - 10^{-5}$ \citep{Abbott-etal2021b, Abbott-etal2021c, Covas-etal2022}. Tighter constraints are obtained for studies of known or suspected pulsars, such as young \acp{NS} \citep{Abbott-etal2020a, Abbott-etal2022a, Abbott-etal2022b} (including those in supernova remnants \citep{Lindblom2020, Abbott-etal2021a, Abbott-etal2022d}) or accreting compact objects in binaries \citep{Zhang-etal2021, Abbott-etal2022c}. For example, for the Crab pulsar the spin-down limit \eqref{eqn:spindownlimit_Crab} has been beaten by a factor $100$ \citep{Abbott-etal2022b}. The tightest upper limit on $\varepsilon$ comes from observations of an old \ac{MSP} (PSR J0711--6830), which is $\varepsilon \lesssim 5.26 \times 10^{-9}$ \citep{Abbott-etal2022b}, corresponding to a mountain of less than $5.3\,$mm.

These constraints are important for several reasons. First, a number of physical mechanisms have been suggested as the cause for asymmetric deformations, but underlying physics are highly uncertain \citep[e.g.,][]{Lasky2015, Glampedakis-etal2018, Gittins-et-al-2021b}. This includes, e.g., intrinsic asymmetries due to non-axisymmetric magnetic-field configurations (so-called \textit{magnetic mountains} \citep{Bonazzola-etal1996, Haskell-etal2008}), where we can estimate the ellipticity as
\begin{align}
    \varepsilon \simeq \frac{B^2 R_{\rm NS}^4}{G M_{\rm NS}^2}
        &\approx 1.9 \times 10^{-12} \left(\frac{B}{10^{12} \, {\rm G}}\right)^2
                \nonumber \\[1.4ex]
        &\times \left(\frac{R_{\rm NS}}{10 \, {\rm km}}\right)^4
            \left(\frac{M_{\rm NS}}{1.4 M_{\odot}}\right)^{-2}.
\end{align}
This value falls well below current \ac{GW} limits. Deformations might, however, be larger for young \acp{NS} with strong toroidal components and magnetars, where $\varepsilon$ can reach up to $10^{-6}$ \citep{Cutler2002, Mastrano-etal2011}. Another potential source of intrinsic asymmetry is the presence of superfluid and superconducting vortices in the stellar interior, which may lead to $\varepsilon \simeq 10^{-8}$ \citep{Jones2010, Melatos-etal2015, Haskell-etal2022} (see also Sec.~\ref{sec:sf}). However, relevant physics remain highly uncertain. Furthermore, accretion of matter from a companion star where composition gradients (strongly influenced by the magnetic-field configuration and accretion rate \citep{Cutler2002, Melatos-etal2005, Singh-etal2020}) can lead to temperature asymmetries that cause \textit{thermal mountains} \citep{Bildsten1998, Ushomirsky-etal2000}. This effect covers a broad range of ellipticities $\varepsilon \simeq 10^{-11} - 10^{-7}$ \citep{Singh-etal2020}. Observational \ac{GW} constraints can thus provide insights into the underlying mechanisms driving the deformation.

Moreover, upper limits on $\varepsilon$ also link to the maximum ellipticity a \ac{NS} can support \citep{Ushomirsky-etal2000, Haskell-etal2006, Gittins-etal2021, Morales-etal2022}. This directly connects to the highly uncertain dimensionless breaking strain, $\sigma$, of \ac{NS} matter, which, e.g., determines crustal failure and associated multi-messenger phenomena outlined in Sec.~\ref{sec:resonant_shattering_flares}. Assuming that the crust is maximally strained, \citep{Ushomirsky-etal2000} estimate that the largest ellipticity the \ac{NS} crust can support is related to $\sigma$ as follows
\begin{align}
    \varepsilon &\approx 1.6 \times 10^{-6}
        \left( \frac{\sigma}{0.1} \right)
        \left( \frac{R_{\rm NS}}{10 \, {\rm km}} \right)^{6.3} \nonumber \\[1.4ex]
        &\times \left( \frac{I_{\rm NS}}{10^{45} \, {\rm g} \,   {\rm cm}^2} \right)^{-1}
        \left( \frac{M_{\rm NS}}{1.4 M_{\odot}} \right)^{-1.2}.
\end{align}
While molecular dynamics simulations that model the crust as a Coulomb crystal suggest that the crust is very strong with $\sigma \sim 10^{-1}$ \citep{Horowitz-etal2009, Caplan-etal2018}, scaling these small-scale simulations to larger volumes remains uncertain. Observational \ac{GW} constraints are, thus, valuable in determining independent information on $\sigma$ and assessing the crustal composition.

Note that the same considerations apply to deformations of possible crystalline phases of hybrid stars or solid quark stars \citep{Owen2005, Haskell-etal2007, Glampedakis-etal2012, Johnson-McDaniel-etal2013}, which are generally able to support larger ellipticities than nucleonic \ac{NS} crusts. For example, \citep{Pereira-etal2023} recently discussed the sensitivity of crustal breaking on the nature of the phase transition between hadronic and quark matter in hybrid stars. Ellipticity limits, hence, also connect to the uncertain dense-matter \ac{EOS} and the stellar core.

To summarize, continuous \ac{GW} signals not only constrain crustal properties of \acp{NS} but can also provide information about the stellar magnetic field and the composition of matter in the \ac{NS} core. However, as outlined above, while \ac{GW} detections provide a direct upper limit on the stellar ellipticity and third-generation detectors have the potential to finally detect these continuous space-time disturbances \citep{Pagliaro-etal2023}, the processes generating \ac{NS} mountains are still highly uncertain and degeneracies between the source distance and moment of inertia remain (see however Sec.~\ref{sec:cgw}), leading to the classification highlighted in Tab.~\ref{tab:crust_comp}.


\subsubsection{Resonant shattering flares}
\label{sec:resonant_shattering_flares}

As highlighted in Sec.~\ref{sec:tidal_def}, \acp{NS} in a binary system with a second \ac{NS} or a low-mass \ac{BH} are subject to tidal deformations close to the merger. In addition to the enhancement of \ac{GW} emission, this effect could result in the shattering of the \ac{NS} crust, which, in turn, may power \ac{EM} emission before the merger, indirectly probing internal physics. This process has been proposed as an explanation for gamma-ray flashes, known as \ac{GRB} \emph{precursors} \citep{Murakami1991, Lazzati2005, Troja2010, Zhu2015}, sometimes observed several seconds before short-duration \acp{GRB}.\footnote{We stress, though, that \ac{GRB} precursors are mainly observed in long \acp{GRB} \citep{Lazzati2005}, which are not generally associated with \acp{CBM}. Thus, crustal shattering due to tidal effects cannot be the only path to generating these precursors. We, however, mention that recent observations suggest that a subset of long \acp{GRB} may also also originate from \acp{CBM}. For an example of a precursor in one of these ``anomalous'' long \acp{GRB} see \citep{Dichiara2023}.} While a single tidal interaction is likely too weak to shatter the crust several seconds prior to merger, periodic tidal forces can enter into resonances with stellar oscillation modes during the inspiral phase \citep{Tsang2012, Tsang2013, Neill2022}. \citep{Tsang2012} first identified the spheroidal $l=2$ $i$-mode as an ideal candidate for this resonance, which causes the mode to grow and releases energy into the crust. Once the stressed solid reaches its elastic limit (see Sec.~\ref{sec:GW-crustpropr}), the crust shatters. The corresponding energy is transferred to the \ac{NS} magnetic field, perturbing it in such a way as to generate an intense electric field that is able to accelerate the particles powering the gamma-ray flare. A necessary condition to attain the required acceleration is the presence of strong magnetic fields ($\gg 10^{13}\, \mathrm{G}$) at the surface of the \ac{NS} \citep{Neill2022}. This requirement prevents precursors to occur in all \ac{GRB} events (in agreement with observations). We also note that confirming resonant shattering flares as origins of \ac{GRB} precursors would imply the persistence of strong $B$-fields in old \acp{NS}, thus, setting constraints on the long-term evolution of the stellar magnetic field.

\emph{$i$-modes}, or \emph{interface modes}, originate from the presence of the interface between the liquid core and the solid crust. As a result, they are strongly influenced by the parameters describing the  dense-matter \ac{EOS}. In particular, $i$-modes are sensitive to the \emph{nuclear symmetry energy},\footnote{For a recent review on the nuclear symmetry energy see, e.g., \citep{BaldoBurgio2016}.} which quantifies the change in binding energy per nucleon as the proton-neutron asymmetry increases. This dependence is noteworthy as macroscopic quantities, like the mass and radius, and characteristic frequencies of other quasi-normal modes like the $f$-mode, $p$-modes or $g$-modes (see Sec.~\ref{sec:gw_asteroseismology_mass}), primarily depend on the \ac{EOS} at the largest \ac{NS} densities. As the composition in these regions is highly uncertain (and unlikely to be purely nucleonic), these observables provide only limited information about the nuclear symmetry energy. In contrast, at the crust-core interface, matter is purely nucleonic and resonant shattering excited by $i$-modes could, thus, probe this uncertain aspect of the \ac{EOS} \citep{Neill2021, Neill2023}. However, to constrain the symmetry energy, detailed information about the mode frequency is necessary. This could be obtained by measuring the \ac{GW} frequency of the chirp signal precisely at the moment when the resonant shattering flare occurs \citep{Neill2021, Neill2023}. The requirement of a simultaneous detection of \ac{GW} and gamma-ray signals establishes this method as a truly multi-messenger technique.

Clearly, this method requires that short \ac{GRB} precursors are powered by the mechanism outlined above. However, different mechanisms have been proposed to explain this phenomenon that do not involve the shattering of the \ac{NS} crust (see, e.g., \citep{Bernardini2013} for an alternative scenario that invokes intermittent accretion onto a newly born magnetar). A detection of the \ac{EM} signal preceding a \ac{GW} merger signal would, however, serve as confirmation of pre-merger models (like shattering flares models) for \ac{GRB} precursors over models based on activity from the merger remnant. Moreover, even if powered by resonant crust shattering, different modes, such as the $f$-mode or $r$-modes might also be responsible for crustal failure \citep{SuvorovKokkotas2020} further complicating the comparison between data and theoretical models. Finally, we also point out that current models of crustal shattering typically neglect other pieces of physics that are likely relevant for the dynamics. This, e.g., includes the presence of nuclear pasta phases close to the crust-core interface (see Sec.~\ref{sec:ns-structure}), which would likely result in a frequency shift of the $i$-mode but leave its qualitative nature unaltered, or the possibility of plastic crustal flow \citep{LevinLyutikov2012, Lander2016, LanderGourgouliatos2019}. The latter, if present, could significantly impact on the $i$-modes dynamics. For these reasons, we classify this method as strongly model dependent in Tab.~\ref{tab:crust_comp}.


\subsubsection{Quasi-periodic oscillations in magnetar giant flares}
\label{sec:giantflares_qpo_crust}

\acp{QPO} are observed during giant flares in magnetars (see also Secs.~\ref{sec:QPOs-compactness} and \ref{sec:QPOs-Bfield}). Based on theoretical calculations, these oscillations were initially thought to be connected with torsional shear modes of the stellar crust. In particular, the frequency of the fundamental mode can be estimated as \citep{Turolla-etal2015}
\begin{equation}
	f_{\rm s}  \simeq \frac{v_{\rm s}}{2 \pi R_{\rm NS}},
\end{equation}
where $v_{\rm s} \equiv \sqrt{\mu_{\rm s} / \rho}$ is the crustal shear speed with $\mu_{\rm s}$ denoting the shear modulus. $\mu_{\rm s}$ is comparable to the Coulomb energy per unit volume of a crustal lattice site, thus introducing a significant dependence on crustal microphysics. Using an estimate from \citep{Piro2005} for the shear modulus based on parameters from \citep{DouchinHaensel2001, Strohmayer-etal1991}, we deduce for the inner crust
\begin{align}
	f_{\rm s} &\approx 17.4 \left( \frac{\rho}{10^{14} \, {\rm g} \, {\rm cm}^{-3}} \right)^{1/6} \left( \frac{Z}{38} \right)
			\left( \frac{A}{302} \right)^{-2/3} \nonumber \\[1.4ex]
		&\times  \left( \frac{1 - X_{\rm n}}{0.25} \right) \left( \frac{R_{\rm NS}}{10 \, {\rm km}}\right)^{-1}  \, {\rm Hz}.
			\label{eqn:shear-QPO}
\end{align}
Here, $Z$ and $A$ denote the proton and total baryon number per lattice site, respectively, and $X_{\rm n}$ the fraction of neutrons that forms the free gas around the lattice nuclei (Sec.~\ref{sec:ns-structure}).

The value~\eqref{eqn:shear-QPO} is close to observed low-frequency \acp{QPO}. The association between measured frequencies and crustal shear modes, thus, provides constraints on dense matter. It, for example, allows us to probe the pasta phases at the base of the crust (see Fig.~\ref{fig:ns_structure}), since the shape of the pasta affects the solid's elastic properties, i.e., the shear modulus, $\mu_{\rm s}$. In the case of cylindrical crustal nuclei, \citep{Sotani2011}, e.g., finds a reduction in $\mu_{\rm s}$ and the shear mode frequency (see however \citep{PassamontiPons2016}). Moreover, by identifying the $29 \,$Hz \ac{QPO} frequency in the 2004 giant flare of SGR 1806--20 \citep{StrohmayerWatts2006} with the fundamental shear mode, \citep{SteinerWatts2009, Sotani-etal2012, Sotani-etal2018} set constraints on the nuclear symmetry energy. Further \ac{EOS} constraints have been obtained by modeling crustal shear modes in \ac{GR}. As highlighted in Sec.~\ref{sec:QPOs-compactness}, this introduces additional $R_{\rm NS}$ and $M_{\rm NS}$ dependencies, specifically connecting relevant mode frequencies with the radial extent of the \ac{NS} crust, $\Delta R$, \citep{SamuelssonAndersson2007}, which in turn is controlled by the \ac{EOS}.

More recent studies have, however, shown that crustal vibrations are strongly damped as the $B$-field couples crust oscillations to the Alfv\'en continuum in the \ac{NS} core \cite{Levin2006, Glampedakis-etal2006, Gabler-etal2012} (see also Sec.~\ref{sec:QPOs-Bfield}). As a result, pure crustal shear modes are likely suppressed in magnetars, and replaced by combined magneto-elastic oscillations. Although these are similar in frequency to the shear modes outlined above, corresponding mode frequencies no longer depend on crustal properties only but also the stellar magnetic field. \citep{Gabler-etal2018}, for instance, find for the lowest-order low-frequency oscillation in the combined crust-core system (for a fixed \ac{EOS} and \ac{NS} mass):
\begin{align}
	f_{0} &\simeq 2.8 X_{\rm SF}^{-0.55}
		\left( \frac{v_{\rm s}}{1.37 \times 10^8 \, {\rm cm} \, {\rm s}^{-1}} \right)^{1/2} \nonumber \\[1.4ex]
		&+ 0.66 X_{\rm SF}^{-0.33}
		\left( \frac{B}{10^{14} \, {\rm G}} \right) \, {\rm Hz},
		\label{eqn:me_fundamental}
\end{align}
where the parameter $X_{\rm SF}$ accounts for \ac{NS} superfluidity (see Sec.~\ref{sec:magnetarqpos_sf} for details). Note that the numerical values here are obtained from fits to a specific set of simulations. Associating the observed high-frequency \acp{QPO} with magnetically modified crustal overtones as highlighted in Eq.~\eqref{eqn:Bfield_mod_fcrust} sets further combined constraints on the magnetic field and the crustal \ac{EOS}. Connecting both expressions~\eqref{eqn:me_fundamental} and \eqref{eqn:Bfield_mod_fcrust} with SGR 1806--20 observations, \citep{Gabler-etal2018} use a Bayesian framework to deduce a relatively high shear speed of $v_{\rm s} \gtrsim 1.4 \times 10^8 \, {\rm cm} \, {\rm s}^{-1}$ and a large crustal thickness of $\Delta R \approx 1.61$ at $68\%$ credibility level together with $B \approx 2.1 \times 10^{15} \, {\rm G}$ for the mean magnetic-field strength.

Although this approach holds promise to constrain highly uncertain crust physics, the crucial issue remains the correct identification of an observed mode frequency with the underlying oscillation mechanism. Moreover, crust and core physics are likely strongly coupled in magnetars, implying that crustal microphysics are difficult to disentangle from magnetic field properties. Combined with potential issues in extracting \ac{QPO} properties from observations (Sec.~\ref{sec:QPOs-compactness}), this leads us to classify this method as orange in Tab.~\ref{tab:crust_comp}.


\subsubsection{Pulsar glitches}
\label{sec:glitches-crustprop}

We introduced the glitch phenomenon in Sec.~\ref{sec:glitches-MoI} in the context of moment-of-inertia measurements. However, in addition to probing macroscopic quantities, sudden spin-ups also provide information about the small-scale properties of the inner \ac{NS} crust. This is due to the fact that glitches are intimately linked to the dynamics of superfluid vortices (see Sec.~\ref{sec:sf} for superfluid microphysics). These structures with diameters on the order of $100 \,$fm exist since superfluids cannot rotate like classical fluids but instead quantize their circulation \citep{Onsager1949, Feynman1955}. Consequently, interactions of individual vortices and their environment are indirectly reflected in observed glitch properties, best measured in the radio band.

The key assumption for building up a superfluid angular momentum reservoir is that vortices can \text{pin} to the crustal lattice (and potentially also to superconducting proton fluxtubes in the \ac{NS} core \citep[e.g.,][]{Sidery2009, WoodGraber2022}), effectively impeding their motion and preventing the superfluid from spinning down on large scales. As the rest of the star slows down, the rotational lag between superfluid and normal matter grows, leading to a hydrodynamical lift force (the so-called \textit{Magnus force} \citep[e.g.,][]{Andersson-etal2006, Glampedakis-etal2011}) on small scales. Vortices unpin and move outwards once the lag is sufficiently large for the lift force to overcome the pinning threshold. Although the exact details of what follows are uncertain, this process drives an avalanche-like event unpinning millions of vortices in the inner crust. This allows the superfluid to spin down and exchange angular momentum with the crust, producing the glitch. The maximum pinning force between a vortex and a lattice site is, hence, closely connected to the size of the angular momentum reservoir and the maximum glitch a pulsar can exhibit \citep{Pizzochero2011, Haskell-etal2012, Pizzochero-etal2017}. Monitoring glitch statistics \citep{Melatos-etal2008, Fuentes-etal2019}, thus, informs theoretical calculations of relevant pinning forces, depending for example on the sizes of the crustal lattice nuclei and the vortices \citep[e.g.,][]{Alpar1977, Epstein1988, Donati2006, Link2009, Seveso-etal2016}. Spin-up statistics are also an important tool to refine our understanding of the glitch trigger, which has been connected with, e.g., inhomogeneous vortex distributions \citep{Cheng-etal1988} and superfluid instabilities \citep{Andersson-etal2003, Mastrano2005}.

Once vortices are unpinned, frictional forces acting on the freed vortices and mechanisms governing their repinning control the post-glitch response on various timescales. Within hydrodynamical models, the coupling between different stellar components and, consequently the glitch morphology, are regulated by \textit{mutual friction}, the coarse-grained dissipation between individual vortices and their environment \citep[e.g.,][]{Andersson-etal2006, Glampedakis-etal2011}. Using a two-fluid approach has allowed the qualitative description of the full evolution of the Vela pulsar and other \acp{NS} showing giant glitches for realistic assumptions on the strength of mutual friction \citep{Haskell-etal2012}. A three-component variant was recently proposed by \citep{Graber-etal2018} (see also \citep{Pizzochero-etal2020}), specifically targeting the glitch rise on minute-long timescales. A comparison of this model with the first observation that caught a glitch in the act (the 2016 Vela pulsar glitch \citep{Palfreyman-etal2018}) provided new information on internal coupling strengths \citep{Ashton-etal2019}, highlighting the potential of this approach in constraining the unknown properties of the \ac{NS} crust (see also \citep{Gugercinoglu-etal2020, Montoli-etal2020, Sourie2020}).

Finally, observed post-glitch relaxation timescales on the order of weeks to months \citep[e.g.,][]{Dodson-etal2007, Espinoza-etal2011, Yu-etal2013, Lower-etal2021} have been used to estimate crustal temperatures within so-called \textit{vortex-creep} models \citep[e.g.,][]{Anderson1975, Alpar-etal1984a, Link1996, Gugercinoglu2014, Gugercinoglu2016}. Instead of focusing on mutual-friction forces as the driver for post-glitch dynamics, these models assume that superfluid vortices are imperfectly pinned and thermal activation allows a certain fraction to hop outwards to neighboring pinning sites as the star spins down. Following a glitch, the creep dynamics in the crust are altered, affecting the stellar spin-down rate and driving the observable relaxation. As these dynamics are sensitive to the temperature of the creep region, a comparison between the model and exponential relaxation timescales observed after Vela pulsar glitches has allowed rough constraints of the internal temperature \citep{Alpar-etal1984a}. Long-term relaxations are, however, not only accounted for within vortex-creep models. \citep{Haskell-etal2020} recently showed that the post-glitch behavior of the Vela pulsar and PSR J0537--6910 can similarly be explained by the dynamics of a turbulent tangle of superfluid vortices. This highlights how sensitive current constraints on \ac{NS} properties are to the underlying models.

In conclusion, while pulsar glitches are ultimately controlled by the interactions of individual vortices and their environments on small scales, thus indirectly probing the properties of the \ac{NS} crust (and to a lesser extent those of the \ac{NS} core \citep[e.g.,][]{Sidery2009, Haskell-etal2012, Gugercinoglu2016, Graber-etal2018, Sourie2020}), the glitch phenomenon encodes the dynamics of the entire star. Theoretical glitch models hence rely on averaging many orders of magnitude to bridge the gap between vortex scale and stellar scale. As a result, inferred parameters are strongly model dependent, leading to the classification shown in Tab.~\ref{tab:crust_comp}.


\subsection{Superfluidity}
\label{sec:sf}

\begin{figure}
    \centering
    \includegraphics[width = 0.45\textwidth]{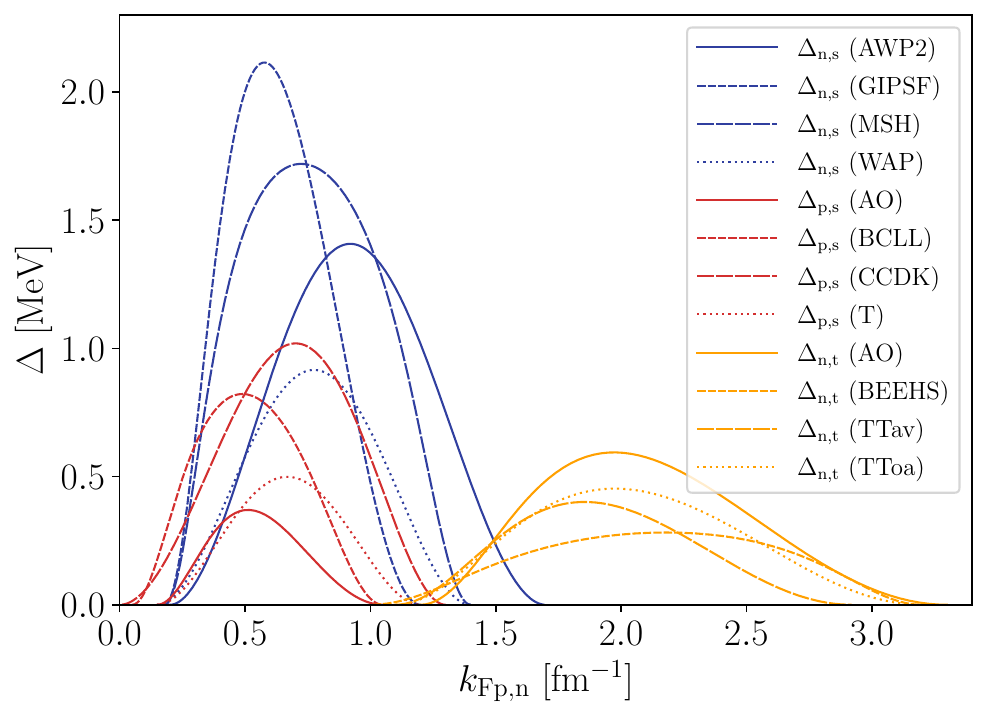}
    \includegraphics[width = 0.48\textwidth]{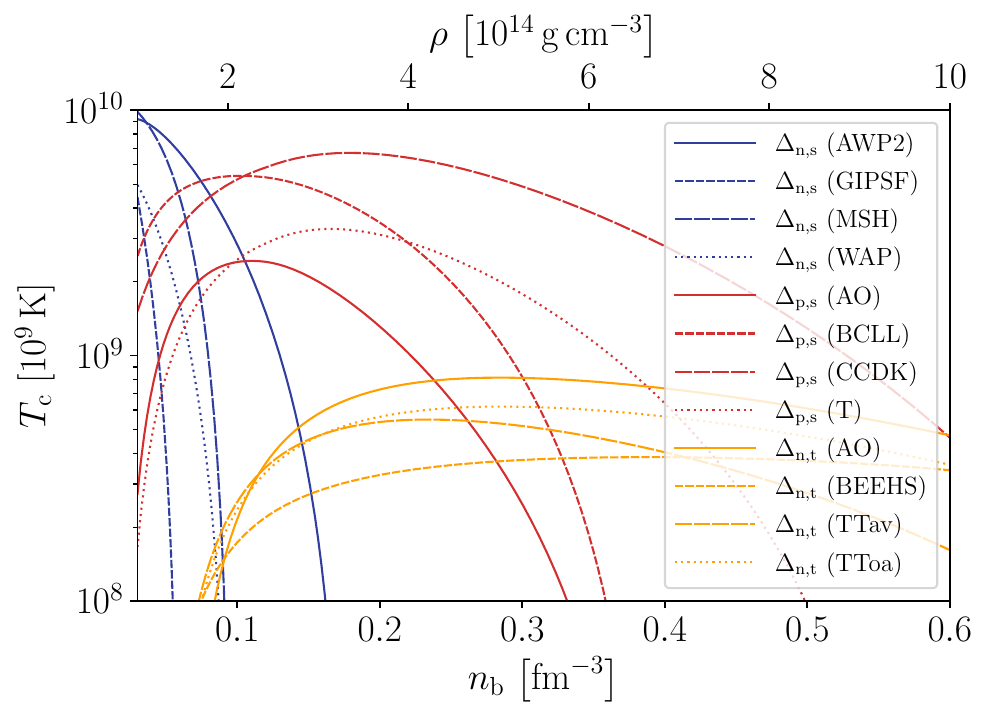}
    \caption{Superfluid energy gaps, $\Delta$, as a function of Fermi wave number, $k_{\rm F}$, (top panel) and the corresponding transition temperatures, $T_{\rm c}$, as a function of baryon density, $n_{\rm b}$, and mass-energy density, $\rho$, (bottom panel) for the three quantum phases in the \ac{NS} interior. We illustrate a representative range of parameterized energy gaps with fit parameters from \citep{Ho-etal2015b} (see this reference for more information on specific gaps shown in the legends): Curves for the singlet-paired neutrons in the \ac{NS} crust are shown in blue (labeled n,s), gaps for the singlet-paired protons in the core in red (labeled p,s) and those for the triplet-paired core neutrons in orange (labeled n,t). Within BCS theory, the energy gap can be directly related to the transition temperature. In particular, in the zero-temperature limit, the two quantities are approximately related as $k_{\rm B} T_{\rm c} \approx 0.567 \Delta$ for singlet pairing and $k_{\rm B} T_{\rm c} \approx 0.118 \Delta$ for triplet pairing \citep{Ho-etal2015b}. To show the density dependence of $T_{\rm c}$ throughout the \ac{NS} core, we specifically considered the NRAPR \ac{EOS} \citep{Steiner-etal2005}.}
    \label{fig:gaps_and_Tcs}
\end{figure}

\begin{table*}[]
    \centering
    \begin{tabular}{|c||c|c|c|c|c|}
        \hline
        \textbf{Method} & \textbf{GWs} & \textbf{Radio} & \textbf{Optical}  & \textbf{X-ray} & $\mathbf{\gamma}$\textbf{-ray}  \\
        \hline
        \hline
        \hyperref[sec:glitches-sf]{Pulsar glitches} & & \cellcolor{orange!40} &  & \cellcolor{orange!40} & \cellcolor{orange!40} \\
        \hline
        \hyperref[sec:cooling_sf]{Cooling} &  & &  & \cellcolor{orange!40}  &    \\
        \hline
        \hyperref[sec:accretingNScooling_sf]{Thermal post-outburst relaxation} &   & &  & \cellcolor{orange!40}  &   \\
        \hline
        \hyperref[sec:magnetarqpos_sf]{Giant-flare QPOs} &   & &  & \cellcolor{orange!40}  &   \\
        \hline
    \end{tabular}
    \caption{Superfluidity measurements. Individual rows represent the methods that probe \ac{NS} superfluidity, while columns denote different messengers/EM wavebands. Individual colors are equivalent to those in Tab.~\ref{tab:mass}.}
    \label{tab:sf}
\end{table*}


We previously highlighted the role of superfluid pulsar glitches when constraining the \ac{NS} moment of inertia (Sec.~\ref{sec:glitches-MoI}) and properties of the stellar crust (Sec.~\ref{sec:glitches-crustprop}). We now provide additional information on superfluid microphysics and relevant observational constraints thereof.

The suggestion that nucleons in \ac{NS} interiors could form large-scale quantum condensates was put forward in \citep{Migdal1959} almost a decade before the first detection of radio pulsars \citep{Hewish-etal1968}. While the properties of matter at such high densities remain uncertain, we suspect that \ac{NS} older than a few hundred years are sufficiently cold \citep{Yakovlev-etal2001, YakovlevPethick2004, Page-etal2006, Potekhin-etal2015} for the neutrons in the inner crust and the neutrons and protons in the \ac{NS} core to become unstable to the formation of Cooper pairs. This is analogous to electronic superconductors on Earth \citep{Bardeen-etal1957}. Generalizing the corresponding theory to \acp{NS} suggests that crustal neutrons and core protons experience pairing in a spin-singlet, $s$-wave state with vanishing angular momentum ($^1 S_0$), while the core neutrons pair in a spin-triplet, $p$-wave channel of non-zero angular momentum ($^3 P_2$) \citep[e.g.,][]{Sauls1989}. The quantities of interest are the so-called \textit{pairing gaps}, $\Delta$ (half the energy required to break a Cooper pair; see Fig.~\ref{fig:gaps_and_Tcs}), which are directly related to the transition temperatures, $T_{\rm c}$, for the onset of superfluidity. These gaps control a number of astrophysically relevant properties of \acp{NS} (see \citep{Haskell2018} for a recent review). However, detailed calculations of pairing properties of superfluid neutrons and superconducting protons are difficult due to unknown many-body interactions and the importance of additional physics such as in-medium effects above saturation density \citep[e.g.,][]{Amundsen1985, LombardoSchulze2001, Schwenk-etal2003, Dean2003, Gezerlis-etal2014, Drischler-etal2017, Lim2021}. 

In this section, we, therefore, focus on those \ac{NS} measurements that provide not only qualitative evidence for \ac{NS} superfluidity but also give quantitative constraints on pairing and superfluid properties even if these are strongly model dependent (see Tab.~\ref{tab:sf}).


\subsubsection{Pulsar glitches}
\label{sec:glitches-sf}

Superfluidity in the interior of \acp{NS} was initially invoked by \citep{Baym-etal1969} to explain the month-long relaxation timescales observed after the first Vela glitch \citep{Radhakrishnan1969, Reichley1969}, its inviscid nature providing a natural justification for weak internal interactions manifest in the long-term relaxation. This notion gained further credence with laboratory experiments using superfluid helium demonstrating similar behavior \citep{Tsakadze1980}.

As discussed in Sec.~\ref{sec:glitches-crustprop}, glitch dynamics are ultimately controlled by the physics of individual vortices and, hence, sensitive to pairing properties. In the standard theory for laboratory superconductors \citep{Bardeen-etal1957}, the diameter of a vortex corresponds to the coherence length, $\xi$, of the superfluid condensate and is, therefore, connected to the energy gap as
\begin{align}
    \xi = \frac{\hbar v_{\rm F}}{\pi \Delta} &\approx 1.2 \times 10^{-12}
            \left( \frac{\Delta}{1 \, {\rm MeV}} \right)^{-1} \nonumber \\[1.4ex]
    &\times \left( \frac{v_{\rm F}}{6 \times 10^{-9} \, {\rm cm} \, {\rm s}^{-1}} \right)
            \, {\rm cm},
\end{align}
where $v_{\rm F}$ is the Fermi velocity. However, measured glitch parameters are typically insensitive to details of the underlying microphysics. Instead, spin-up models invoke the existence of internal superfluids to motivate a multi-component description as outlined in Sec.~\ref{sec:glitches-MoI}. In this coarse-grained framework, small-scale dynamics are averaged out in favor of a model for the entire star, resulting in the loss of information of pairing physics.

As a result, pulsar glitches present strong qualitative evidence for the existence of \ac{NS} superfluidity but current measurements and models are insufficient to provide quantitative constraints on microscopic superfluid properties. We, therefore, classify this method as orange in Tab.~\ref{tab:sf}.


\subsubsection{Cooling of isolated neutron stars}
\label{sec:cooling_sf}

Additional constraints on superfluid pairing can be obtained by observing the cooling of isolated \acp{NS}, whose thermal radiation from their surfaces is observable in the X-rays \citep[e.g.,][]{YakovlevPethick2004, Page-etal2006, Potekhin-etal2020}. As discussed in Sec.~\ref{sec:ns_cooling}, the emission of neutrinos from the stellar interior dominates the cooling of \acp{NS} younger than $\sim 100 \, {\rm kyr}$, while photon emission from the surface plays a crucial role at later times. Extracting temperatures from the X-ray emission of young objects \citep{Ozel2013, Potekhin2014, Potekhin-etal2015, CotiZelati-etal2018}, thus, has the potential to provide insights into the composition of the cooling matter.

The young ($\approx 330 \, {\rm yr}$ old) \ac{CCO} in the supernova remnant Cassiopeia A has been of particular interest in this context, because its surface temperature appears to decrease faster than expected within the standard neutrino cooling scenario \citep{Heinke2010, Yakovlev-etal2011}, which invokes modified Urca processes \citep{Yakovlev-etal2001, YakovlevPethick2004, Page-etal2006, Potekhin-etal2015}. Although, the effect is difficult to detect and requires detailed understanding of telescope systematics and emission processes \citep{PosseltPavlov2022, Plucinsky-etal2022}, its presence hints at unmodeled physics. While a number of interpretations have been put forward \citep[e.g.,][]{Blaschke-etal2012, Noda-etal2013, Negreiros-etal2013, Leinson2014}, such an anomalous temperature drop can be explained naturally by the recent onset of large-scale superfluidity \citep{Page-etal2011, Shternin-etal2011}. While the presence of superfluidity suppresses Urca reactions, it also reduces the matter's heat capacities and results in additional neutrino emission due to the continuous breaking and forming of Cooper pairs close to the superfluid transition temperature. The net effect is faster cooling compared to the scenario without macroscopic quantum phases. Specifically, the size of the observed Cassiopeia A temperature drop allowed a constraint of the maximum critical temperature of the core neutrons to $\sim 5 - 10 \times 10^{8} \, {\rm K}$, while also requiring the protons to already be superconducting, i.e., have an even higher $T_{\rm c, max}$ \citep{Page-etal2011, Shternin-etal2011, Shternin-etal2021, Shternin-etal2023}. Applying the same modeling approach to other young \acp{NS} at the centers of supernova remnants (notably the compact object hypothesized to have formed in the  Galactic supernova SN 1987A) leads to broadly consistent results \citep{Ho-etal2021, Page-etal2020}.

In summary, closely monitoring the cooling behavior of young, isolated \acp{NS} gives quantitative information about the onset of superfluidity in \acp{NS}, provided that the temperature drop is indeed real and caused by a superfluid transition. This approach, hence, allows us to constrain uncertain energy gaps. However, systematics in extracting accurate temperature measurements currently hinder tighter gap constraints, leading us classify this method as orange in Tab.~\ref{tab:sf}.


\subsubsection{Thermal relaxation of quasi-persistent X-ray transients}
\label{sec:accretingNScooling_sf}

The concepts underlying thermonuclear X-ray bursts were discussed in detail in Secs.~\ref{sec:pre} and \ref{sec:X-rayburst_cooling_crustprop}. Monitoring the quiescent emission of those systems that undergo outbursts after extended periods of accretion (quasi-persistent transients) are of particular interest in the context of extracting superfluid parameters.

In particular, if the neutrons in the inner crust are superfluid, their heat capacity is significantly suppressed (as it is the case in isolated \ac{NS}; see Sec.~\ref{sec:cooling_sf}), resulting in faster cooling. In contrast, in those layers where neutrons are normal (non-superfluid), they constitute the dominant contribution to the heat capacity, store more heat and, therefore, delay the thermal relaxation. The shape of the cooling curves of quasi-persistent transients on intermediate timescales is thus sensitive to the nature of the crustal neutrons and observations of enhanced cooling in this intermediate regime suggest that crustal superfluidity is indeed present \cite[e.g.,][]{Brown2009, Shternin-etal2007}. In particular, cooling curves are dependent on `how many' of the crustal neutrons are superfluid, which in turn probes the shape of the pairing gap. As illustrated in the bottom panel of Fig.~\ref{fig:gaps_and_Tcs}, it is unknown at which densities the singlet-pairing gap closes (i.e., when $T_{\rm c}$ drops to zero) and whether it does so before the crust-core boundary located at $\sim 10^{14} {\, \rm g \, cm}^{-3}$ or at higher densities in the core. If the former is the case, this would imply the presence of a normal layer of neutrons between the crustal and the core superfluids, effectively delaying the cooling. \citep{Deibel-etal2017} indeed found evidence for such a late-time thermal relaxation.

Though quasi-persistent transients become increasingly faint and thus difficult to detect, sensitive X-ray telescopes have observed the cooling behavior of a few objects several years post outbursts \citep[e.g.,][]{Cumming-etal2017, Degenaar-etal2021}. Such measurements hold particular promise: the longer we observe the thermal relaxation, the deeper we probe into the stellar interior, i.e., the cooling curve on long timescales is sensitive to \ac{NS} core physics. This is especially powerful if the outburst itself and the subsequent cooling are monitored, as this allows an estimate of the energy deposited into the core and constraints on the \ac{NS} envelope composition, a major model uncertainty. Combined with temperature and neutrino-luminosity estimates, we can therefore get a handle on the core's heat capacity, which in turn sets constraints on the properties of dense matter. While \citep{Cumming-etal2017} used this approach to rule out \ac{NS} cores dominated by a quark color-flavor-locked phase in several systems, \citep{Degenaar-etal2021} were able to show that observations of another source are difficult to reconcile with a significant presence of core superfluidity and superconductivity.

In conclusion, although models are complex and several systematics, such as the \ac{NS} envelope compositions and distances, often difficult to assess (leading to the classification in Tab.~\ref{tab:sf}), long-term thermal relaxations of \acp{NS} that show outbursts after years to decades of accretion provide a unique view into the stellar interior. In particular, monitoring the cooling curve with time allows us to look deeper and deeper into the star providing unique quantitative constraints on the matter's pairing properties.


\subsubsection{Quasi-periodic oscillations in magnetar giant flares}
\label{sec:magnetarqpos_sf}

Superfluidity has a significant impact on the \ac{NS} oscillation spectrum \citep[e.g.,][]{Andersson-etal2009, Chamel2008}. Magnetar \acp{QPO}, hence, also encode information about macroscopic quantum phases in \ac{NS} interiors. In particular, superfluidity plays a crucial role in explaining observations of high-frequency \acp{QPO} \citep{StrohmayerWatts2005}, which are difficult to explain with the physical mechanisms explained in Secs.~\ref{sec:QPOs-compactness}, \ref{sec:QPOs-Bfield} and \ref{sec:giantflares_qpo_crust} alone. Although the crust can, in principle, vibrate at frequencies of several hundred Hz, these oscillations should be strongly damped due to coupling to the core, making it difficult to reconcile observations of long-lasting high-frequency \acp{QPO} (see however \citep{Huppenkothen-etal2014} for a discussion of potential detection biases).

Treating the core superfluid in an effective manner, \citep{Gabler-etal2013, PassamontiLander2014} find that magneto-elastic models with superfluidity can produce oscillations at high frequencies. As a result of superfluidity, core neutrons and protons are coupled, involving a larger fraction of mass in the stellar oscillations. \citep{Gabler-etal2016, Gabler-etal2018} encapsulate this interaction in an effective parameter $X_{\rm sf} \equiv \epsilon_{\star} X_{\rm p}$, where $\epsilon_{\star}$ measures the coupling between the neutrons and protons via the \textit{entrainment effect} (see, e.g., \citep{Andersson-etal2006, Glampedakis-etal2011}) and $X_{\rm p}$ denotes the fraction of protons. In this picture, the mechanism responsible for generating a giant flare excites initial crustal modes, which subsequently resonantly excite a few specific torsional Alfv\'en overtones in the core. This could explains why only a few high-frequency modes are observed and not the full spectrum excited. Moreover, the range of possible high-frequency \acp{QPO} decreases for increasing $X_{\rm sf}$ \citep{Gabler-etal2018}, highlighting that observations of these modes put constraints on superfluid properties.

This is particularly powerful in combination with the impact of superfluidity on the low-frequency magneto-elastic modes, discussed in Sec.~\ref{sec:giantflares_qpo_crust}. Specifically, the parameter $X_{\rm sf}$ also affects the frequency of the fundamental oscillation as given in Eq.~\eqref{eqn:me_fundamental}, obtained from fits to numerical simulations. Combining low- and high-frequency \acp{QPO} from SGR 1806--20 \citep{StrohmayerWatts2005}, \citep{Gabler-etal2018} obtain $X_{\rm sf} \approx 0.09$ at $68\%$ confidence level. This value is close to $0.046$ which is representative of a fully superfluid model \citep{DouchinHaensel2001} and, thus, hints at a significant fraction of the \ac{NS} star being in a superfluid state.

In addition to the uncertainties around accurately identifying oscillation physics with measured \acp{QPO} frequencies, degeneracies between model parameters and observational biases (see Secs.~\ref{sec:QPOs-compactness}, \ref{sec:QPOs-Bfield} and \ref{sec:giantflares_qpo_crust}), current models of superfluid magneto-elastic oscillations neglect a number of relevant physics. The effective prescription based on the parameter $X_{\rm sf}$ introduced above describes the star as a single fluid, neglecting the multi-component nature of \ac{NS} interiors. Moreover, the influence of superfluid vortices as well as the presence of superconductivity have been ignored to date. Advanced simulations that incorporate a more realistic treatment of superfluids in magneto-elastic media (e.g., based on \citep{Carter-etal2006, CarterSamuelsson2006}) will be needed to drive this approach forward. As a result of these caveats, we classified this method as strongly model dependent in Tab.~\ref{tab:sf}.


\subsection{Further multi-messenger measurements}
\label{sec:other_measurements}

To conclude, we focus on a few additional approaches that provide insights into \ac{NS} characteristics but which did not fit with our categories presented in the previous sections. In particular, multi-messenger astrophysics provides the opportunity to study a single celestial source with different messengers, including photons, \acp{GW}, and neutrinos. The advantage of employing multiple messengers is that they offer complementary information, providing unique insights into the source's properties, as exemplified in Sec.~\ref{sec:nsbh_grb}. In this section, we briefly outline two additional angles for deducing \ac{NS} properties from multi-messenger observations.


\subsubsection{Kilonovae}
\label{sec:kilonovae}

The advent of the multi-messenger era offers us several other ways to study \acp{NS} through the different \ac{EM} transients associated with \ac{BNS} and \ac{NS}-\ac{BH} mergers. For recent reviews of \ac{EM} counterparts to \ac{GW} signals, we refer the reader to \citep{Nakar2020, Ascenzi2021}. In particular, during a \ac{CBM}, a fraction of the \ac{NS}'s matter, typically $< 0.1\, M_\odot$ \citep[e.g.,][]{Radice2018}, can be ejected and become gravitationally unbound from the system. This matter, heated by internal radioactive decay, powers the transient optical and near-infrared \ac{CBM} counterpart known as \ac{KN} \citep{Li_Paczynski98, Metzger2010, Roberts2011, Goriely2011, Barnes_Kasen2013} (see \citep{Metzger_kilonovae} for a comprehensive review of theoretical and observational aspects).

The ejection of matter can occur through different channels like tidal forces, shocks at the interface between the two stars (for a \ac{BNS} merger), or accretion disk winds. These ejecta components are characterized by different masses, velocities, and chemical compositions, and consequently different opacities, which determine the observational properties of \acp{KN}, such as their peak luminosity, temperature, and time at which the emission peaks. As these observable parameters, and the relative importance of the different ejecta components, are influenced by characteristics of the binary system, such as its total mass, the components' mass ratio and the \ac{NS} \ac{EOS}, observations of \acp{KN}, thus, encode information on the properties of dense matter \citep{Metzger_kilonovae}.

Moreover, the nature of the merger remnant is affected by the \ac{EOS} and impacts on the \ac{KN}. In particular, whether the merger leads to the prompt formation of a \ac{BH}, the formation of a metastable \ac{NS} that subsequently experiences delayed collapse into a \ac{BH}, or the formation of a stable \ac{NS}, depends not only on the two binary masses but also on the dense-matter \ac{EOS}. For example, stiffer \acp{EOS} produce longer-lived (and eventually stable) \ac{NS} remnants, while softer \acp{EOS} are associated with prompt collapse to a \ac{BH} \citep[e.g.,][]{Piro2017, PatricelliBernardini2020}.

If the remnant is indeed a metastable or stable \ac{NS} rather than a \ac{BH}, emitted neutrinos irradiate the ejecta matter, which increases its proton fraction and prevents the production of lanthanides. Without these lanthanides, which are characterized by high opacities, the \ac{KN} would be bluer in color and evolve faster \citep{MetzgerFernandez2014, Perego2014}. Moreover, the presence of a \ac{NS} remnant is also expected to lead to an increase of ejected matter because additional material can be ablated from the stellar surface due to neutrino irradiation \citep{Duncan1986, QianWoosley1996,Thompson2004, Metzger2007, Dessart2009} or the magnetic pressure from the interior of the \ac{NS} remnant \citep{Siegel2014, Ciolfi2019}. This additional matter-ejection channel enhances the blue component of the \ac{KN} emission at early times (i.e., within $\sim 1 \, {\rm day}$ after the merger) \citep{Metzger2018}. To assess the presence of a \ac{NS} remnant, observing the transient at early times is thus crucial. To this aim, \acp{GW} can act as a trigger to guide optical follow-up observations and help catch the first phase of this transient.

We note that multi-messenger observations of GW170817 have already been exploited to constrain \ac{NS} parameters. For example, \citep{MargalitMetzger2017} combined limits on the ejecta mass from observations of the \ac{GRB} and \ac{KN} following the merger with the system's total mass inferred from the \ac{GW} signal to obtain a limit on the maximum mass of a \ac{NS}. They find $M_{\rm max} = 2.17\,M_\odot$ at $90\%$ confidence level. The idea is that, for a given total mass of the binary, the remnant would have undergone prompt collapse into a \ac{BH} if the maximum mass is not sufficiently large. Such a prompt collapse is inconsistent with the large fraction of lanthanide-free ejecta observed from the \ac{KN}. On the other hand, if the maximum mass is too high, the collapse is sufficiently delayed for the remnant to inject its rotational energy through magnetic dipole emission (see Sec.~\ref{sec:MD_braking}) into the ejecta and/or the \ac{GRB} jet, which is also inconsistent with observations. Interestingly, the above constraint is similar to the mass limit of $M_{\rm max} = 2.16 \,M_\odot$ obtained by \citep{Rezzolla2018} from the \ac{GW} signal along, i.e., without relying on modeling the \ac{EM} counterpart.\footnote{In a follow-up study, \citep{Most2018} combined the limit from \citep{Rezzolla2018} with the tidal deformability inferred from GW170817 (see Sec.~\ref{sec:tidal_deformability_cbm}) to derive a constraint on $R_{\rm NS}$ for a star with $M_{\rm NS} = 1.4\, M_\odot$. They find values between $12.00-13.45 \, \rm km$ at $2-\sigma$ confidence level for purely hadronic \acp{EOS}.}

Finally, the \ac{EOS} also affects the so-called \emph{\ac{KN}-afterglow}, which is an emission powered by the shock of the ejecta onto the interstellar medium \citep{NakarPiran2011, Hotokezaka2013, Hotokezaka2018, Nedora2023}. The properties of this emission, such as the peak flux and peak time, are greatly influenced by the presence and the characteristics of the fast ($\gtrsim 0.7\, c$), mildly-relativistic component of the \ac{CBM} ejecta. This ejecta component, in turn, is launched by shocks induced either by remnant oscillations or at the contact interface between the two merging \acp{NS} \citep{Hotokezaka2013, Bauswein2013}. Both mechanisms are dependent on the \ac{NS} \ac{EOS}, with softer relations associated with faster ejecta \citep{Hotokezaka2018, Nedora2023}, thus, resulting in \ac{EOS}-sensitive \ac{KN}-afterglows.

To summarize, the study of \ac{EM} transients that accompany \acp{CBM} allows us to probe the unknown \ac{EOS} of \acp{NS}. Future \acp{CBM} multi-messenger observations will be crucial to fully explore the potential of this approach, which to date has been limited by the small number of observed events. In the spirit of previous sections, we, however, point out that \acp{KN} are complex and corresponding frameworks strongly model dependent, i.e., many individual ingredients influence their observable features, complicating \ac{EOS} constraints.


\subsubsection{Neutrinos}
\label{sec:neutrinos}

As a final remark, we note that while we have touched upon the indirect impact of neutrinos on \ac{NS} properties (see Secs.~\ref{sec:ns_cooling}, \ref{sec:X-rayburst_cooling_crustprop}, \ref{sec:cooling_sf}, \ref{sec:accretingNScooling_sf}), we have neglected their role as direct messengers for constraining \ac{NS} physics. In particular, we remind the reader that neutrinos are crucial for the secular evolution of \acp{NS}, because they drive \ac{NS} cooling during the first $\sim \mathcal{O}(10\, \mathrm{kyr})$ of the star's life (see \citep{Yakovlev-etal2001, Page-etal2006} for detailed reviews). As different processes are involved in this secular cooling, the ability to distinguish between them would allow us to gain insights into the stellar composition and the onset of superfluidity (see Sec.~\ref{sec:sf}). However, direct detection of these neutrinos is not possible with present and future facilities, because on one hand neutrino energies (which are of the order of the \ac{NS} temperature, i.e., $< 1\,\rm MeV$) are too low to be detected, and on the other they are likely too low in flux (see \citep{Yakovlev-etal2001} for typical neutrino emissivities). Instead, the best way to constrain these processes is indirectly, through the observation of thermal emission in the \ac{EM} wavebands as described in previous sections.

Neutrinos are also emitted during violent events involving \acp{NS} such as \ac{BNS} mergers. Whereas thermal neutrinos are difficult to detect \citep{Kyutoku2018}, the neutrinos produced by non-thermal merger processes may be detectable by facilities like IceCube or KM3net, even though to date no such neutrinos have been observed \citep{Abbasi2022}. However, the origins of these non-thermal neutrinos are most likely connected to the acceleration of hadrons in shocks occurring in relativistic outflows. Although of interest for constraining the physics of particle acceleration and the outflow's composition, we do not expect these neutrinos to directly probe the physics of \acp{NS}.

In the \ac{NS} context, the most interesting sources for the emission of detectable neutrinos are core-collapse supernovae. Supernova SN1897A in the Large Magellanic Cloud, which was detected via its \ac{EM} and neutrino emission \citep{Kunkel1987, Bionta1987}, was the first astrophysical transient accessible with multiple messengers. Such a supernova neutrino signal can provide us with information about the explosion mechanism, and the resulting proto-\ac{NS}. However, covering this topic was beyond the scope of this review and we instead refer the interested reader to \citep{Janka2012} and references therein.


\section{Summary and Conclusions}
\label{sec:conclusion}

The study of \acp{NS} presents fascinating opportunities for scientific exploration. These compact objects, characterized by strong gravity, high densities, large magnetic fields and fast rotation, offer insights into the properties of matter and strong interactions under extreme conditions that we cannot replicate in terrestrial environments. However, as we demonstrated in this review, measuring the characteristics of \acp{NS} and using these to constrain dense-matter physics is a challenging task. We specifically outlined techniques and underlying model assumptions and systematics used to extract global observables, focusing on the \ac{NS} mass (Sec.~\ref{sec:mass}), radius (Sec.~\ref{sec:radius}), moment of inertia (Sec.~\ref{sec:MoI}), tidal deformability (Sec.~\ref{sec:tidal_def}), and compactness (Sec.~\ref{sec:compactness}). This was followed by a discussion of measuring non-global quantities, specifically the \ac{NS} magnetic field (Sec.~\ref{sec:Bfields}), crustal physics (Sec.~\ref{sec:crust_prop}) and superfluidity (Sec.~\ref{sec:sf}).

To overcome individual measurement challenges and advance our understanding of \acp{NS}, the development of independent methods targeting different observables and relying on distinct assumptions is crucial. This approach allows for cross-validation and reduces the reliance on specific models. The exploration of different \ac{NS} classes and associated phenomena (Sec.~\ref{sec:NS-zoo}) using a multi-wavelength and multi-messenger approach, which involves complementary information from observations across the \ac{EM} spectrum as well as the \ac{GW} domain, is particularly useful in forming a more comprehensive view of these objects.

Among the methods discussed, radio timing of pulsars in binary systems stands out. These observations have proven particularly effective in constraining \ac{NS} masses, and might also provide moment-of-inertia measurements in the future. This technique, which leverages the incredible stability of radio pulsar signals, offers several advantages, including high precision and reduced model dependence, compared to other methods discussed. Note that monitoring how young, isolated pulsars divert from their stable behavior and analyzing these glitches in details, provides further information on internal physics. However, underlying model uncertainties currently limit us to using these irregularities primarily for qualitative constraints. With the Square Kilometre Array (expected to detect all visible Galactic radio pulsars and to monitor them with unprecedented precision) having finally begun construction in 2022, we expect radio timing to continue to play an essential role in determining \ac{NS} parameters.

We further highlighted that the \ac{NS} radius is the macroscopic quantities that is the hardest to measure. Many of the methods we presented suffer from strong model dependencies. However, we concluded that the most promising and least model-dependent approach is to measure the radius through X-ray pulse-profile modeling of \acp{MSP}. By analyzing thermal emission patterns on the surfaces of these rapidly rotating \acp{NS}, valuable insights into their sizes can be gained. Monitoring the X-ray emission of other \ac{NS} classes, such as accreting \acp{NS} after outbursts or young isolated \acp{CCO}, to track their thermal evolution, provides complementary information on the stellar interior, although significant uncertainties remain due to the complexity of modeling internal \ac{NS} physics. The same holds for decoding observations of strongly magnetized magnetars after giant flares or individual emission features. Future X-ray missions, such as ATHENA, will  expand the sensitivity limits of existing observatories and allow us to provide better radius constraints and look deeper into the \ac{NS} crust and core. 

In addition to \ac{EM} observations, the recent detections of \ac{BNS} and \ac{NS}-\ac{BH} mergers in the \ac{GW} band have opened up an entirely new avenue to study \acp{NS}. \acp{GW} have specifically enabled new measurements of \ac{NS} masses and first constraints on the tidal deformability. Further dense-matter constraints (although prone to strong uncertainties) have been possible due to the observations of \ac{EM} merger counterparts including short \acp{GRB} and \acp{KN}. In the future, the increased sensitivity of next-generation \ac{GW} interferometers, such as the Einstein Telescope and Cosmic Explorer, holds great promise for further enhancing our understanding of \ac{NS} physics. Achieving higher sensitivity will allow for the detection of more distant signals, significantly increasing the number of detected merger events. This will enable more precise tidal-deformability and mass measurements, complemented by ongoing multi-messenger efforts. Advanced instruments will also finally enable the analysis of currently unexplored aspects of \ac{NS} physics, such as the emission of continuous \acp{GW}, and allow us to take full advantage of asteroseismology to measure global and non-global properties of \acp{NS}.

In summary, studying \acp{NS} is a complex and challenging endeavor that offers remarkable opportunities for scientific advancement. By employing a variety of measurement techniques, embracing a multi-wavelength approach, and taking advantage of \ac{GW} astronomy, we can deepen our understanding of these intriguing objects. The development of new observational capabilities, combined with a push for better theoretical models, holds great promise for shedding light on the (as yet unknown) physics of these extreme stars.


\section*{Author contributions}

S.~A. and V.~G. contributed equally to this work.


\section*{Acknowledgments}

The authors thank Emilie Parent and Andrea Possenti for helpful conversations about the timing of radio pulsars, and Francesco Iacovelli for fruitful discussion about measuring the tidal deformability with third generation \ac{GW} detectors. The authors also thank Emmanuel Fonseca, Michael Gabler, Fabian Gittins, Bryn Haskell, Wynn Ho, Christian Kr\"uger, Andrea Maselli, Gor Oganesyan and David Tsang for providing feedback on the manuscript. N.~R. also thanks the organizers (Ingo Tews, Samaya Nissanke, Bruno Giacomazzo, and Jerome Margueron) and the participants of the ECT* workshop ``Neutron stars as multi-messenger laboratories for dense matter'' (June 2021) for triggering the idea behind this review, and Anna Watts for contributing to an early version of the \ac{NS} measurement tables presented in this work. Finally, S.~A., V.~G. and N.~R. acknowledge support from the European Research Council (ERC) via the Consolidator Grant ``MAGNESIA'' under grant agreement No. 817661 (PI: Rea), partial support from grant SGR2021-01269 (PI: Graber) and the program Unidad de Excelencia Mar\'ia de Maeztu CEX2020-001058-M. S.~A. also acknowledges the Pro3 grant from MUR. V.~G. further acknowledges support from a Juan de la Cierva Incorporaci\'on Fellowship.
\appendix


\section*{Acronyms and variable definitions}

We have defined the following acronyms in this work: \vskip 0.2cm

\tablefirsthead{\toprule}
\tabletail{}
\begin{supertabular}{ll}
    \hline
         \ac{BH} & Black Hole \\
         \ac{BNS} & Binary Neutron Star \\
         \ac{CBM} & Compact Binary Merger \\
         \ac{CCO} & Central Compact Object \\
         \ac{CFS} & Chandrasekhar-Friedman-Schutz \\
         \ac{CRSF} & Cyclotron Resonance Scattering Feature \\
         \ac{EM} & Electromagnetic \\
         \ac{EOS} & Equation of State \\
         \ac{GR} & General Relativity \\
         \ac{GRB} & Gamma-Ray Burst \\
         \ac{GW} & Gravitational Wave \\
         \ac{HMXB} & High-Mass X-ray Binary \\
         \ac{ISCO} & Innermost Stable Circular Orbit \\
         \ac{KN} & Kilonova \\
         \ac{LMXB} & Low-Mass X-ray Binary \\
         \ac{LOS} & Line of Sight \\
         \ac{MSP} & Millisecond Pulsar \\
         \ac{NICER} & Neutron Star Interior-Composition Explorer \\
         \ac{NS} & Neutron Star \\
         PN & Post-Newtonian \\
         \ac{PRE} & Photospheric Radius Expansion \\
         \ac{QPO} & Quasi-Periodic Oscillation \\
         \ac{RRAT} & Rotating Radio Transient \\
         \ac{SNR} & Signal-to-Noise Ratio \\
         \ac{SuNR} & Supernova Remnant \\
         \ac{WD} & White Dwarf \\
         \ac{XDINS} & X-ray Dim Isolated Neutron Star \\
    \hline
\end{supertabular} 

\vspace{0.5cm}

We further defined the following variables in the review:  \vskip 0.2cm

\tablefirsthead{\toprule}
\tabletail{}
\begin{supertabular}{ll}
    \hline
    $\rho_0$ & nuclear saturation density\\
    $P$ & \ac{NS} spin period\\
    $\dot{P}$ & time derivative of \ac{NS} spin period\\
    $\tau_c$ & \ac{NS} characteristic age\\
    $B_\mathrm{dip}$ & \ac{NS} surface dipolar magnetic field strength\\
    $E_\mathrm{rot}$ & \ac{NS} rotational energy\\
    $L_\mathrm{rot}$ & \ac{NS} rotational luminosity ($|\dot{E}_\mathrm{rot}|$)\\
    $I_\mathrm{NS}$ & \ac{NS} moment of inertia\\
    $T_\mathrm{BB}$ & black-body temperature\\
    $\rho_\mathrm{NS}$ & neutron drip density\\
    $M_\mathrm{NS}$ & \ac{NS} (gravitational) mass\\
    $f_\mathrm{bin}$ & binary system mass function\\
    $M_\mathrm{c}$ &  mass of the binary companion\\
    $i$ & binary system inclination angle\\
    $K_\mathrm{c}$ & radial velocity of the binary companion\\
    $P_\mathrm{orb}$ & binary system orbital period\\
    $e$ & binary system eccentricity\\
    $K_\mathrm{NS}$ & \ac{NS} radial velocity\\
    $a$ & binary orbital separation\\
    $\dot{\omega}$ & periastron advance\\
    $\dot{P}_\mathrm{orb}$ & rate of orbital decay\\
    $\gamma$ & Einstein delay\\
    $M$ & total mass of the binary system  \\
    $r$ & Shapiro delay range\\
    $s$ & Shapiro delay shape\\
    $a_1$ & semi-major axis of the orbit\\
    $x$ & projected semi-major axis\\
    $f_\mathrm{GW}$ & \ac{GW} frequency\\
    $\dot{f}_\mathrm{GW}$ & \ac{GW} frequency time derivative\\
    $\mathcal{M}$ & chirp mass\\
    $\omega_{f,w,r}$ & ($f,w,r$)-mode frequency\\
    $\tau_{f,w}$ & ($f,w$)-mode damping time\\
    $T$ & temperature\\
    $C_v$ & heat capacity \\
    $f_\mathrm{amp}$ & amplified flux\\
    $R_\mathrm{e}$ & Einstein ring radius\\
    $R_\mathrm{NS}$ & \ac{NS} radius\\
    $R_\infty$ & \ac{NS} apparent radius\\
    $D$ & source distance\\
    $F_\infty$ & bolometric flux\\
    $C_\mathrm{NS}$ & \ac{NS} compactness\\
    $F_\mathrm{Edd}$ & Eddington flux\\
    $L_\mathrm{Edd}$ & Eddington luminosity\\
    $X_{\rm e}$ & electron fraction\\
    $T_\mathrm{col}$ & color temperature\\
    $g_\mathrm{eff}$ & effective surface gravity\\
    $A_\infty$ & apparent area\\
    $f_\mathrm{c}$ & color factor\\
    $N_H$ & hydrogen column density\\
    $E_\mathrm{GRB}$ & \ac{GRB} energy\\
    $M_\mathrm{disk}$ & accretion disk mass\\
    $a_\mathrm{BH}$ & dimensionless \ac{BH} spin parameter\\
    $\epsilon_\mathrm{jet}$ & jet-launching efficiency\\
    $M_\mathrm{BH}$ & \ac{BH} mass\\
    $\mathbf{S}$ & spin 
    angular momentum\\
    $\mathbf{L}$ & orbital 
    angular momentum \\
    $P_\mathrm{precession}$ & precession period \\
    $\Omega$ & angular velocity \\
    $I_\mathrm{sf}$ & superfluid moment of inertia \\
    $\omega_\mathrm{c}$ & critical angular velocity lag\\
    $t_\mathrm{glitch}$ & inter-glitch time\\
    $\langle \mathcal{A} \rangle$ & average glitch activity parameter\\
    $I_3$ & moment of inertia along rotation axis\\
    $\epsilon$ & \ac{NS} ellipticity\\
    $h_0$ & \ac{GW} amplitude\\
    $n$ & braking index\\
    $\mathcal{E}_{i,j}$ & tidal field\\
    ${Q}_{i,j}$ & quadrupole moment\\
    $\lambda_\mathrm{NS, c}$ & tidal deformability\\
    $k_{2}$ & $l=2$ tidal love number\\
    $\tilde{\Lambda}$ & binary's tidal deformability parameter\\
    $\Lambda_\mathrm{NS,c}$ & dimensionless tidal deformability\\
    $\chi$ & dimensionless spin parameter\\
    $z$ & gravitational redshift\\
    $f$ & QPO frequency during giant flare\\
    $\psi$ & angle between magnetic and rotation axis\\
    $E_\mathrm{cyc}$ & energy of cyclotron line\\
    $n_L$ & Landau level number\\
    $B$ & magnetic field strength \\
    $f_\mathrm{A}$ & frequency of Alfv\'en oscillation\\
    $v_\mathrm{A}$ & Alfv\'en velocity\\
    $\rho_\mathrm{c}$ & (mass) density of charged particles\\
    $f_\mathrm{crust}$ & frequency of crustal oscillations\\
    $L_\mathrm{neutrino}$ & neutrino luminosity\\
    $L_\mathrm{photon}$ & photon luminosity\\
    $L_\mathrm{crustal\, heating}$ & heat generated in X-ray bursts \\
    $Q$ & heating per accreted nucleon\\
    $\langle \dot{M} \rangle$ & time-averaged accretion rate\\
    $f_\mathrm{s}$ & frequency of torsional shear mode\\
    $v_\mathrm{s}$ & crustal shear speed\\
    $\mu_\mathrm{s}$ & shear modulus\\
    $\rho$ & mass density\\
    $Z$ & atomic number\\
    $A$ & mass number\\
    $X_\mathrm{n}$ & neutron fraction\\
    $\xi$ & superfluid coherence length\\
    $v_F$ & Fermi velocity\\
    $\Delta$ & pairing energy gap\\
    $T_\mathrm{c}$ &critical temperature for superfluidity\\
    $k_\mathrm{F}$ &Fermi wave number\\
    $n_\mathrm{b}$ & baryon number density\\
    $\epsilon_\star$ & entrainment parameter\\
    $X_\mathrm{p}$ & proton fraction\\
    $M_\mathrm{max}$ & maximum mass of a \ac{NS}\\
    \hline
\end{supertabular}


\bibliographystyle{model1a-num-names_mod}

\bibliography{bib_review_new_long_journals}


\end{document}